\documentclass[fleqn,usenatbib,usedcolumn]{mnras}

\usepackage[british]{babel}
\usepackage{newtxtext}
\usepackage[slantedGreek]{newtxmath}
\let\la=\lesssim

\usepackage[T1]{fontenc}
\usepackage{ae,aecompl}

\usepackage{graphicx}
\usepackage{amssymb}
\usepackage{amsmath}

\title[Downsizing of haloes hosting infrared sources]{Downsizing of Star Formation Measured from the Clustered Infrared Background Correlated with Quasars}

\author[K. R. Hall et al.]{Kirsten R. Hall,$^{1}$
Devin Crichton,$^{2}$
Tobias Marriage,$^{1}$
Nadia L. Zakamska,$^{1}$
\newauthor{Rachel Mandelbaum$^{3}$}
\\
$^{1}$Department of Physics and Astronomy, Johns Hopkins University, Baltimore, MD 21218, USA \\
$^{2}$Astrophysics and Cosmology Research Unit, School of Mathematics, Statistics and Computer Science, \\
University of KwaZulu--Natal, Durban 4041, South Africa \\
$^{3}$McWilliams Center for Cosmology, Department of Physics, Carnegie Mellon University, Pittsburgh, PA 15
213, USA}

\date{2 August 2018}

\pubyear{2018}

\begin{document}
\label{firstpage}
\pagerange{\pageref{firstpage}--\pageref{lastpage}}
\maketitle

\begin{abstract}
Powerful quasars can be seen out to large distances. 
As they reside in massive dark matter haloes, they provide a useful tracer of large scale structure. We stack \textit{Herschel}-SPIRE images at 250, 350 and 500 microns at the location of 11,235 quasars in ten redshift bins spanning $0.5\leq z \leq 3.5$.
The unresolved dust emission of the quasar and its host galaxy dominate on instrumental beam scales, while extended emission is spatially resolved on physical scales of order a megaparsec. 
This emission is due to dusty star-forming galaxies clustered around the dark matter haloes hosting quasars. 
We measure radial surface brightness profiles of the stacked images to compute the angular correlation function of dusty star-forming galaxies correlated with quasars. 
We then model the profiles to determine large scale clustering properties of quasars and dusty star-forming galaxies as a function of redshift. 
We adopt a halo model and parameterize it by the most effective halo mass at hosting star-forming galaxies, finding $\log(M_\mathrm{eff}/M_\odot) = 13.8^{+0.1}_{-0.1}$ at $z=2.21-2.32$, and, at $z=0.5-0.81$, the mass is  $\log(M_\mathrm{eff}/M_\odot)=10.7^{+1.0}_{-0.2}$. 
Our results indicate a downsizing of dark matter haloes hosting dusty star-forming galaxies between $0.5 \leq z \la 2.9$. 
The derived dark matter halo masses are consistent with other measurements of star-forming and sub-millimeter galaxies. 
The physical properties of dusty star-forming galaxies inferred from the halo model depend on details of the quasar halo occupation distribution in ways that we explore at $z>2.5$, where the quasar HOD parameters are not well constrained.
\end{abstract}


\section{Introduction}

The current paradigm of structure formation entails initial density fluctuations growing into dark matter haloes inside of which galaxies form \citep{whit78}. 
The clustering of dark matter haloes is now well understood (e.g., \citealt{tink08}), and we can use this information to study the clustering of galaxies and to connect visible galaxies to their underlying dark matter haloes. 
Studies of galaxy clustering as a function of brightness, color, morphology, star formation, and dark matter halo mass (\citealt{zeha02}; \citealt{zeha05}; \citealt{coil08}; \citealt{beth14}) can provide invaluable information concerning galaxy evolution as a function of cosmic time. 
Furthermore, with large surveys of galaxies and large maps of the sky at various wavelengths, we are beginning to probe galaxy clustering at large cosmic distances as well as nearby. 
The variety of available wavelength coverage is useful for probing a wide range of galaxy populations and galaxy properties as a function of redshift.

Thanks to the large volume of the Sloan Digital Sky Survey (SDSS, \citealt{york00}; \citealt{abaz09}; \citealt{eise11}; \citealt{daws13}; \citealt{alam15}) hundreds of thousands of quasars have been catalogued with well-known spectroscopic redshifts, enabling the study of their clustering properties as a function of cosmic time. Though quasars have a low spatial density, they are excellent tracers of large scale structure. 
Quasars are active galactic nuclei (AGN) with bolometric luminosities $\ga 10^{45}$ erg/s that are detected at large distances out to $z\sim7$. 
They inhabit massive galaxies in small groups and clusters (\citealt{wold00}; \citealt{cold06}; \citealt{trai12}), and in the local Universe they trace superclusters of galaxies \citep{eina14}. 
Large statistical samples from the SDSS and also the 2dF QSO Redshift Survey \citep{croo04} enabled the measurement of the quasar auto-correlation function from $z\sim0.5$ (e.g., \citealt{porc04}; \citealt{croo05}) out to $z\ga2.9$ \citep{shen07}.

The measured clustering information of spectroscopic quasar catalogs make them especially useful for cross-correlation studies  (\citealt{shen13}; \citealt{wang15}; \citealt{schm15}), particularly in combination with other tracers of large-scale structure. 
The cosmic infrared background (CIB) is one such tracer. 
It results from the thermal re-emission of optical and UV star light in the far-infrared by dust located inside of galaxies. 
The re-emitted light is observed as a diffuse background emission spanning 1-1000~$\mu$m. 
At least 70\% of the cosmic star formation rate density is attributable to the dust-enshrouded, IR-luminous galaxies that make up the CIB \citep{char01}.
Therefore, understanding the sources, redshift distribution, and overall nature of this background is essential for a complete understanding of the formation and evolution of galaxies. 
The total intensity of the CIB has been derived using data from the Far Infrared Absolute Spectrometer (FIRAS) on the Cosmic Background Explorer (COBE) satellite (\citealt{puge96}; \citealt{fixs98}; \citealt{dwek98}), but determining luminosities and redshifts of the contributing sources remains a challenge.

The clustered component of sources making up the CIB, the clustered infrared background, was first detected a decade ago by \textit{Spitzer} at 160 $\mu$m \citep{laga07}. 
The power spectrum of clustering was measured at the peak intensity of the CIB (250, 350, and 500 $\mu$m) using the Balloon-borne Large Aperture Sub-millimeter Telescope (BLAST, \citealt{vier09}), and measurements at millimeter wavelengths over a wide range of angular scales were soon followed by studies using data at longer wavelengths from the South Pole Telescope (SPT, \citealt{hall10}), the Atacama Cosmology Telescope (ACT, \citealt{dunk11}), and \textit{Planck} (\citealt{plan11}; \citealt{ambl11}). 
Surveys from instruments such as the Photoconductor Array Camera and Spectrometer (PACS) and the Spectral and Photometric Imaging REceiver (SPIRE) aboard the \textit{Herschel Space Observatory} brought hopes of resolving the dusty galaxies responsible for the total emission, but they have been hindered by source confusion. 
For example, \citet{beth12} and \citet{oliv10} show that only 15\%\ of sources can be individually detected at 250~$\mu$m. 

Stacking analyses are complementary to the attempt to identify individual sources of the CIB because they yield the average flux densities of a population. 
For example, \citet{albe14} implement a stacking analysis of sources to compare galaxies residing in clusters and in the field. 
They find that an increasing number of DSFGs reside in clusters with increasing redshfit.
This is consistent with both theory \citep{wech98} and observations \citep{adel05} that clustering of star forming galaxies increases with increasing redshift.
Stacking analyses can also account for sources that are too faint to be detected in a confusion-limited map, provided that some other suitable tracer that correlates with those sources is available to stack.
Many stacking analyses (e.g., \citealt{dole06}; \citealt{mars09}; \citealt{devl09}) have revealed which sources produce the bulk of the CIB emission. 
\citet{vier15} show that $>$ 90\%\ of CIB can be accounted for with known galaxies and their faint companions. 
But, accounting for the flux does not answer all of the questions relating these objects to galaxy formation. 
Some of the remaining questions include: In what mass haloes do these highly star forming galaxies primarily reside? What are their average star formation rates? Do these values evolve with redshift? 

Other auto- and cross-correlation studies have been useful in probing the properties of cosmic infrared background sources. 
In particular, it is useful to invoke halo occupation distribution (HOD) models to probe the masses of dark matter haloes hosting star forming galaxies (\citealt{peac00}; \citealt{scoc01}; \citealt{berl02}; \citealt{coor02}; \citealt{shan12}; \citealt{vier13}). 
This can be accomplished by first using our current understanding of the large-scale spatial distribution of dark matter; that is, assuming that all dark matter is constrained within spherically collapsed structures. 
The haloes are then populated with galaxies according to the galaxy HOD function, which statistically defines the way galaxies populate haloes as a function of their dark matter halo mass and redshift. 
Predictions for the galaxy distribution (e.g., through galaxy correlation functions) from the halo models can then be confronted with data to constrain how galaxies populate dark matter haloes and determine other physical properties.
Due to the wavelength coverage of \textit{Herschel} and \textit{Planck} over the peak intensity of the CIB, datasets from these satellites are extremely useful for modelling the dark matter halo masses of dusty star forming galaxies (DSFGs) that make up the clustered infrared background. 
Because of the difficulties of conducting a complete census of star forming galaxies in the infrared, this method of modeling provides one of the few available probes of the physical properties of this population of galaxies as a function of redshift.

One challenge is that the results are dependent on the choice of HOD model. 
\citet{plan11} and \citet{ambl11} invoked HOD models in which the luminosities of IR bright galaxies do not depend on halo mass. 
Soon after, \citet{shan12} developed a new prescription for how luminosity depends on halo mass and redshift, and this has been invoked in more recent years by \citet{vier13}, \citet{plan14}, and \citet{wang15}. 
All of these studies depend on a prescribed redshift evolution of the normalization of the luminosity-mass relation, and they derive a single characteristic mass at which haloes are most efficient at forming stars for a broad range in redshift spanning the peak of cosmic star formation. 

Alternatively, it has been postulated for quite some time that star formation at high redshift occurs in more massive galaxies than today. 
A decrease in the preferred mass scale of star forming galaxies with redshift is a process known as downsizing, and was first observed by \citet{cowi96} as a trend in decreasing K-band luminosity of highly star forming galaxies with cosmic time.   
There has since been many different definitions of downsizing relating to luminosity, stellar mass, and dark matter halo mass (see \citealt{font09} and \citealt{conr09} for reviews), and there has been some question under what conditions the trend exists. There is also an implication that the downsizing trend is connected to the build up of the red-sequence of galaxies (\citealt{font09} and references therein). 
Probing the HOD of DSFGs as a function of redshift can shed light on the existence of downsizing in the dark matter halo masses hosting DSFGs.

In this paper we investigate the clustering of cosmic infrared background (CIB) sources around quasars via a stacking analysis. We stack \textit{Herschel}-SPIRE data on the locations of a spectroscopic sample of optically-selected quasars from the Sloan Digital Sky Survey (SDSS). 
The ability to determine the quasars' redshifts accurately is advantageous for studying the redshift evolution of DSFGs that are correlated with the quasar hosts. 
Assuming a general understanding of the clustering of quasar hosts and dark matter haloes, we present a halo occupation distribution model of the angular cross correlation of quasars and infrared sources.
This model assumes fixed values for the parameters related to the quasar population and constrains the evolution of the physical parameters describing the sources contributing to clustered infrared background signal from z=.5-3.5.
We parameterize the model in a similar manner as \citet{shan12} to estimate the most efficient halo mass at hosting DSFGs as a function of redshift, enabling a test of cosmic downsizing. Through the model, the data also constrain the mean infrared light to halo mass ratio as a function of redshift, enabling a calculation of the mean star formation rate density of the clustered DSFGs.

In Section~\ref{data} we describe the data sets used for the analysis, as well as the implementation of the stacking algorithm. Section~\ref{measurements} outlines the measurement/detection of the angular correlation function.
In Section~\ref{model} we describe the modeling of the stacked data. Section~\ref{results} describes the results of the model, their dependencies on assumptions, and the physical interpretation of the model with comparisons to other works. In Section~\ref{discuss} we summarize, and state our conclusions. This paper adopts the default Planck 2015 cosmology \citep{plan15} with $H_0 = 67.81$ km s$^{-1}$ Mpc$^{-1}$, $\Omega_m = 0.308$, $\Omega_{\Lambda} = 0.692$, and $\sigma_8 = 0.814$.

\begin{figure}
\includegraphics[width=3.25in]{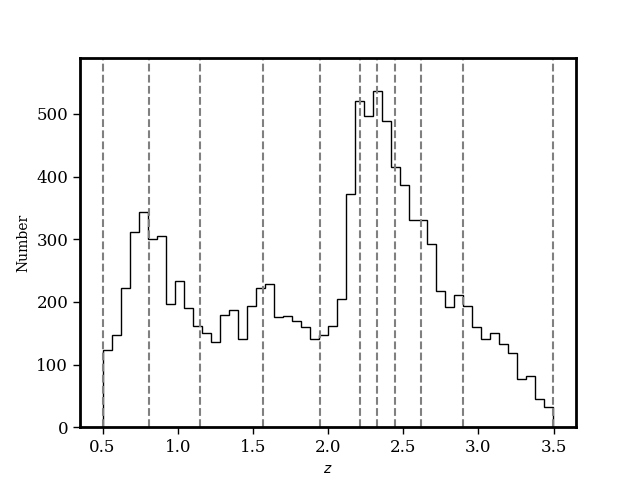}
\caption{Histogram of SDSS quasars used in this study with vertical dashed lines at the redshifts of our bin edges. The total sample includes 11,235 quasars and the bins contain approximately equal numbers of quasars, $N\sim1000$. \label{dndzq}}
\end{figure}

\begin{figure*}
\includegraphics[width=6.5in]{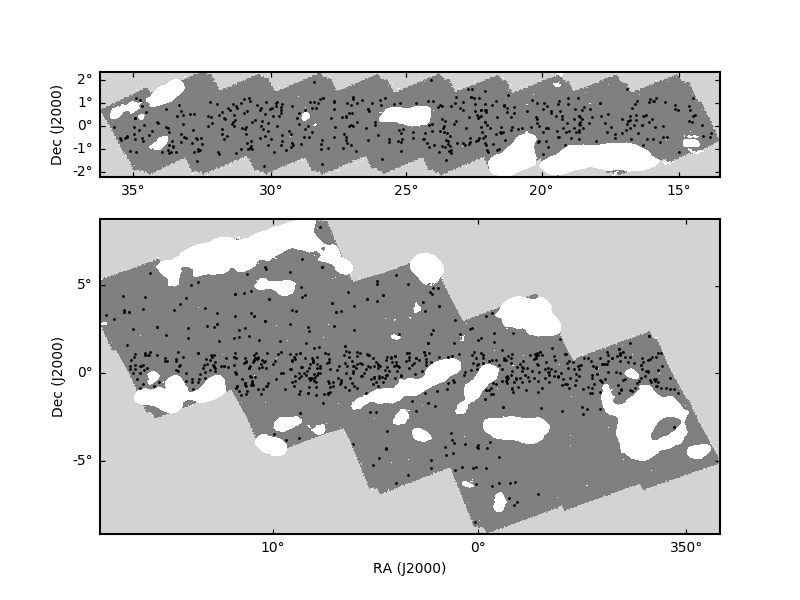}
\caption{HerS (top) and HeLMS (bottom) 250 $\mu m$ coverage maps with Galactic cirrus masked in white and with black dots marking the coordinates of quasars contained in our first redshift bin $z=1.56-1.95$. There is a higher concentration of quasars along the SDSS region Stripe 82 that covers the Celestial Equator, and the region of the HeLMS map at (RA $\la 357^{\circ}$, DEC $\la -3^{\circ}$) has no SDSS coverage. \label{maps}}
\end{figure*}

\begin{figure*}
\includegraphics[width=6.5in]{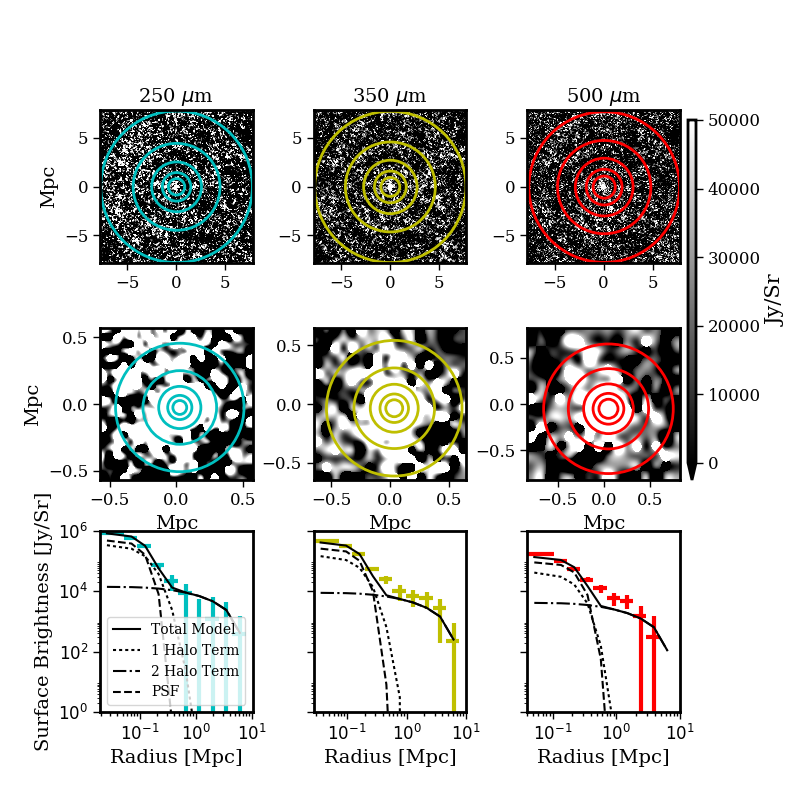}
\caption{Quasar-correlated infrared emission in the redshift bin spanning $z=1.56-1.95$, with negative values saturated to the color black. Top row: 2D stacked images with concentric rings marking the annuli at the outer five  radii of the surface brightness. The full images are $30' \times 30'$, and the approximate physical scale is shown here at the average redshift within this bin. Middle row: Zoom in on the central five radial bins (circles) of the 2D stacked images corresponding to $\sim 0.5$ Mpc in each band. The pixels are interpolated using a spline, and the 500 $\mu$m images are scaled by 1.5 to be on the same color scale as the other bands. Bottom row: Binned radial profiles with best-fitting PSF (dashed line) and halo model (solid line) broken into its one halo (dotted line) and two halo (dot-dashed line) components.  \label{stack}}
\end{figure*}

\section{Data and Stacking}
\label{data}

We make use of the publicly available \textit{Herschel} Stripe 82 Survey\footnote{http://www.astro.caltech.edu/hers/Data\_Product\_Download.html} (HerS; \citealt{vier14}) and the HerMES Large-Mode Survey\footnote{http://hedam.lam.fr/HerMES/} (HeLMS, \citealt{oliv12}) from the \textit{Herschel} Space Observatory instrument SPIRE. 
SPIRE observes at 250~$\mu$m, 350~$\mu$m, and 500~$\mu$m \citep{grif10}. HerS covers 79 square degrees overlapping the celestial equator ($13^{\circ} < \alpha < 37^{\circ}, -2^{\circ} < \delta < 2^{\circ}$), and HeLMS covers 270 square degrees (roughly $-10^{\circ} < \alpha < 18^{\circ}, -8^{\circ} < \delta < 8^{\circ}$). 
The SPIRE beam full widths at half maximum (FWHM) are derived from observations of Neptune and are 18.1$''$, 25.2$''$, and 36.6$''$ for the 250~$\mu$m, 350~$\mu$m, and 500~$\mu$m bands, respectively \citep{grif13}. The flux sensitivity of the \textit{Herschel}-SPIRE bands is 13.0, 12.9, and 14.8 mJy/beam for the 250~$\mu$m, 350~$\mu$m, and 500~$\mu$m bands, respectively, with the confusion noise contributing 8 mJy/beam at all wavelengths.

The quasar catalog used in this study is derived from the Sloan Digital Sky Survey (SDSS) Data Release 7 (DR7, \citealt{abaz09}), Data Release 10 (DR10, \citealt{pari14}), and Data Release 12 (DR12, \citealt{pari17}) spectroscopic quasar catalogs. 
Any objects present in more than one catalog are included only once. 
We remove all quasars outside of the redshift range $0.5<z<3.5$ as this range contains 95\%\ of the quasars and allows us to avoid low population tails of the distribution. 
The sample includes 11,235 sources that lie within the \textit{Herschel}-SPIRE HerS and HeLMS survey footprints. 
Figure~\ref{dndzq} shows the redshift distribution of the quasars with dashed vertical lines at the redshift boundaries of each bin.

\citet{mars09} and others (e.g. \citealt{vier13}) have shown that stacking is the equivalent of taking  the cross-correlation of the intensity map with the catalog. In the case of a confusion dominated or Gaussian noise dominated map, the cross-correlation is a maximum likelihood estimate of the average flux of the catalog.
Therefore, the stacking analysis is implemented to measure the mean far-infrared flux density correlated with the quasars as a function of angular distance from the quasar. 
We make the choice of studying the correlated flux as a function angular distance as opposed to physical distance in order to avoid constructing a model that must use a scale-dependent point spread function (PSF). 
Instead, we implement a redshift-dependent model for the angular correlation function.

Prior to stacking the \textit{Herschel} maps we mask regions of Galactic cirrus by first smoothing the images with a Gaussian function that has a standard deviation of 10 pixels ($\sim 3\times$ FWHM of the beam at each wavelength). 
At this scale, the signal from the sources is smoothed down to low noise, while the extended emission from Galactic cirrus is still significant in the smoothed maps.
We remove the cirrus by subtracting areas of the unsmoothed map which have pixel  values in the smoothed map that are greater than the rms of the unsmoothed map. 
We smooth the masks with Gaussian functions with widths equal to the FWHM of each beam in order to apply a smooth mask to the map edges. 
In the end, the masked regions constitute 13\% and 19\% of the HerS and HeLMS maps, respectively.
From this masked image, we create a coverage map that we use to determine exactly which quasar coordinates lie within the boundaries of the map. 
Figure~\ref{maps} plots the 250 $\mu m$ HerS and HeLMS coverage maps with Galactic cirrus masked out in white and with black dots plotted at the coordinates of the quasars contained in the redshift bin $z=1.56-1.95$. 
The region of the HeLMS map within ($RA\la357^\circ, DEC\la-3^\circ$) has no SDSS coverage, and there is a higher concentration of quasars within the SDSS region Stripe 82 that has deeper coverage along the Celestial Equator. 
The average number of quasars per square degree along Stripe 82 varies between redshift bins from 2-6 quasars per square degree. 
The redshift bin shown has 5 quasars per square degree, or $\sim$1.25 quasars per 30 arcmin$^2$ (the size of our stacked thumbnails). 
The peak average brightness from a quasar at 250~$\mu$m is at most 10$^6$~Jy/sr. The total flux is then $\sim10$~mJy. If you have $\sim1$ extra quasar in the 0.25 square degree patch, then they will  contribute $\sim$1.4e2~Jy/sr. Thus, we are not concerned that the average large scale flux of $\ga$5e3~Jy/sr is coming primarily from quasars. 

We perform the stacking of the \textit{Herschel} maps on the locations of quasars in ten redshift bins, divided such that there are approximately the same number of quasars in each bin, $N\sim1100$.
This yields redshift bins of different widths based on the number density of quasars as a function of redshift and selection effects from SDSS. 
We mean subtract the total map before stacking, and apply a local background subtraction to each $30'\times30'$ thumbnail that goes in to a stack in order to subtract away the confusion-limited background that results from source blending due to the size of the SPIRE beams \citep{vier13b}. 
The local background subtraction is performed in order to eliminate any fluctuations on scales of order the size of our thumbnails that is not accounted for in the mean subtraction of the entire map.
The local background subtraction is performed by computing the mean value of the map in a three arcminute wide annulus at the edge of each thumb nail, avoiding masked regions.
This local background subtraction eliminates any potential bias on the stacked flux density due to the confusion-limited background emission \citep{mars09}.
We test for any potential bias in the stacked flux via a "null" stacking analysis in Section \ref{psf}.
The same background subtraction is applied to the model.
For each redshift bin, the stacked signal is constructed by taking the inverse variance weighted average of the $30'\times30'$ thumbnails centered on all quasars in that redshift bin:
\begin{equation}
\bar{D_{\nu}^{\alpha}} = \frac{\sum\limits_{i}^{N_{\alpha}}w_{i,\nu}D_{i,\nu}}{\sum\limits_{i}^{N_{\alpha}}w_{i,\nu}}.
\label{eq:stackmap}
\end{equation}
Equation~(\ref{eq:stackmap}) gives the mean stacked intensity map $\bar{D_{\nu}^{\alpha}}$ of the redshift bin $\alpha$ containing $N_{\alpha}$ sources, each contributing individual map intensity $D_{i,\nu}$ in the frequency band $\nu$. 
The inverse variance weights $w_{i,\nu}$ are determined from the \textit{Herschel} error maps. The error maps vary between bands and are derived from the diagonal component of the pixel-pixel covariance matrices generated in the map making pipeline \citep{pata08}. The weights and data from masked regions of the map are excluded from the mean calculation.

\begin{figure}
\includegraphics[width=3.25in]{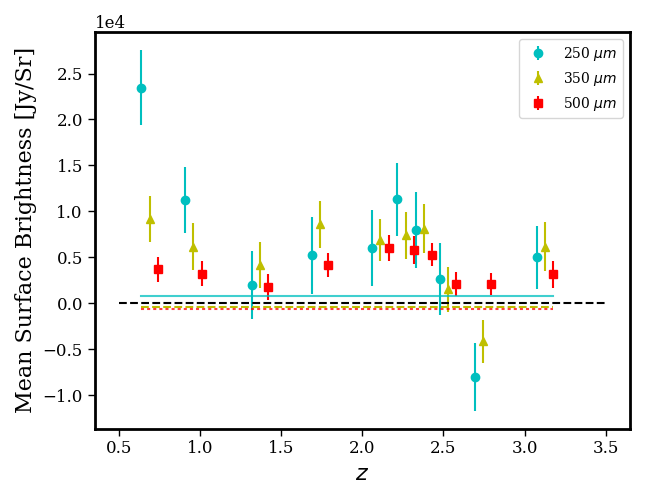}
\caption{Mean surface brightness of large scale (R $\ga$ 0.3 Mpc) clustered signal as a function of redshift and wavelength. Uncertainties are derived from propagating the uncertainty in each radial bin including the contribution from the covariance between radial bins. The 350~$\mu$m data is plotted at the positions of the average redshift of the bin, while the 250~$\mu$m and 500~$\mu$m data are shifted slightly horizontally for clarity. The solid, dashed, and dotted lines indicate the mean surface brightness of the null stacks outside the same radius at 250, 350, and 500 $\micron$, respectively. \label{ampvz}}
\end{figure}

\begin{figure}
\includegraphics[width=3.25in]{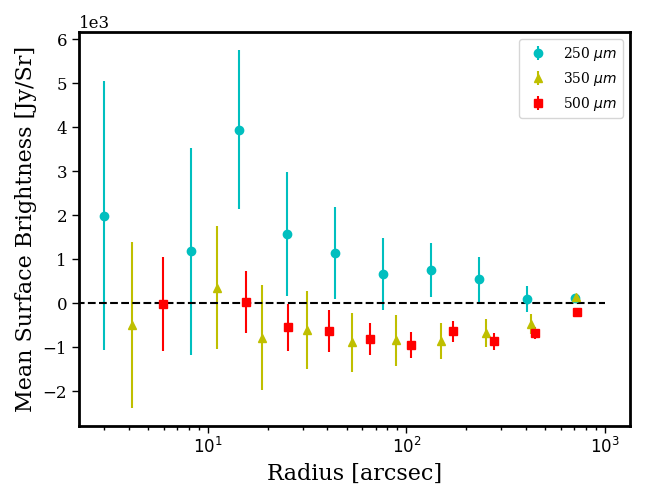}
\caption{Mean surface brightness profiles of 100 null stacks in each wavelength band and redshift bin. Error bars denote the standard error on the mean of the null stacks. \label{nullprofs}}
\end{figure}

\section{Measurements and analysis}
\label{measurements}

\subsection{Angular correlation function}
\label{radprofiles}

In this analysis we are using quasars as tracers of large scale structure in order to probe the correlated, clustered background emission resulting from dusty star-forming galaxies in the same redshift bins as our stacked quasars. 
We generate radial profiles from the stacked images for each of the ten redshift bins in each wavelength band. 
The stacked images have the same units as the map, Jy/beam, and are converted to Jy/sr by dividing by the beam areas, 0.9942, 1.765, and 3.730 $\times 10^{-8}$ sr for the 250~$\mu$m, 350~$\mu$m, and 500~$\mu$m bands, respectively \citep{grif13}. 
The radial profiles are generated by determining the average surface brightness in annuli as a function of radius, with radial bins that are equally spaced on a logarithmic scale. 
Figure~\ref{stack} shows an example of the 2D images and radial profiles for the redshift bin spanning $z = 1.56-1.95$. 
The top row depicts the full 30$'\times$30$'$ 2D images, with axes converted to a physical scale at the average redshift of the bin. 
The second row is a zoom of the images to better depict the images within $\sim$0.5 Mpc. 
On the bottom row, we plot the radial profiles of the stacked images with the best-fitting model overplotted as a solid black line.
At radii within the beam width, the observed signal is a result of the unresolved dust emission associated with quasars and a contribution from any correlated emission on small scales ($\la 30''$). 
At larger radii, there is excess signal associated with correlated clustered infrared background emission.
The radial profiles generated from the stacked images are equivalent to the angular cross correlation function of quasars and the clustered infrared background.
It is this correlation that we wish to model as it is a signature of the sources responsible for the background.

\begin{figure*}
\centering
\includegraphics[width=6.5in]{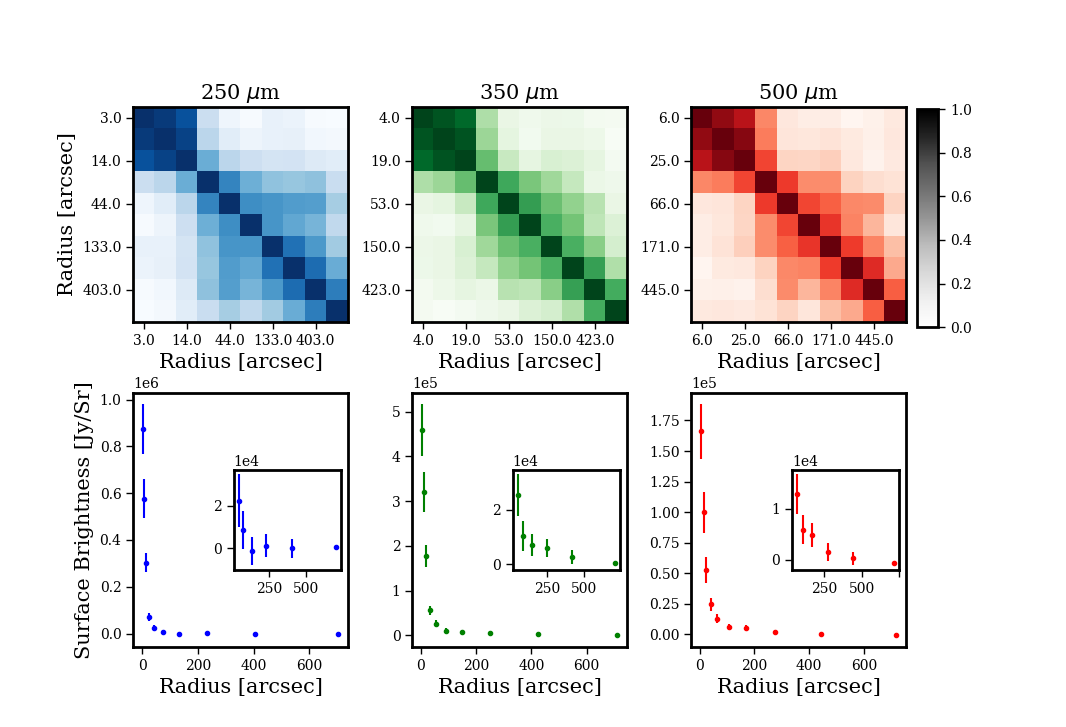}
\caption{Top row: Correlation matrices for the radial profiles at 250~$\mu$m (left), 350~$\mu$m (center), and 500~$\mu$m (right) bands in the redshift bin $z = 1.56-1.95$ computed from the bootstrap analysis. We use the covariance matrices for each redshift bin when computing the maximum likelihood in the MCMC analysis. The correlations in the first three bins are due to the variance from the quasar emission scatter, and the large scale correlations are due to variance of the clustered infrared background. 
Bottom row: Surface brightness with uncertainty taken from the diagonal of the covariance matrices. Inset plots the outer six data points. The radial profiles demonstrate the decreasing uncertainty as a function of radius. \label{cov}}
\end{figure*}

\subsection{Analysis of the quasar emission and the extended emission}
\label{psf}

The quasar emission is spread out over the beams of the \textit{Herschel}-SPIRE instrument, and is modeled using Gaussian PSFs with FWHMs equivalent to those of the beams described by \citet{grif13} for each of the SPIRE bands. 
Row 3 of Figure~\ref{stack} shows the Gaussian PSFs overplotted with the data. 
When modelling the surface brightness profiles, we simultaneously fit for the amplitude of the flux from the quasar using these PSFs and fit a model that describes the extended emission. 
The complete model must be convolved with the beam before fitting to the data, and this convolution is best calculated in Fourier space. 
Transforming a Gaussian in two dimensions is efficient and easy to compute accurately. 
That is the first reason for using the Gaussian beams. 
Secondly, we find that our Markov Chains show better convergence when using the Gaussian beams. 
We incorporate the 10$\%$ uncertainty associated with using the Gaussian beams\footnote{http://herschel.esac.esa.int/Docs/SPIRE/spire\_handbook.pdf}, and we find no significant difference in the resulting best-fitting parameters when using the Gaussian vs. the numerical PSF defined by \citet{grif13}.

The mean amplitude of the signal beyond 2$\times$FWHM of the PSF (r $\ga$ 0.3 Mpc) in each redshift bin is shown in Figure~\ref{ampvz} for each of the observed wavelength bands.
The 350~$\mu$m data is plotted at the positions of the average redshift of the bin, while the 250~$\mu$m and 500~$\mu$m data are shifted slightly horizontally for clarity. The error bars reflect the propagated uncertainty of the mean surface brightness including the contribution from the covariance between radial bins (Section~\ref{covsec}).  

``Null'' stacked thumbnails have been generated using a synthetic catalog drawn from random positions uniformly distributed over the sky coverage. 
The stacking method remains the same as for the quasar positions.
A series of 100 null stacks are generated, each using a catalog of the same size as the mean number of quasars contained in each redshift bin. The null stacks are treated in the same manner as the data, and errors are determined by taking the standard deviation on the mean of the 100 null average flux densities in each radial bin. 
Figure~\ref{nullprofs} demonstrates the mean radial surface brightness profile of the 100 null stacks for each wavelength band with the standard deviation on the mean plotted as error bars.
Note that the errors are correlated between bins.
We observe a bias in the null stacks with amplitudes that are $\la 10\%$, $20\%$, and $50\%$ of the error on the correlation functions at 250~$\mu$m, 350~$\mu$m, and 500~$\mu$m, respectively.
The source of these biases is unknown, but we find they are small enough not to affect our results.
The average values of the null profiles are computed over the same extended radial region as the data ($\geq 2\times$ PSF FWHM) and these values are plotted as solid, dashed, and dotted lines for each wavelength in Figure~\ref{ampvz}. 
We compute the $\chi^2$ values of the mean surface brightnesses of the data with respect to the mean null surface brightnesses. 
For each increasing wavelength band the respective $\chi^2$ values are 59, 76, and 112 with 10 degrees of freedom (one for each redshift bin) yielding p-values of 5e-9, 3e-12, and 2e-19.
This indicates that the data are inconsistent with the null hypothesis.
We conclude that the amplitude of the surface brightness of the extended IR emission around quasars is significant beyond the signal produced by random large scale fluctuations. 

We implement an additional test of the surface brightness profiles to determine if a significant portion of the signal is due to detected \textit{Herschel} sources. 
We mask all of the detected sources in the 500 $\micron$ maps using the source catalog described in \citet{schu17}. 
We then run the same stacking analysis as in our science analysis and determine the source-subtracted angular cross correlation functions. 
The residuals of the original angular cross correlation functions minus the source-subtracted angular cross correlations functions indicate there is no significant deviation in the amplitudes of the profiles except in the two lowest redshift bins. 
In the two lowest redshift bins the three smallest radial bins are decreased by $\le$1$\sigma$ in the source-subtracted profiles as compared to the original profiles. 
This is likely due to the detected \textit{Herschel} sources being preferentially at low redshift due to detection limits. 
We assert that the original non-source-subtracted profiles are to be used in our analysis as it is the clustered emission that we model in order to obtain the physical parameters of dusty sources.

\subsection{Covariance Estimation}
\label{covsec}

We generate a covariance matrix for each wavelength band (see noise correlation matrices in Figure~\ref{cov}) in each redshift bin from 100 bootstrap realizations of the stacked images. We compute them by drawing randomly with replacement from the source locations within each redshift bin to create 100 bootstrapped realizations of the 30'$\times$30' stacked images. We then create the radial surface brightness profiles of the stacked images in the same way as for the stacked data described in Section~\ref{radprofiles}, and we compute the surface brightness covariance within and between radial bins of the correlation functions.

The uncertainty in the null stacks is $\sim 10\%$ of the uncertainty calculated from the diagonal elements of the  covariance matrix in each wavelength band and redshift bin. We thus determine that the noise in the stacked images is dominated on large angular scales ($\ga 30''$) by cosmic variance, the statistical fluctuations in the brightness of the clustered CIB around quasars. 
Though some of the cosmic variance is dampened by our large sample size, the sample is not large enough to smooth all fluctuations.
The resulting radial profiles therefore show significant bin-to-bin covariance due to their dependence on the inherent cosmic fluctuations. 
Furthermore, the radial bins are smaller than the beam size at small radius ($\la 18'', 25'', 36''$ for the 250, 350, 500 $\mu$m bands, respectively), so the variance from the quasar emission scatter is captured in the first three bins of the covariance matrices. 
We use the resulting covariance matrices to perform the maximum likelihood calculation.
We account for the fact that the inverse of the covariance matrix estimated from our bootstrapping technique is a biased estimator of the inverse covariance by using the \citet{hart07} method to increase the covariance.
The error bars on the data points in the plotted radial profiles are obtained from the variance values provided by the diagonal elements of this covariance matrix. The relative amplitude of the uncertainties with respect to the data increases with radius from $\sim 6\%$ to up to $\ga 100\%$. The data with uncertainties for $z = 1.56-1.95$ are plotted in the bottom row of Figure~\ref{cov}. The matrices shown here are representative of those associated with the other redshift bins.   

\section{The Clustered Infrared Background}
\label{model}

In this section, we explain the necessity for using the halo model formalism (Section~\ref{whyhalo}) and present a theoretical model for our data. 
The model of the clustered infrared background power spectrum is described in Section~\ref{hodmodel}, followed by the quasar model power spectrum in Section~\ref{qsohod}, and the final combined angular cross-correlation function in Section~\ref{xcorr}.

\subsection{Why halo occupation models are necessary}
\label{whyhalo}

Figure~\ref{ampvz} presented in Section~\ref{measurements} demonstrates that the surface brightness of the infrared background clustered around optically-selected quasars is evolving as a function of redshift. 
There is evidence for a peak in the clustered signal in all three frequencies between $z\sim1.5$ and $z\sim2.5$, and a decline in signal at $z>2.5$. 
This behavior as a function of redshift mimics that of the overall luminosity density evolution of star-forming galaxies \citep{mada14}, but the relationship between our clustered CIB brightness and cosmic star-formation history depends on several factors.

\begin{figure}
\includegraphics[width=3.25in]{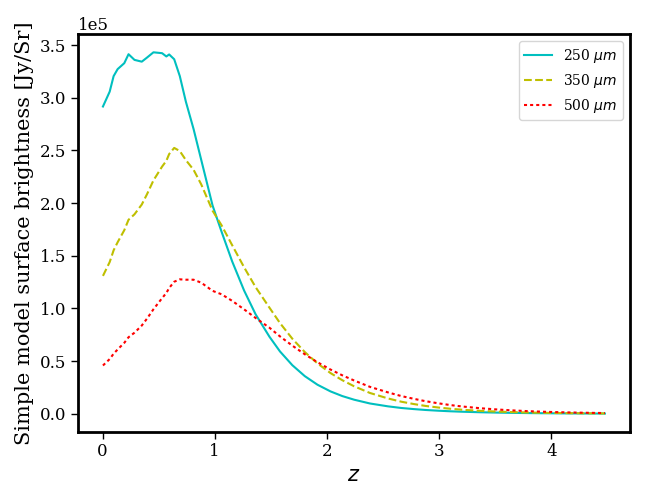}
\caption{Model of expected surface brightness of the clustered infrared background computed by redshifting an example star-forming galaxy SED from \citet{riek09}, convolving it with the \textit{Herschel} bandpasses, normalizing by the average star formation rate density as a function of redshift, and accounting for the cosmological surface brightness dimming. The expected distribution of signal from this model does not adequately describe the shape of our extended emission as a function of redshift. \label{toymodel}}
\end{figure}

Our first attempt at understanding the origin of our detection is a model in which dusty star-forming galaxies cluster around quasars in exactly the same way at every redshift, but their infrared luminosities evolve with redshift in agreement with the cosmic co-moving luminosity density of star formation $\rho(z)$, which we take from \citet{hopk08}. 
To compute the redshift evolution of the resulting signal, we take several spectral energy distributions (SEDs) of star-forming galaxies from \citet{riek09}, place them at a range of redshifts, convolve them with Herschel bandpass functions, normalize the overall luminosity using $\rho(z)$ and take into account the cosmological surface brightness dimming.
We find that regardless of the adopted spectral energy distribution, the resulting signal peaks at redshifts $z\la 1$. 
This result can be understood as a product of two strong functions of redshift: while the luminosity density $\rho(z)$ increases as a function of redshift for $0<z<2.5$ by a factor of $\ga 10$, the cosmological surface brightness (SB) dimming is a strongly decreasing function of redshift, so the two effects combine to produce a peak of the signal at $z\sim 1$. 
Figure~\ref{toymodel} shows the surface brightness as a function of redshift computed using this straightforward model and an example SED template. 
The method yields the surface brightness per comoving Mpc. 
To arrive at the curves in Figure~\ref{toymodel}, which can be directly compared to our data, we multiply the calculated surface brightness per comoving Mpc by the differential comoving distance as a function of redshift $d\chi/dz$. 

The differences arising due to k-corrections (i.e., because different intrinsic parts of the SEDs fall into the same Herschel bands at different redshift) result in differences between the evolution of the signal as seen in different bands, but these variations are less strong than those due to $\rho(z)$ and SB($z$) dimming. 
In particular, in the range $1<z<3.5$ k-corrections at 500~\micron\ are negative for all SEDs in the \citet{riek09} library \citep{blai93}, whereas k-corrections at 250~\micron\ and 350~\micron\ can be positive or negative, depending on the wavelength of the peak of the template, in this redshift range. 
Thus, in our model of the clustered background co-evolving with the cosmic luminosity density the apparent surface brightness peaks at somewhat higher redshifts for longer wavelengths, but still not close to $z\sim2$ and with a steep decline toward higher redshifts.

Therefore, we conclude that our observations are inconsistent with infrared-bright galaxies clustering around quasars in exactly the same way at all redshifts. 
Several effects might enhance the clustered background signal at $z\ga 2$. 
For example, if high-redshift quasars occupy more massive dark matter haloes than the low-redshift ones, they will have more galaxies clustered around them. 
Alternatively, even if the numbers of clustered galaxies are similar, we still can have a peak of infrared signal at high redshifts if the DSFGs clustered around quasars are more efficient at forming stars and therefore have a higher luminosity to mass ratio. 
For example, massive galaxies at low redshift tend to be passive ellipticals and are therefore not forming stars in proportion to $\rho(z)$, whereas massive galaxies at high redshifts are exactly those that contain a significant fraction of the star formation activity \citep{mart16}. 
This effect will clearly enhance the strength of the clustered infrared signal at high redshifts.

In other words, the observed redshift dependence of the clustered infrared background around quasars suggests that our observations are sensitive to the redshift evolution of the clustering of dark matter haloes hosting quasars and / or to the mass and luminosity-to-mass evolution of dark matter haloes hosting DSFGs. 
In an attempt to understand these effects, we explore the full halo occupation distribution models in application to our data. 
In this attempt at understanding, we fix the parameters describing the mean number of central and satellites that describe the quasar halo occupation function and explore the most effective mass and luminosity-to-mass ratio of haloes hosting DSFGs. We then explore the effects of any assumptions we make on the derived parameters.

\subsection{Halo Model of clustered infrared background sources}
\label{hodmodel}

The theoretical model is presented by first considering the auto correlation of the CIB anisotropies and of the quasars, separately, in order to introduce relevant terminology and give the reader a better understanding of the HOD model before combining them into the theoretical cross-correlation function. 
The halo model for cosmic infrared background anisotropies was developed in \citet{shan12} and used in several other works (e.g. \citealt{vier13}; \citealt{plan14}; \citealt{serr14}; \citealt{wang15}).
The model used here is essentially the same as that used in \citet{wang15}, but we focus on determining different physical parameters. 

Early halo models of large scale structure (e.g. \citealt{coor02}) and of CIB anisotropies (e.g. \citealt{ambl11}) assumed that all galaxies contribute equally to the emissivity, independently from the mass of the dark matter halo.
The key difference between the halo model we adopt from \citet{shan12} and those of the previous works is that it does not assume that all of the dusty sources have the same luminosity, but instead makes use of a parametric relationship between infrared luminosity and halo mass.
Using this $L-M$ relation in our halo occupation distribution model allows us to weight the number of central and satellite galaxies that occupy a dark matter halo by the luminosity density at a given host halo mass $M$ as a function of redshift. 
We add more complexity to our model by not solely assuming a prescribed redshift evolution, but allowing the peak halo mass and normalization of the $L-M$ relation to vary in each redshift bin.

We start our halo modeling by postulating that the distribution of dark matter is known, and therefore the halo mass function and halo bias are also known (\citealt{tink08}; \citealt{tink10}). 
The galaxy power spectrum is described by the sum of the linear two-halo term which dominates the correlated emission on large scales between galaxies in separate dark matter haloes, the non-linear one-halo term which dominates on small scales due to galaxy pairs within the same halo, and a Poisson term due to the far infrared emission of an unclustered population.
\begin{equation}
P(k,z) = P^{1h}(k, z) + P^{2h}(k, z) + P^{Poisson}(k, z)
\end{equation}

The two-halo term of the auto power spectrum of the dusty sources at a given frequency is given by the linear matter power spectrum weighted by the square of the bias of the dusty star forming galaxies
\begin{equation}
\begin{aligned}
    P^{2h}(k, z) = b_{DSFG}^2(k,z)P_{lin}(k=2\pi k_{\theta}/\chi, z),
\end{aligned}
\end{equation}
where $\chi(z)$ is the comoving distance, $b_\mathrm{DSFG}(k,z)$ is the bias of the dusty sources making up the clustered background, and $P_{lin}(k,z)$ is the linear dark matter power spectrum. 
We explicitly write $P_{lin}$ in terms of the angular or projected wave number $k_\theta$ as that is the coordinate being transformed into real angular space, $\theta$. 
The linear bias of the DSFGs is unknown, as is the number counts of the sources. The general expression for the linear bias of a particular group of galaxies \citep{coor02} applies also to that of the DSFGs
\begin{equation}
\label{bias}
    b_{gal}(k,z) = \frac{\int dM \frac{dn}{dM}(z) b_h(k,M,z) \langle N(M) \rangle}{\int dM \frac{dn}{dM} (z) \langle N(M) \rangle},
\end{equation}
where $\frac{dn}{dM}$ is the halo mass function \citep{tink08}, $b_h$ is the linear dark matter halo bias \citep{tink10}, and $\langle N \rangle$ denotes the mean number of sources in a halo of mass $M$ and is the sum of the mean number of central galaxies and the mean number of satellite galaxies. 

The extended emission in our angular cross correlation function comes from DSFGs that lie within the same redshift bin as that of the quasar population, but the number counts as a function of redshift of dusty galaxies is unknown. 
Because of this, we must parameterize the DSFG bias in terms of the mean emissivity per comoving volume
\begin{equation}
    b_{DSFG}(k,z) = \frac{\int dM \frac{dj_\nu}{dM}(z) b_h(k,M,z)}{\int dM \frac{dj_\nu}{dM} (z)}
\end{equation}
where the emissivity per comoving volume is
\begin{equation}
\label{jeq}
    \bar{j}_{\nu}(z) = \int dL \frac{dn}{dL}(L,z) \frac{L_{(1+z)\nu}}{4 \pi},
\end{equation}
The luminosity density $L_{\nu}$ is the infrared luminosity density at the observed frequencies and $\frac{dn}{dL}$ is the comoving luminosity function of the sources. 
This parameterization uses the fact that galaxies' emissivities in a given dark matter halo depend on redshift, halo mass, and frequency through their luminosity density. 
Within this formalism, the infrared luminosity density is
\begin{equation}
L_{(1+z)\nu} = L_0 \Phi(z) \Sigma(M)\Theta[(1+z)\nu],
\label{Lnu}
\end{equation}
where $L_0$ is the normalization factor with units of luminosity per solar mass $[L_\odot/M_\odot]$. $L_0$ is used to determine the amplitude of the average luminosity density of DSFGs in each redshift bin per unit halo mass. $\Phi(z)$ describes the redshift evolution of the $L-M$ relation, $\Sigma(M)$ describes the luminosity-halo mass relation, and $\Theta[(1+z)\nu]$ describes the shape of the SED.

The exact form of the redshift evolution of the $L-M$ relation used in this model is still a matter of some debate. 
It is primarily based on the scaling of SFR and bolometric infrared luminosity \citep{kenn98} and the expected evolution of specific star formation rate with redshift, which is observed to be a power law $(1+z)^\eta$ with a plateau at $z\approx2$ (\citealt{bouc10}; \citealt{wein11}). 
Thus, this is what we assume for $\Phi(z)$ with a power law index $\eta=2.19$ in agreement with \citet{vier13}. 

The shape of the SED is described by a normalized modified blackbody function,
\begin{equation}
    \Theta[(1+z)\nu] \propto (\nu(1+z))^{\beta}B_{\nu}(\nu(1+z), T_d)
\end{equation}
at the rest frame frequency, where $B_{\nu}(\nu, T_d)$ is the Planck blackbody function at dust temperature $T_d$. We fix $\beta$ to be 1.45 in agreement with \citet{vier13} and \citet{wang15} and allow only the dust temperature to vary. 
We ran trials in which $\beta$ is allowed to vary and fixing it is inconsequential.

The function $\Sigma(M)$ relates the infrared luminosity to the dark matter halo mass. 
Galaxy luminosity functions and halo mass functions indicate that star formation only occurs in a range of halo masses, with suppression at both high and low masses due to physical mechanisms stunting star formation such as photoionization, supernovae feedback, AGN feedback, and virial shocks (e.g. \citealt{birn03}, \citealt{bowe06}, \citealt{crot06}).
Thus, we describe this relationship between $L$ and $M$ via a log-normal distribution of the form
\begin{equation}
\label{Sig}
\Sigma(M) = \frac{M}{(2\pi \sigma_{L/M}^2)^{1/2}}e^{-(\log_{10}(M) - \log_{10}(M_\mathrm{eff}))^2/2\sigma_{L/M}^2},
\end{equation}
where $M_\mathrm{eff}$ is the peak of the specific IR emissivity, or the most efficient halo mass at hosting dusty star forming galaxies. 
A similar relation is observed between SFR and halo mass via abundance matching \citep{bethdore12}.
The width of the halo mass distribution is $\sigma_{L/M}$, which is a unitless quantity and is difficult to constrain when left as a free parameter.
We fix the value $\sigma_{L/M}^2 = 0.5$ as in \citet{shan12}, \citet{plan14}, and \citet{serr14}.
This is consistent with the range of values found by \citet{vier14} who leave it as a free parameter in their model.

In the same manner as other CIB anisotropy HOD models, we assume that the same $L-M$ relation (and redshift evolution thereof) applies to both the central and satellite galaxies. Our method is different from the other halo models of the CIB anisotropies in that we fit our model to the data in each redshift bin instead of applying the model across one or two broad redshift ranges. 
We are aiming to estimate values of $L_0$, $T_d$ and $M_\mathrm{eff}$ in each redshift bin, probing any additional changes in these parameters as a function of redshift that is not accounted for by the prescribed evolution, $\Phi(z)$, which describes the observed evolution of specific star formation rate with redshift.

\begin{table*}
\centering
\caption{Quasar bias and HOD parameters used in our model for each redshift bin. HOD parameters are derived from \citet{wang15}. \label{bqso}}
    \begin{tabular}{lcccccr}
    \hline
        z bin & $b_Q$, (Reference) & $\log(M_{min,q} [M_\odot])$ & $\log(M_1 [M_\odot])$ & $\log(M_{cut} [M_\odot])$ & $\alpha$ & $\sigma_{\log M}$ \\
        \hline
        0.5-0.81 & 1.38, \citep{shen13} & 12.4 & 13.6 & 13.2 & 3.0 & 1.1 \\
        \hline
        0.81-1.15 & 2.17, \citep{hick11} & 12.4 & 13.6 & 13.2 & 3.0 & 1.1 \\
        \hline
        1.15-1.56 & 2.17, \citep{hick11} & 12.4 & 13.6 & 13.2 & 3.0 & 1.1 \\
        \hline
        1.56-1.95 & 2.17, \citep{hick11} & 12.4 & 13.6 & 13.2 & 3.0 & 1.1  \\
        \hline
        1.95-2.21 & 3, \citep{myer06} & 12.4 & 13.6 & 13.2 & 3.0 & 1.1 \\
        \hline
        2.21-2.32 & 3.54, \citep{efte15} & 12.4 & 13.6 & 13.2 & 3.0 & 1.1 \\
        \hline
        2.32-2.45 & 3.54, \citep{efte15} & 12.4 & 13.6 & 13.2 & 3.0 & 1.1 \\
        \hline
        2.45-2.62 & 3.54, \citep{efte15} & 12.38 & 13.52 & 13.14 & 3.04 & 1.14 \\
        \hline
        2.62-2.90 & 3.54, \citep{efte15} & 12.35 & 13.37 & 13.02 & 3.10 & 1.20   \\
        \hline
        2.90-3.50 & 8, \citep{shen07} & 12.3 & 12.8 & 12.7 & 3.21 & 1.3 \\
        \hline
    \end{tabular}
\end{table*}

Using the definition of luminosity density in Equation~(\ref{Lnu}), we can define the number of central and satellite galaxies weighted by the luminosity density of each population as a function halo mass and redshift:
\begin{equation}
\label{fcen}
    f_{\nu}^{cen}(M,z) = N^{cen}\frac{L^{cen}_{(1+z)\nu}(M,z)}{4\pi} 
\end{equation}
and
\begin{equation}
\label{fsat}
    f_{\nu}^{sat}(M,z) = \int dm \frac{dn}{dm}(M,z) \frac{L^{sat}_{(1+z)\nu}(m,z)}{4\pi} 
\end{equation}
where $N^{cen}$ is the number of dusty galaxies that are the central galaxy in the halo, taken to be 0 below and 1 above a minimum halo mass $10^{10} M_{\odot}$ (consistent with \citealt{vier13} and \citealt{serr14}), and the effective number of satellites is found by integrating the subhalo mass function $dn/dm$ over subhalo mass $m$ at the time of accretion for a given host halo mass $M$ \citep{tinkwetz10}. 
We model the subhaloes in this way because the luminosity and mass of a satellite galaxy are best correlated at the time of accretion before the mass has been stripped by the host halo (\citealt{wetz10} and references therein).
Equations~(\ref{Lnu})-(\ref{fsat}) allow the emissivity per covoming volume to be recast as an integral over halo mass
\begin{equation}
\label{jnu}
    \bar{j}_{\nu}(z) = \int dM \frac{dn}{dM}(z) \left[f_{\nu}^{cen} + f_{\nu}^{sat} \right].
\end{equation}

We can further define the DSFG bias in terms of halo mass and luminosity density weighted number of galaxies as
\begin{equation}
\begin{aligned}
\label{bDSFG}
b_{DSFG}(k, z) = \frac{1}{\bar{j}_{\nu}(z)}\int dM \frac{dn}{dM}(z) b_h(k,M,z) \\
\times u_{gal}(k,z,M) \left[ f_{\nu}^{cen}(M, z) + f_{\nu}^{sat} \right],
\end{aligned}
\end{equation}
where $u_{gal}$ is the Fourier transform of the density distribution of matter inside a halo of mass $M$ at redshift $z$, normalized to one in the center and taken to be the Navarro-Frank-White (NFW; \citealt{NFW96}) density profile.
We use a redshift-dependent concentration-mass relation for $u_{gal}(k,z,m)$ with $c(M,z)=10/(1+z)\times(M/M*)^{-0.2}$ (\citealt{whit01}; \citealt{bull01}), though we checked that we are not sensitive to the choice of $c(M,z)$.

Combining all of this information, we arrive at the following expression for the two-halo term of the power spectrum for dusty galaxies
\begin{equation}
\begin{aligned}
P^{2h}(k,z) = P_{lin}(k,z) \left(\frac{1}{\bar{j}_{\nu}(z)}\int dM \frac{dn}{dM}(z) b(k,M,z) \right. \\
\left. u_{gal}(k,z|M) \times [f_{\nu}^{cen}(M, z) + f_{\nu}^{sat}(M,z)] \right)^2.
\end{aligned}
\end{equation}
with $u_{gal}(k,z|M)$ included for completeness, though we find that it does not make a significant difference in the model, and the squared portion in parenthesis is $b_\mathrm{DSFG}$. 

The one-halo term dominates the power spectrum on small scales and in the case of a cross correlation at two different frequencies $\nu1$ and $\nu2$ is expressed
\begin{equation}
\begin{aligned}
    P_{\nu1 \nu2}^{1h}(k,z) = \frac{1}{\bar{j}_{\nu1} \bar{j}_{\nu2}}\int dM \frac{dn}{dM} \\ \times [f_{\nu1}^{cen}(M, z)f_{\nu2}^{sat}(M,z)u_{gal}(k,z|M) \\ + f_{\nu2}^{cen}(M, z)f_{\nu1}^{sat}(M,z)u_{gal}(k,z|M) \\ + f_{\nu1}^{sat}(M,z)f_{\nu2}^{sat}(M,z)u_{gal}^2(k,z|M)].
\end{aligned}
\end{equation}

Using the \citet{limb53} approximation, the clustering term of the angular power spectrum at a given frequency is given by
\begin{equation}
\begin{aligned}
    P_{\nu}(k_{\theta}) = \int \frac{dz}{\chi^2} \left(\frac{d\chi}{dz}\right)^{-1}P_{\nu}^{clust}(k, z) \left(\frac{dS}{dz}(z, \nu)\right)^2
\end{aligned}
\end{equation}
The redshift distribution of the emissivity of the clustered sources can be described as
\begin{equation}
\begin{aligned}
\frac{dS_{\nu}}{dz} = \frac{c}{H(z)(1+z)}\bar{j}_{\nu}(z),
\end{aligned}
\end{equation}
where $H(z)$ is the hubble constant at redshift $z$. 

In this work, we fit the angular cross correlation function that is defined by the Hankel transform of the cross correlation between the power spectrum defined here with the quasar power spectrum defined by the quasar HOD. 
The parameters for which we fit are the most effective halo mass at hosting DSFGs, $M_\mathrm{eff}$, which is the peak of the $L-M$ relation defined by Equation~(\ref{Sig}), the overall normalization of the cross correlation function, and the dust temperature of the SED that describes the mean emission of the DSFGs.

\subsection{Halo model of quasars}
\label{qsohod}
The clustering term of quasars may be written
\begin{equation}
    P(k_{\theta}) = \int \frac{dz}{\chi^2} 
     \left(\frac{dN}{dz}(z)\right)^2P_{QSO}(k=2\pi k_{\theta}/\chi, z)
\end{equation} 
where $\frac{dN}{dz}$ is the normalized redshift distribution of the sample of quasars within each redshift bin. 
The two-halo term for the quasar power spectrum is given by
\begin{equation}
P^{2h}_{QSO}(k,z) = b_{QSO}^2(k,z)P_{lin}(k=2\pi k_{\theta}/\chi, z).
\end{equation}

The quasar portion of the DSFG-quasar cross power spectrum is written in terms of the halo occupation distribution of quasars. We make use of the following form of the mean halo occupation number of quasars, split into the mean number of central and satellite haloes of mass $M$ hosting quasars.
\begin{equation}
\label{Nqso}
    \langle N_q(M) \rangle = \langle N_{cen}(M) \rangle + \langle N_{sat}(M) \rangle
\end{equation}
The number of central haloes hosting quasars is given by a softened step function in the literature such that above a minimum mass ($M_{min,q}$) of haloes hosting quasars on average half of them will contain a central quasar. This function quickly increases to one.
\begin{equation}
\label{Ncenqso}
    \langle N_{cen}(M) \rangle = \frac{1}{2}\left(1+erf\left(\frac{\log(M)-\log(M_{min,q})}{\sigma_{logM}}\right)\right).
\end{equation}
The mean number of satellite quasars in dark matter haloes hosting quasars is given by the common functional form of a rolling-off power law,
\begin{equation}
\langle N_{sat}(M) \rangle = \left( \frac{M}{M_1} \right)^{\alpha} \exp\left( \frac{-M_{cut}}{M} \right),
\end{equation}
where $M_1$ is the halo mass scale at which haloes host one satellite quasar on average and $M_{cut}$ is the halo mass scale below which the number of satellite quasars decreases exponentially.
Following the findings of \citet{chat12}, we treat the HOD of central and satellite quasars separately. That is, the presence of a quasar in a satellite within a given host halo does not require the central galaxy in that halo to also host a quasar. 

We assume values for the quasar HOD parameters as a function of redshift based on the findings of \citet{wang15}. We fix the quasar bias $b_{QSO}$ in each redshift bin based on previous studies of auto-correlation functions of quasars (\citealt{myer06}; \citealt{shen07}; \citealt{hick11}; \citealt{shen13}; \citealt{efte15}). 
The values of the quasar bias that we use and the studies from which they are taken are listed in Table~\ref{bqso}. 
The values of the quasar HOD parameters $M_{min,q}$, $M_1$, $M_{cut}$, $\alpha$, and $\sigma_{\log M}$ are fixed to values found for the \citet{wang15} low redshift sample for our redshift bins spanning $z=0.5-2.45$. 
We then implement a linear transition to the high redshift values over our final three redshift bins spanning $z=2.45-3.5$.
Using these HOD parameters the satellite fractions vary in each redshift bin, with all values below 0.1. 
The HOD parameter values are also recorded in Table~\ref{bqso}. 
In the text, we refer to the use of the parameters recorded in Table~\ref{bqso} as the fiducial model. 
We compare this model to one in which we fix the quasar HOD parameters to the low redshift values across all redshift bins, and in the plots comparing these we refer to the fiducial model as the ``Evolving QHOD" model and the alternative as the ``Constant QHOD" model. 


\subsection{Angular cross correlation function of quasars and the clustered infrared background}
\label{xcorr}

Combining the above formalisms for the power spectra of quasars and the DSFGs that make up the CIB, we find for the angular cross correlation power spectrum
\begin{equation}
\begin{aligned}
\label{halomodel}
    P(k_{\theta}) = \int \frac{dz}{\chi^2} \left(\frac{d\chi}{dz}\right)^{-1}(P^{1h}_{Q,DG}(k, z) \\ + P^{2h}_{Q,DG}(k,z)) 
     \frac{dS}{dz}(z, \nu)\frac{dN}{dz}(z)
\end{aligned}
\end{equation}
with the one-halo term
\begin{equation}
\begin{aligned}
P^{1h}_{Q,DG}(k,z) = \frac{1}{\bar{j}_{\nu}\bar{n}}\int dM \frac{dn}{dM} \\ \times [f^{cen}_{\nu}(M,z)N^{sat}_{Q}u_{gal}(k, z|M) \\ +  f^{sat}_{\nu}(M,z)N^{cen}_{Q}u_{gal}(k, z|M)  \\ + f^{sat}_{\nu}(M,z)N^{sat}_{Q}u^2_{gal}(k, z|M)].
\end{aligned}
\end{equation}
This formalism allows for scenarios in which the central object is either a quasar or a DSFG, and the satellite population can be made up of a combination of DSFGs and quasars, though the satellite fraction of quasars is small. 
We assume that the correlated emission on small scales is a result of a combination of emission from clustered infrared background sources and far-infrared emission from the quasar itself, which we model separately using the PSF. 
Thus, this formalism does not allow for contributions to the one halo term from quasar host galaxy IR emission,  but instead the quasar contribution to the amplitude of the one halo term is from the halo occupation function $\langle N(M) \rangle$ of quasars. 
In this way, this cross-power spectrum model does not account for any emission on small scales that may be coming from the quasar host galaxy and contributing to the correlated flux.
The two-halo term is represented by the product of the quasar bias, the DSFG bias given in Equation~(\ref{bDSFG}), and the linear dark matter power spectrum:
\begin{equation}
P^{2h}_{Q,DG}(k,z) = rb_{QSO}(z)b_{DSFG}(k,z)P_{lin}(k, z)
\end{equation}
where we've included the cross-correlation coefficient r, which we assume to be 1, for completeness.

The relationship between the angular correlation function and the angular power spectrum is an inverse Hankel transform: 
\begin{equation}
w(\theta) = \int \frac{dk_{\theta} k_{\theta}}{2\pi}P(k_{\theta})J_0(k_{\theta}\theta),
\end{equation}
where $J_0(x)$ is the zeroth order Bessel function. We also need to convolve our theoretical correlation function with the PSF. We first Hankel transform the real-space PSF, then multiply this by the power spectrum such that,
\begin{equation}
    PSF(k_\theta) = \int d\theta' \theta' PSF(\theta') J_0(k_{\theta}\theta')
\end{equation}
\begin{equation}
\label{wconv}
\begin{aligned}
w_{conv}(\theta) = \frac{1}{2 \pi}\int dk_{\theta} k_{\theta} P(k_{\theta})PSF(k_\theta)J_0(k_{\theta}\theta) 
\end{aligned}
\end{equation}
where the first integral is the Hankel transform of the PSF and the second is the convolution on Fourier space that is inverse transformed back into real space.

\begin{figure*}
\centering
  \includegraphics[width=6in]{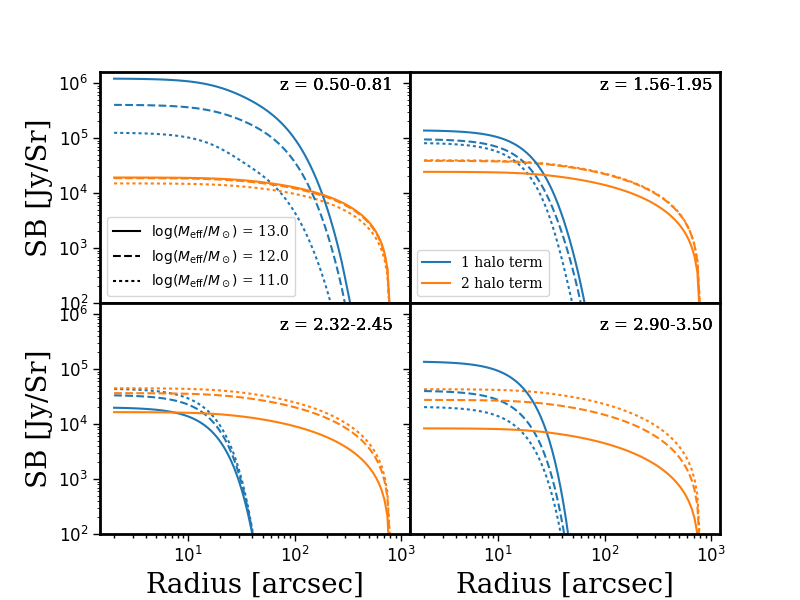}
\caption{Illustrative evolution of the one-halo (blue/small scale) and two-halo (orange/large scale) terms of the model at 350 $\mu$m as a function of redshift and $M_\mathrm{eff}$ using the fiducial model with the evolving quasar HOD parameters. The dotted, dashed, and solid lines indicate $\log(M_\mathrm{eff}/M_\odot) =$ 11.0, 12.0, and 13.0, respectively. The dust temperature and overall normalization $L_0$ are fixed at 35 K and 0.2 $L_\odot/M_\odot$, respectively. $L_0$ directly changes the amplitude, as does $T_d$ to a good approximation.}
\label{modelevolution}
\end{figure*}

\subsection{Halo Model Implementation}
\label{implementation}

In the implementation of our HOD model, all parameters relating to the HOD of quasars is fixed, and we fit for physical parameters pertaining to the HOD of Dusty Star Forming Galaxies. 
We fit each redshift bin independently with the HOD model convolved with the PSF, Equation~(\ref{wconv}). 
Each redshift bin is divided into four smaller bins at which $P(k,z)$ is calculated, and then we integrate to arrive at $P(k_\theta)$. 
In the model there are six free parameters.
Three parameters determine the DSFG HOD model: $M_\mathrm{eff}$, the most effective halo mass at hosting the clustered DSFG sources, $L_0$, the global normalization parameter for the $L-M$ relation, and $T_d$, the dust temperature describing the shape of the DSFG's SEDs. 
We simultaneously fit the data for the amplitude of the mean far-infrared emission from the quasars using a Gaussian with FWHM equal to that of the \textit{Herschel}-SPIRE beams at each respective wavelength. 
The three amplitudes are not a part of the halo model, but we fit the full model as the amplitudes times the PSF plus the halo model curve.

We assume flat priors for all of our fitting parameters except the dust temperature. The DSFG HOD parameter priors are  $10\leq\log(M_\mathrm{eff}/M_\odot)\leq14$, 1e-3 $\leq L_0/(L_\odot/M_\odot) \leq 5$, and we put a Gaussian prior on the dust temperature $T_d$ with a mean of 30~K and standard deviation of 10~K. 
This prior on the temperature is in line with expectations for the mean dust temperatures of DSFG SEDs \citep{case14}. 
The minimum $M_\mathrm{eff}$ is set by the minimum mass for hosting a DSFG, while the maximum is a cluster-scale mass that provides a physical upper limit for the range of redshifts probed. 
Beyond 10$^{14} M_\odot$, the $z>2$ halo mass function drops rapidly.
The PSF amplitudes are constrained to be between 1-100\% of the amplitude of the first bin of the surface brightness profile for each respective wavelength.

We add an additional prior on the star formation rate density $\rho_{SFR}$. 
Using the HOD parameters, we derive the average cosmic star formation rate density of the DSFGs. 
We constrain the value of $\rho_{SFR}$ using the formula from \citet{mada14} for the cosmic star formation rate density $\rho_{SFR}(z)$ as a function of redshift. 
We impose a Gaussian prior with mean equal to the $\rho_{SFR}(z)$ computed at the average redshift of each bin with a standard deviation of 0.05~M$_\odot$/yr. 
We find that even with the Gaussian prior on the dust temperature, the freedom of $L_0$ allows the model parameters $L_0$ and $T_d$ to vary considerably while producing the same model amplitude, but adding the prior on $\rho_{SFR}$ helps to constrain these values. 
The freedom without the prior on $\rho_{SFR}$ can be problematic for the convergence of the MCMC in some redshift bins. 
In Section \ref{quasartests}, we discuss the consequences of not using the prior on $\rho_{SFR}$. 
The derived $\rho_{SFR}$ from the best-fitting model parameters is discussed in Section~\ref{interpret}.

The contributions from the one-halo and two-halo terms of the galaxy correlation function change as a function of redshift and halo mass. 
The relative amplitude of the one- and two-halo terms and the contribution that each component makes at a particular radius is affected.
\citet{wats11} demonstrate that at a constant minimum halo mass, the one-halo term evolves strongly with redshift and the two-halo term evolves weakly. 
Furthermore, both the one- and two-halo terms grow in amplitude with increasing minimum halo mass. Our mass parameter, $M_\mathrm{eff}$, is similar in this regard, but it also manipulates the overall contribution from the one-halo term relative to the contribution from the two-halo term as a function of redshift. This affects the halo model evolution in a manner that deviates from the classical definition of the HOD in which all galaxies occupy haloes in the same manner regardless of luminosity.

\begin{table*}
\centering
\caption{Marginalized parameter constraints from MCMC analysis of the PSF plus HOD model. The reported values are the 50th percentile plus and minus the 84th and 16th percentiles, respectively. The reduced $\chi^2$ values are reported for the full model and for the model outside the PSF. \label{bestfit}}
\begin{tabular}{||l c c c c c c r||} 
\hline
z bin & $\log(M_\mathrm{eff}/M_{\odot})$ & $L_0 ~[L_\odot/M_{\odot}]$ & $T_d [K]$ & $\log(A_{250} ~[Jy/sr])$ & $\log(A_{350} ~[Jy/sr])$ & $\log(A_{500} ~[Jy/sr])$ & $\chi^2_{red, outside}$\\ 
\hline
0.5-0.81 & $10.7^{+1.0}_{-0.2}$ & $0.11^{+0.05}_{-0.06}$ & $31^{+7}_{-3}$ & $5.6^{+0.1}_{-0.1}$ & $5.1^{+0.1}_{-0.2}$ & $4.5^{+0.1}_{-0.2}$ & $2.1$, $1.0$ \\
\hline
0.81-1.15 & $11.3^{+0.6}_{-0.3}$ & $0.06^{+0.03}_{-0.02}$ & $32^{+4}_{-4}$ & $5.62^{+0.08}_{-0.1}$ & $5.30^{+0.08}_{-0.1}$ & $4.7^{+0.1}_{-0.2}$ & $1.7$, $0.7$\\
\hline
1.15-1.56 & $13.3^{+0.3}_{-0.3}$ & $0.08^{+0.04}_{-0.04}$ & $42^{+6}_{-7}$ & $5.73^{+0.08}_{-0.1}$ & $5.36^{+0.09}_{-0.1}$ & $4.90^{+0.08}_{-0.1}$ & $3.3$, $1.5$\\
\hline
1.56-1.95 & $13.5^{+0.2}_{-0.2}$ & $0.18^{+0.07}_{-0.05}$ & $38^{+5}_{-4}$ & $5.6^{+0.1}_{-0.1}$ & $5.3^{+0.1}_{-0.1}$ & $4.8^{+0.1}_{-0.2}$ & $4.2$, $1.6$\\
\hline
1.95-2.21 & $13.8^{+0.1}_{-0.1}$ & $0.49^{+0.1}_{-0.08}$ & $32^{+3}_{-3}$ & $5.5^{+0.1}_{-0.1}$ & $5.0^{+0.2}_{-0.3}$ & $4.5^{+0.2}_{-0.4}$ & $2.7$, $2.6$\\
\hline
2.21-2.32 & $13.8^{+0.1}_{-0.1}$ & $1.8^{+0.7}_{-0.6}$ & $41^{+4}_{-4}$ & $5.2^{+0.2}_{-0.3}$ & $5.0^{+0.2}_{-0.3}$ & $4.7^{+0.1}_{-0.2}$ & $1.4$, $1.8$\\
\hline
2.32-2.45 & $13.8^{+0.1}_{-0.2}$ & $1.7^{+0.7=9}_{-0.6}$ & $42^{+4}_{-5}$ & $5.3^{+0.2}_{-0.2}$ & $5.0^{+0.2}_{-0.2}$ & $4.7^{+0.1}_{-0.2}$ & $2.5$, $2.3$\\
\hline
2.45-2.62 & $13.9^{+0.1}_{-0.1}$ & $2.8^{+1.0}_{-0.9}$ & $45^{+4}_{-4}$ & $5.3^{+0.1}_{-0.2}$ & $5.0^{+0.1}_{-0.2}$ & $4.8^{+0.1}_{-0.1}$ & $2.4$, $2.3$\\
\hline
2.62-2.9 & $13.9^{+0.1}_{-0.2}$ & $1.4^{+0.9}_{-0.6}$ & $41^{+7}_{-9}$ & $5.4^{+0.1}_{-0.2}$ & $5.1^{+0.1}_{-0.1}$ & $4.8^{+0.1}_{-0.2}$ & $1.7$, $1.4$\\
\hline
2.9-3.5 & $13.2^{+0.3}_{-0.2}$ & $0.32^{+0.1}_{-0.09}$ & $39^{+5}_{-4}$ & $5.2^{+0.2}_{-0.3}$ & $4.7^{+0.3}_{-0.4}$ & $4.6^{+0.2}_{-0.3}$ & $1.1$, $0.7$ \\
\hline
\end{tabular}
\end{table*}

\begin{figure*}
\begin{center}
    \includegraphics[width=6.5in]{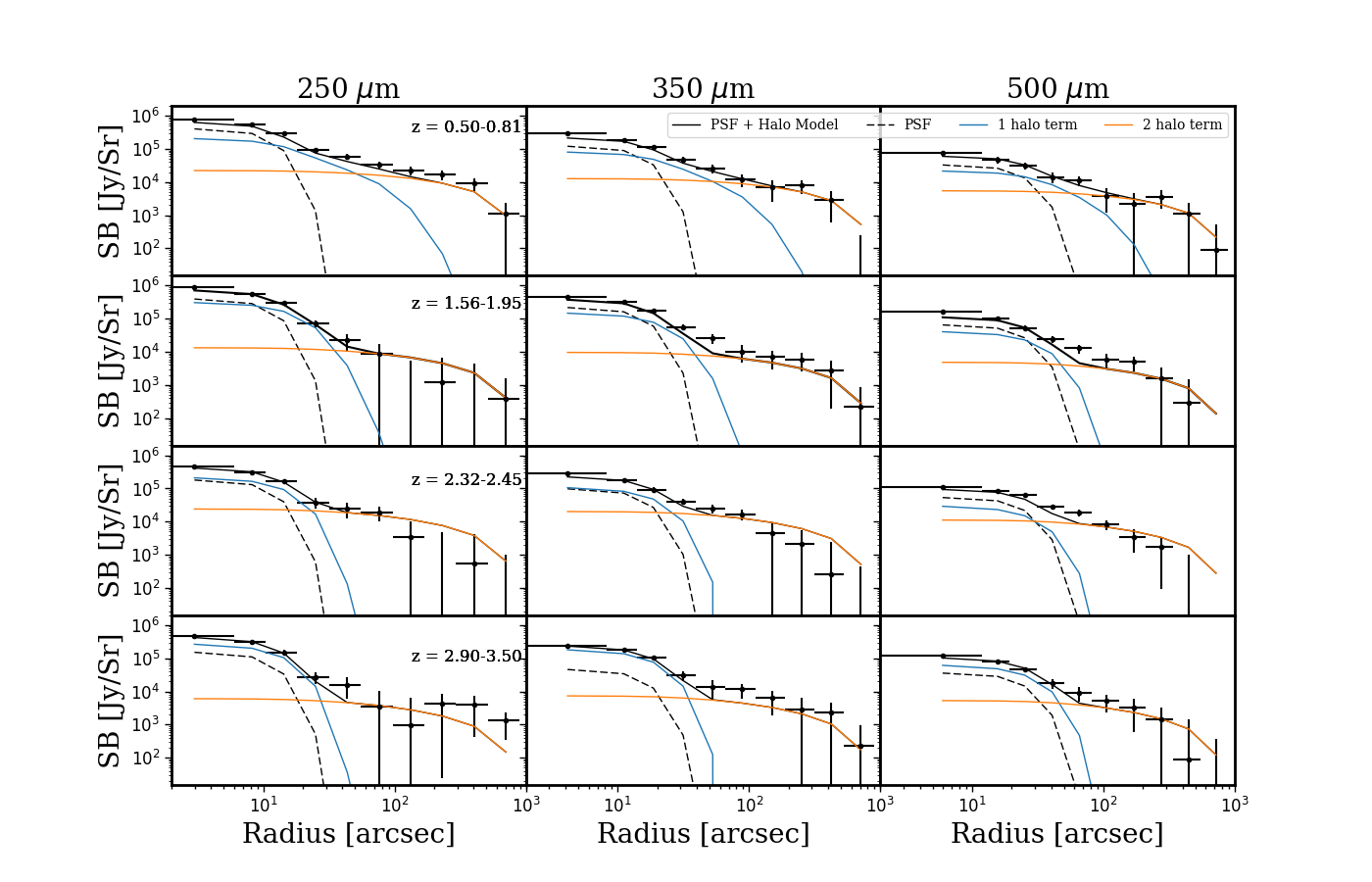}
    \caption{Angular cross correlation functions for four representative redshift bins spanning the full range in redshift probed with the complete HOD + quasar emission amplitude model overplotted as a black line. The dashed line is the best-fitting Gaussian PSF. The solid blue lines (small scales) and solid orange lines (large scales) display the one-halo and two-halo terms of the HOD model, respectively. Models are derived from the 50$^{th}$ percentile of the 1D marginalized parameter posterior distributions. Recall that the radial bins have large correlations and therefore the preferred model sometimes lies entirely above or below the data.}
    \label{profilez0369}
\end{center}
\end{figure*}

Figure~\ref{modelevolution} illustrates the evolution of the one and two-halo terms of our model with $M_\mathrm{eff}$ and with redshift for the 350 $\mu$m wavelength band. 
We sample our model at our lowest redshift bin $z=0.5-0.81$, two intermediate bins $z=1.56-1.95$ and $z=2.32-2.45$, and our highest redshift bin $z=2.9-3.5$. 
In each panel, the dust temperature and overall normalization $L_0$ are fixed to 35 K and 0.2 $L_\odot/M_\odot$, respectively, while the $M_\mathrm{eff}$ increases over two orders of magnitude from 10$^{11}$-10$^{13} M_\odot$. 
The normalization parameter $L_0$ strictly changes the amplitude of the overall halo model, as does $T_d$ to a good approximation since our wavelengths reside in the Rayleigh-Jeans regime of the SED for most temperatures and redshifts. There is a positive correlation between the dust temperature and overall normalization at fixed $M_\mathrm{eff}$.
At fixed $L_0$ an increase in $T_d$ corresponds to a decrease in amplitude of the one and two-halo terms. 
This is because the dust temperature determines the shape of the SED, or the relative amplitude of the different wavelength bands, but it is normalized in the $L-M$ relation such that the integral is equal to 1. 
This allows the parameter $L_0$ to determine the amplitude of the flux in relation to the halo mass. 
Thus, at fixed $M_\mathrm{eff}$ the light-to-mass normalized $L_0$ would have to increase with increasing dust temperature to maintain the same flux density in a given SPIRE band on the Rayleigh-Jeans side of the spectrum.

Figure~\ref{modelevolution} illustrates the effects that the $M_\mathrm{eff}$ and redshift have on the model, both in terms of overall amplitude and final shape, with the one-halo term following the shape of a projected NFW profile but heavily influenced by the convolution with the PSF. 
The broader shape of the one-halo term at low redshift as compared to high redshift is driven by evolution in the halo mass function. 
At low redshift ($z\la2$), the amplitude of the one-halo term increases relative to the two-halo term with increasing $M_\mathrm{eff}$.
This is due to a stronger increase in total luminosity of the halo (central plus satellites) than in number of satellite galaxies for a given halo mass.
As the host halo mass increases, the masses of the subhaloes also increase, and therefore, the brightness of the satellite galaxies and total luminosity of the halo increases.
As $M_\mathrm{eff}$ increases, the slope of the $L-M$ relation increases, which is why the amplitude of the one-halo term changes more than the amplitude of the two-halo term \citep{shan12}. 
This evolution becomes less strong with redshift due to the decrease in the halo mass function such that by $z\sim2$ we no longer observe an increase in one halo term with increasing $M_\mathrm{eff}$. 
In our model, the total amplitude of the one halo term is also influenced by the quasar HOD.
Our choice of quasar HOD parameters from \citet{wang15} have constant values until $z\sim2.5$ at which point they change for the higher redshift quasars. 
The new quasar HOD function beyond $z\sim2.5$ influences the one halo term in such a way that we again observe an increase in the one-halo term with increasing $M_\mathrm{eff}$.
We explore the effects of a different quasar HOD in our derived model parameters in Section~\ref{quasartests}.

\begin{figure*}
\centering

    \includegraphics[width=3.25in]{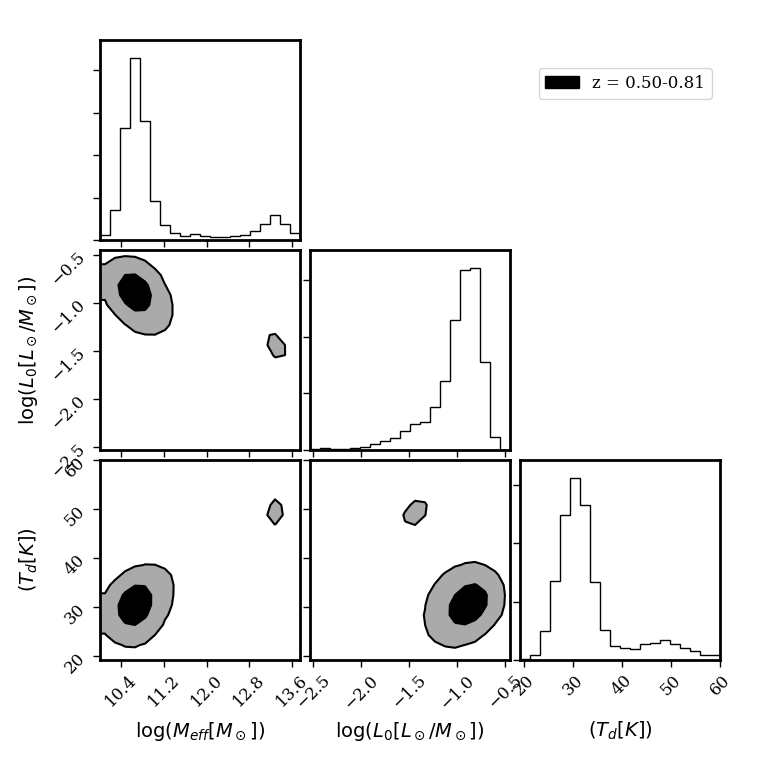}
    \includegraphics[width=3.25in]{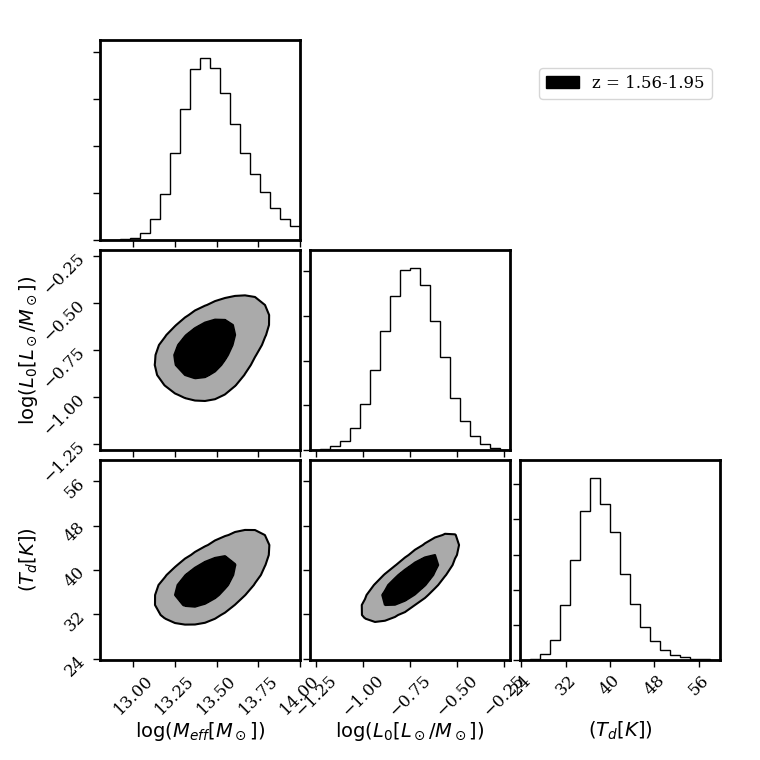}
    \includegraphics[width=3.25in]{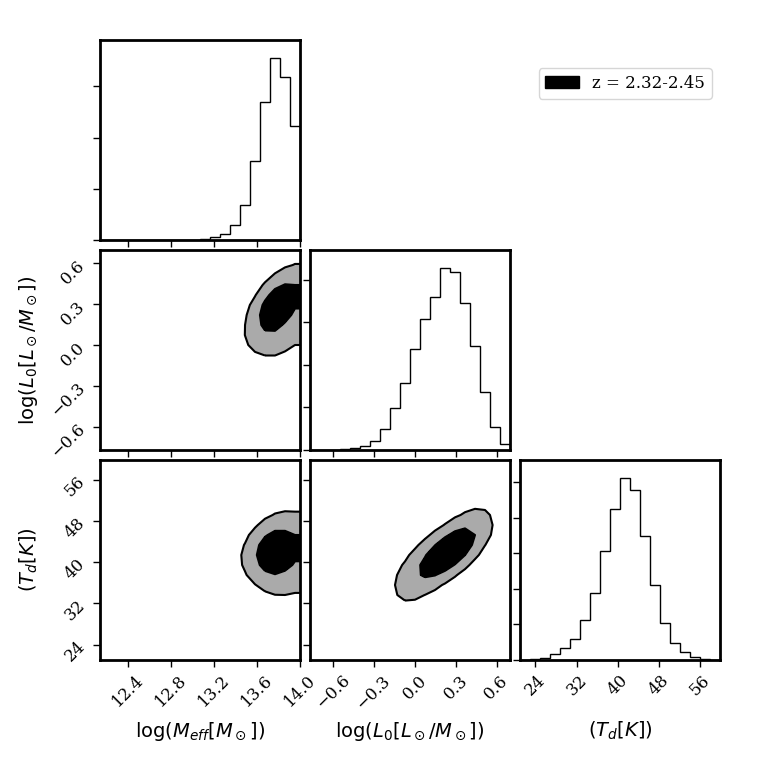}
    \includegraphics[width=3.25in]{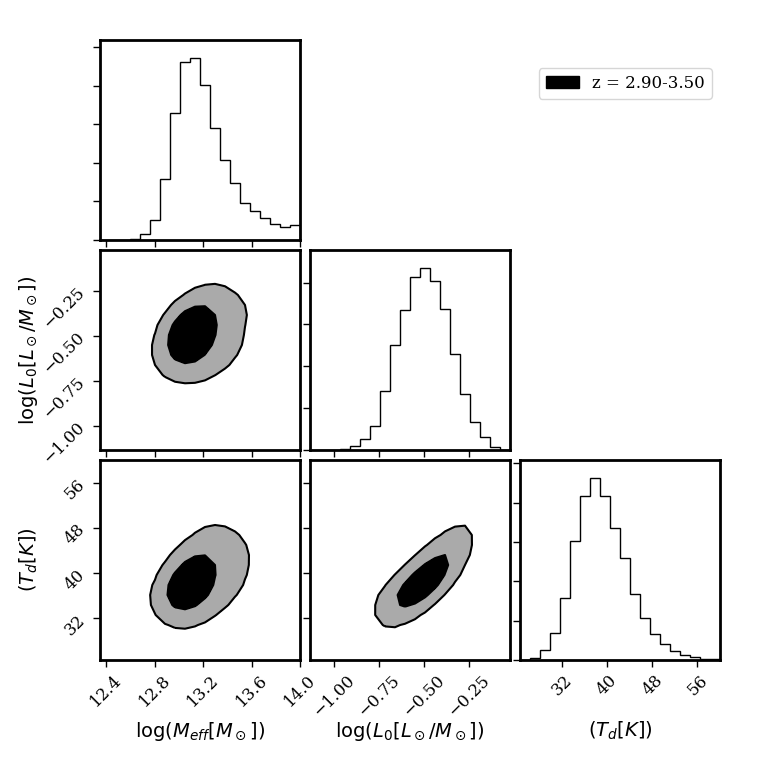}

\caption{Contours of the 68$^{th}$ and 95$^{th}$ percentiles of the posterior distribution of the parameters in our DSFG HOD model using an evolving quasar HOD for four redshift bins spanning the entire range of redshifts that we probe. The histograms show the marginalized posterior distributions. From top left to bottom right the redshift intervals are $z=0.5-0.81$, $z=1.56-1.95$, $z=2.32-2.45$, $z=2.9-3.5$. The contours for all redshift bins including the PSF amplitudes are plotted in Appendix~\ref{appendixA}. \label{cornerz0369}}
\end{figure*}

We use the package emcee \citep{fore13} to perform a Markov Chain Monte Carlo sampling of the Gaussian likelihood function defined using the covariance matrices of the radially binned data in order to understand the posterior distribution of the fit parameters. 
The marginalized constraints on the parameters for each redshift bin are presented in Table~\ref{bestfit} using the 50$^{th}$ percentile plus and minus the 84$^{th}$ and 16$^{th}$ percentiles, respectively.
Reduced $\chi^2$ values are also reported and range from 1.2 - 4.2 depending on the redshift bin. 
In many redshift bins, the cause of the poor $\chi^2_{red}$ is due to the first three data points that are contained within the PSF FWHM. 
Calculated using only data points outside of the PSF FWHM, the $\chi^2_{red}$ values range between 0.7-2.6 and are recorded in Table~\ref{bestfit}. 
There are other factors contributing to the $\chi^2_{red}$ calculated outside the PSF FWHM still indicating a poor fit. 
The available data are challenging to fit with this phenomenological model and its many assumptions.
For example, it is often the case that the transition region from the one-halo to two-halo term is not fit well. 
With respect to the DSFG SED, we use a single temperature modified blackbody to fit the average SED of the dusty galaxies, when the physics is likely more complicated. 
This model assumes optically thin dust, while studies have shown high redshift dusty galaxies have optically thick dust (\citealt{riec13}; \citealt{huan14}; \citealt{su17}). 
Furthermore, the SEDs of DSFGs typically peak around 100 $\micron$ (\citealt{taka03}; \citealt{case14}), meaning that our data points lie on the Rayleigh-Jeans tail of the spectrum until $z\sim1.5$. 
Since our data mostly sample the Rayleigh-Jeans tail of the SED, we do not have strong constraints on the dust temperature. 
We thus impose the priors described above on dust temperature and the cosmic star formation rate density. 
Nonetheless, as we have carefully estimated the covariance in the data (Section~\ref{covsec}) we argue that the errors presented in Table~\ref{bestfit} are representative of the statistical error and, as we show in Section~\ref{quasartests}, are large enough to characterize systematic uncertainties in the model as we allow model assumptions to vary.

\begin{figure}
\includegraphics[width=3.25in]{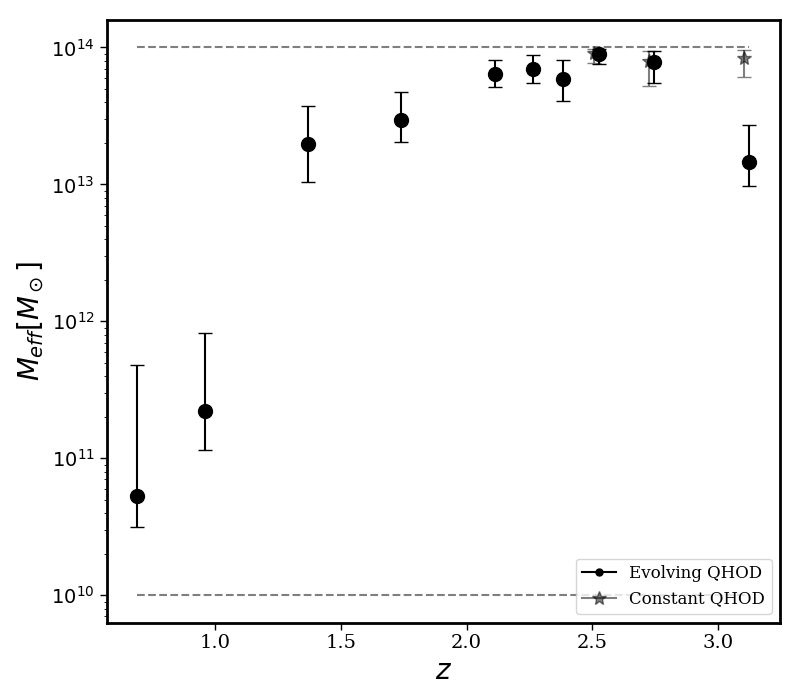}
\caption{Most efficient halo mass at hosting star formation as a function of redshift as determined from the $50^{th}$ percentile value from the posterior distribution of the Markov Chains. Error bars are $16^{th}$ and $84^{th}$ percentiles. Black points are the results from our fiducial model in which the quasar HOD parameters evolve at $z>2.5$ and gray stars are the results of keeping the quasar HOD parameters constant across all redshifts. The gray dashed lines indicate the limits of the prior. The 2$\sigma$ upper limit of the four points spanning $2.2<z<2.9$ is not well constrained. \label{Meffz}}
\end{figure}

\begin{figure}
\includegraphics[width=3.25in]{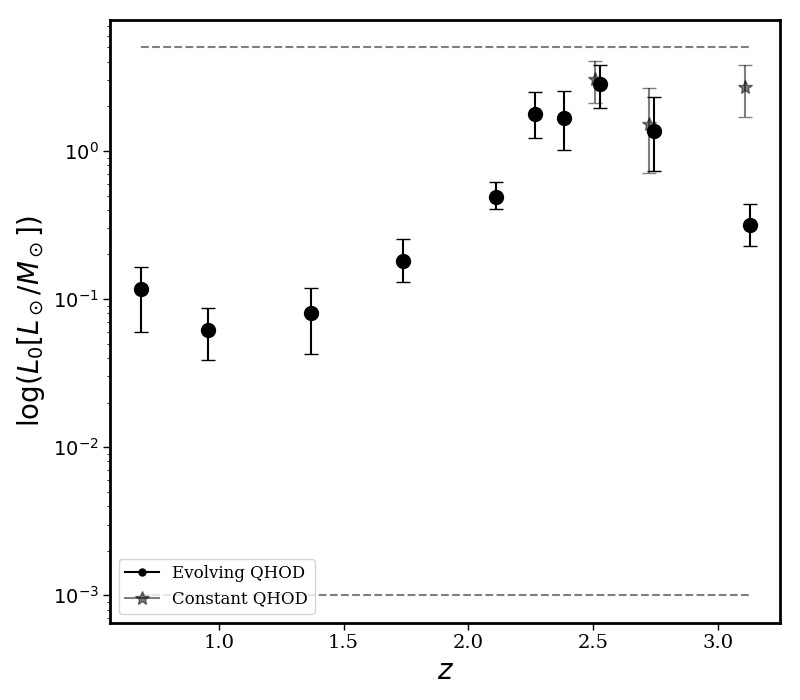}
\caption{Normalization of the $L-M$ relation for DSFGs as a function of redshift as determined from the $50^{th}$ percentile value from the posterior distribution of the Markov chains. Error bars are $16^{th}$ and $84^{th}$ percentiles. Black points are the results from our fiducial model in which the quasar HOD parameters evolve at $z>2.5$ and gray stars are the results of keeping the quasar HOD parameters constant across all redshifts. The gray dashed lines indicate the limits of the prior.  \label{L0z}}
\end{figure}

\begin{figure}
\includegraphics[width=3.25in]{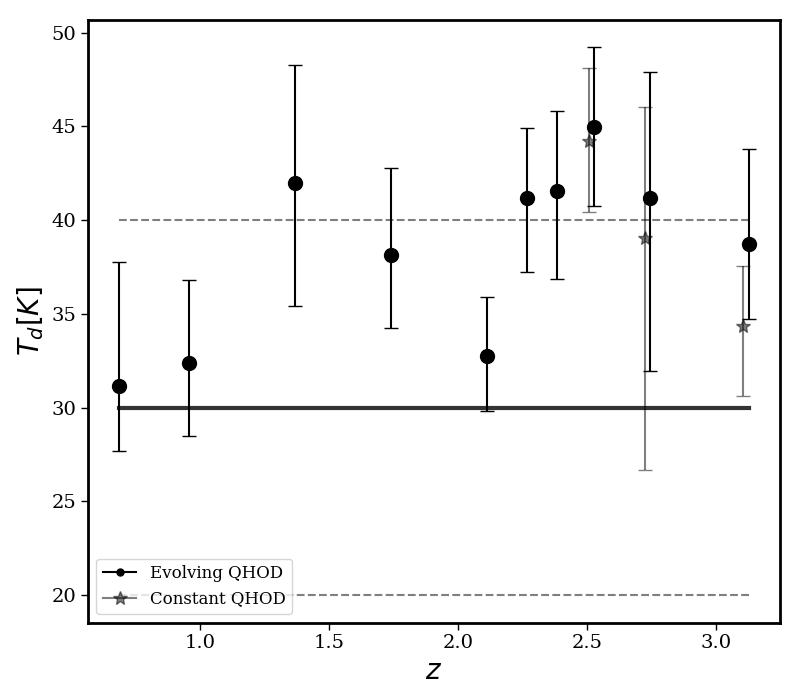}
\caption{Dust temperature of DSFG SEDs as a function of redshift as determined from the $50^{th}$ percentile value from the posterior distribution of the Markov chains. Error bars are $16^{th}$ and $84^{th}$ percentiles. Black points are the results from our fiducial model in which the quasar HOD parameters evolve at $z>2.5$ and gray stars are the results of keeping the quasar HOD parameters constant across all redshifts. The gray dashed lines indicate the limits of the prior.  \label{Tdz}}
\end{figure}

\section{Halo Model Results}
\label{results}

\subsection{Model parameters as a function of redshift}

The data and best-fitting models associated with four of our redshift bins, $z=0.5-0.81$, $z=1.56-1.95$, $z=2.32-2.45$, and $z=2.9-3.50$, are presented in Figure~\ref{profilez0369}. 
The rows show the data from a given redshift bin, labeled in the first column, and the columns show the data from the three wavelength bands.
The black solid line in each plot is the best-fitting complete model: PSF fit to quasar emission plus the halo model. 
The best-fitting model is defined using the 50$^{th}$ percentile values of each parameter taken from the 1D marginalized posterior distribution given by the MCMC. 
The blue solid lines show the small scale, one-halo term calculated using the best-fitting parameters of the complete model. 
The orange solid lines extending to large scales show the two-halo term calculated using the best-fitting parameters of the complete model. 
The dashed line is the PSF fit to the amplitude of the quasar emission. Figure~\ref{cornerz0369} displays the marginalized distribution of the HOD parameters in our model for these four redshift bins. 
The profiles with best-fitting models and marginalized parameter contraints for the other six redshift bins, as well as individual plots for the four bins plotted in Figure~\ref{profilez0369}, are displayed in Appendix~\ref{appendixA}. 

Figures~\ref{Meffz},~\ref{L0z}, and~\ref{Tdz} display the parameters $M_\mathrm{eff}$, $L_{0}$, and $T_d$ as a function of redshift. 
The black data points with error bars indicate the 50$^{th}$, 16$^{th}$, and 84$^{th}$ percentiles from the marginalized distributions. 
We find evidence that $M_\mathrm{eff}$ increases with redshift up to $z\sim2.9$, with values ranging from $\log(M_\mathrm{eff}/M_\odot) = 10.7-13.9$.
This is a sensible result as the data prefer a large amplitude one-halo term in relation to the two-halo term at all redshifts, and the only way to achieve this result as a function of increasing redshift is by increasing $M_\mathrm{eff}$ (see Figure \ref{modelevolution}).
We also see an increase in $L_0$ toward higher redshifts in order  to boost the overall amplitude of the model.
In the four redshift bins spanning $z=2.2-2.9$ the posterior distriubtion of $M_\mathrm{eff}$ is pushing against the upper limit of the prior such that the 2$\sigma$ upper limits are not well constrained. We test increasing the upper limit to 1e15 $M_\odot$, but $M_\mathrm{eff}$ is still not well constrained.
The range of derived $L_0$ and $M_\mathrm{eff}$ in the last three bins reflects the uncertainty in the quasar HOD parameters.
These two parameters decrease in the highest redshift bin in the fiducial model with evolving quasar HOD, while the drop is not present in the alternative model of using constant quasar HOD parameters across all redshifts (gray stars in Figures~\ref{Meffz}-\ref{rhoSFR}). 
As shown in Figure~\ref{Tdz}, the dust temperatures evolve marginally with redshift within their 10-20\% uncertainties (at most $2\sigma$ between the $z\sim0.7$ and the $z\sim2.5$ values). 
We are unable to obtain well constrained values within the limitations of this dataset due to the fact that there are only three wavelength points to define the SED and the dust temperature is positively correlated with the normalization $L_0$. 
The values of $M_\mathrm{eff}(z)$ are robust against the choice of mean and standard deviation of the Gaussian prior on $T_d$.
In the next section, we test the model parameter dependencies on various assumptions made in the fiducial model.

\subsection{Testing model parameter dependencies}
\label{quasartests}
In this section we explore the dependence of our model parameters on the assumptions made in the evolving quasar HOD model, particularly those relating to the quasar portion of the model angular cross correlation function. We also test the dependence of our results on the SED of DSFGs, on the prescribed redshift evolution of the $L-M$ relation, and on the width of the $L-M$ relation. We implement these tests by modifying the variable(s) under question and running the MCMC analysis while keeping the other aspects of the fiducial model fixed.

Because we use the results of quasar auto-correlation studies to fix the quasar bias parameter and quasar HOD parameters (see equations in Section~\ref{qsohod}), we run tests to understand the dependencies of our derived DSFG HOD parameters on the assumptions about quasar clustering. 
In most redshift bins the correlated signal on scales larger than $\sim30''$ is primarily associated with the two-halo term, which determines large scale clustering parameters.
The two-halo term in the correlation function is derived from a re-scaling of the dark matter correlation function by the biases of the observed populations.
The quasar bias at high redshift is still under investigation with \citet{shen07} finding an increase beyond $z\sim2.9$ by a factor of $\ga 2$ and more recent studies such as \citet{efte15} finding that it might remain relatively flat with a value of $b\sim3.5$ up to $z\sim3$.
In our fiducial model, we use the value from \citet{shen07} of $b\sim8$ in our highest redshift bin. 
As a test, we set the quasar bias to a constant value $b=3.5$ at $z>2$, which affects only our highest redshift bin $z=2.9-3.5$.
The result is an increase in the normalization $L_0$ and a decrease in the efficient halo mass $M_\mathrm{eff}$, both by an amount $< 1\sigma$ deviation from the fiducial model. 
Our choices of quasar biases presented in Table~\ref{bqso} are taken from references that use slightly different cosmologies, but we find that the adjustments to $b_{q}$ in our chosen cosmology are smaller than their reported uncertainties. 
The only case in which this may not be true is for the highest redshift bin, in which the adjustment to their value of the quasar bias in our cosmology brings its value down to $b_q=7.38$. 
Because decreasing the bias by a factor $\ga 2$ does not appreciably change our results, we are not concerned about a 7\% change. 

It is not just the quasar bias that impacts the two-halo term, since the normalization of the DSFG-quasar cross-correlation halo model $L_0$ is the same for both the one-halo and two-halo terms.
We must therefore test the effect of changing the values of the parameters that describe the quasar halo occupation function (Equation~\ref{Nqso}). 
As outlined in Section~\ref{qsohod}, the quasar halo occupation function parameters only go directly into the one-halo term of the model, as the relevant parameter for the two-halo term is the quasar bias. 
Of the five parameters that describe Equation~(\ref{Nqso}), the parameters that significantly affect the amplitude of the one-halo term are the minimum mass for hosting a central quasar, the minimum mass for hosting at least one satellite, and the power law index setting the mean number of satellite quasars as a function of halo mass. 
To test the dependence of our results on these three quasar parameters, we alter their values while maintaining the correct order of magnitude for the number density of quasars ($n_q \approx 10^{-6} Mpc^{-3}$) and redo our MCMC analysis. 

We first fix the minimum halo mass for hosting a central quasar ($M_{min, q}$, Equation~\ref{Ncenqso}) to be the same at all redshifts. 
The change we implement increases the values of $M_{min,q}$ from those found by \citet{wang15} ($10^{12.4} M_\odot$ for $z<2.5$ and $10^{12.3} M_\odot$ for $z>2.5$) to $10^{13} M_\odot$.
The effect of increasing $M_{min,q}$ serves to increase the amplitude of the one halo term, and thus causes a corresponding decrease in the best-fitting values of the normalization $L_0$ at all redshifts by $\sim 1\sigma$. 
$M_\mathrm{eff}$ remains consistent with the fiducial values for z bins 1-2 ($0.5\leq z \leq $ 1.15), with similarly large error bars as in the fiducial model. 
$M_\mathrm{eff}$ decreases by $\sim 1\sigma$ in the third redshift bin ($z=1.15-1.56$). 
In the redshift bins 4 and 5, spanning $z=1.56-2.21$, the best-fitting $M_\mathrm{eff}$ is $\sim2\sigma$ lower than in the fiducial model, and in the redshift bins with $z\geq2.21$ there is a $\sim1\sigma$ decrease in $M_\mathrm{eff}$.
There is a slow increase in $M_\mathrm{eff}$ through $z=2.9$, with the values at redshifts spanning $2.45<z<2.9$ consistent with the fiducial model. 
The take away from this test is that in comparison with the fiducial model, with an increase in the minimum mass hosting quasars, we still observe the increase in the effective halo mass hosting clustered DSFGs with redshift, but with a more gradual increase from $\log(M_\mathrm{eff}/M_\odot) = 10.7$ to $\log(M_\mathrm{eff}/M_\odot) = 13.6$ at $z=2.9$.
Furthermore, the reduced $\chi^2$ values for the best-fitting model with $M_{min,q}=10^{13}M_\odot$ are larger than our fiducial model, with the p-value ratios of the fiducial to this model $>5$ for all redshifts, suggesting that this version is a less accurate functional form for the quasar HOD.
We do not test the effect of decreasing the minimum mass of haloes hosting quasars because it is relatively well established that quasars live in haloes of at least $10^{12} M_\odot$ (\citealt{shen13}; \citealt{rich12}; \citealt{porc04}). 

Modifying the quasar satellite halo occupation function, specifically in changing the value of the minimum halo mass to host a satellite quasar $M_1$ or the power law index $\alpha$, also causes significant changes in the amplitude of the one-halo term.
The effect of decreasing the value of $M_1$ at high redshift ($z\ga 2.5$) is to re-establish the increase in the one-halo term relative to the two-halo term with increasing $M_\mathrm{eff}$ as exemplified in Figure~\ref{modelevolution}. 
It is this specific aspect of the fiducial model that results in the decrease in $L_0$ at high redshift and a corresponding decrease in $M_\mathrm{eff}$ in the highest redshift bin. 
If instead we fix the HOD parameters to be constant across all redshift bins to the values of the \citet{wang15} low redshift sample, which increases the value of M1 in the last 3 bins  (see Table~\ref{bqso}), we find a significant increase ($> 5\sigma$) in $M_\mathrm{eff}$ and $L_0$ for the highest redshift bin. 
The results of this change are shown as gray stars in the halo model parameter results as a function of redshift, Figures~\ref{Meffz}-\ref{rhoSFR}. 
Other studies provide evidence that the clustering of quasars can be modeled by constant HOD parameters up to $z\sim2.5$ (\citealt{coil07}; \citealt{ross09}; \citealt{rich12}), while \citet{rich12} and \citet{chat12} provide some evidence that the quasar HOD parameters evolve at $z\ga3.0$ in a manner that is consistent with the \citet{wang15} findings. 
\citet{rich12} implement an HOD model of SDSS quasars for their $\bar{z}=1.4$ sample and for the \citet{shen07} $\bar{z}=3.2$ sample and find that the minimum mass of central quasars decreases in the high-z sample relative to the low-z sample, but they cannot draw any conclusions about the high redshift satellite quasar occupation function. \citet{chat12} find similar results for simulated AGN at $z=1$ and $z=3$ with $L_{bol}\geq10^{42}$ erg s$^{-1}$, and find that the minimum mass for hosting a satellite quasar is smaller at $z=3$, similarly to \citet{wang15}.

The limited information on HOD models of observed high redshift ($z > 2.5$) quasars is a limiting factor in determining our halo model parameters of DSFGs in this regime. 
A decrease in the quasar HOD parameter $M_1$ at $z>2.5$ produces a decrease in the derived $M_\mathrm{eff}$ in the highest redshift bin by $>5\sigma$.
Constant quasar HOD parameters fixed to the \citet{wang15} low $z$ values across our entire redshift range produce a continued increasing or flat distribution in the derived properties of the DSFGs. 
Though some studies have found evidence for evolution of the $z>2.9$ quasar HOD parameters (\citealt{rich12}; \citealt{chat12}), we need more information about the satellite quasar occupation function beyond $z\sim2.5$ in order to use them as a useful tracer at these redshifts and higher. 
Furthermore, it is still possible that the evolution of the quasar HOD parameters is prevalent across all epochs, not just the high redshift regime, and it is not precisely determined that the low redshift values used in our fiducial model are correct.
For example, the \citet{rich12} $\bar{z}=1.4$ halo occupation function differs from that of \citet{wang15} in that both the minimum halo mass for hosting central quasars and the minimum halo mass for hosting a satellite quasar is increased relative to the \citet{wang15} values.
Further investigating the parameters that describe the halo occupation function of quasars is beyond the scope of this paper.

To test the dependence of our results on the DSFG SED, we implement a version of the model in which we do not place a prior on the star formation rate density, but vary the mean of the Gaussian prior on dust temperature.
The choices of mean temperatures are informed by two sources.
The first is a mean of 25~K informed by the best-fitting results of \citet{vier13} and \citet{serr14}. 
The second is a mean of 35~K informed by fitting a modified blackbody to the \citet{beth12} effective SEDs for star forming galaxies. 
These fits yield dust temperatures increasing from $\sim31-37$K as a function of the average redshifts of our bins.
Implementing these tests allows us to explore the degeneracy between the dust temperature and the other parameters in the model, in particular $L_0$.
$T_d$ and $L_0$ are positively correlated at all redshifts, so it is possible to achieve the same model by increasing or decreasing both of these parameters together. 
The results of these tests are plotted in Appendix~\ref{appendixB} with gray triangles indicating the mean 25~K resuls and gray X's indicating the mean 35~K results.
Figures~\ref{Meffz2}-\ref{rhoSFR2} are analogs of Figures~\ref{Meffz}-\ref{rhoSFR} with comparisons of these dust temperature test to the fiducial model. 
The importance of these results is that the constraints on $M_\mathrm{eff}$ are unchanged, except in the first two bins where $M_{eff}$ is $\la 1\sigma$ lower. 
The consquences are that $L_0$ and $T_d$ shift by $\sim 1\sigma$, which in some bins creates $\ga2\sigma$ shifts in $\rho_{SFR}$. 
The resulting $M_\mathrm{eff}$ values remain consistent with the fiducial model (changes $\le$ 1 $\sigma$) in all redshift bins.
Thus, using a Gaussian prior on $\rho_{SFR}$ does not affect the primary result of this paper, but helps to yield more physical values for $T_d$ and $L_0$. 

As described by Equation~(\ref{Lnu}), our model contains a redshift evolution parameter $\Phi(z)$ that describes the observed increase in specific star formation rate, and therefore luminosity, with redshift. 
The exact shape of that evolution is not precisely known. 
In this paper, we are searching for physical parameters that evolve with redshift and produce the observed redshift distribution of our signal, and thus, we test the consistency of our results with changes in the prescribed redshift evolution $\Phi(z)$. 
We compare our fiducial model with a power law index $\eta = 2.19$ (consistent with \citealt{vier13}) to a model with $\eta = 3.6$ (consistent with \citealt{plan14}).
Because the relation flattens at $z\geq2$ the effect of this change on the halo model is to increase the amplitude by a factor $\la 4$, increasing with redshift according to $(1+z)^\eta$ until flattening at $z=2$.
The three PSF Amplitudes, $M_\mathrm{eff}$, and $T_d$ are unaffected by this change.
The normalization of the $L-M$ relation $L_0$ changes by  $\leq1.5\sigma$ relative to the fiducial model.

We also test the results of our model in the scenario in which we increase the dispersion in the $L-M$ relation, $\sigma_{L/M}$. 
Increasing $\sigma_{L/M}$ from 0.6 to 0.7 does not change the results at all. 
Increasing it dramatically from 0.6 to 2 has the effect of increasing the dust temperature and decreasing the overall normalization, $L_0$ both by $\sim1\sigma$, and the changes to  $M_\mathrm{eff}$ are by amounts $<1\sigma$. 
It also significantly decreases the quality of the fits, i.e. the $\chi^2_{red}$ statistic increases by up to a factor of 2 and the models visibly do not fit the data well. 
This test of dramatically increasing $\sigma_{L/M}$ provides supporting evidence that it is necessary to model the data using a range of halo masses that are most efficient at hosting star formation, which is in agreement with the other studies using this parameterization (\citealt{serr14}; \citealt{plan14}; \citealt{vier13}; \citealt{shan12}). 
This parameterization is also consistent with what is expected from the relation between halo mass and SFR using abundance matching \citep{beth12}. 
Though the exact value of $\sigma_{L/M}$ is currently unconstrained, we can say with some confidence that it should be less than 1, and the exact value does not affect constraints on the best-fitting parameters in the HOD model. 

\subsection{Physical interpretation of the halo model results}
\label{interpret}

The results of our HOD model indicate evidence for redshift evolution of the halo mass that is most effective at hosting star forming galaxies. In our analysis, we find that in the epoch of peak star formation in the universe the haloes most effective at forming stars are more massive than they are at $z\sim0.5$. 
$M_\mathrm{eff}$ is the value at the peak of the luminosity-halo mass relation, meaning that the most IR luminous (or most actively star forming) objects are hosted in haloes of this mass scale.
This does not mean that this mass scale should be considered characteristic of the halo masses at a given redshift. 
In fact, the most efficient halo masses that we derive are more massive than typical haloes hosting galaxies, particularly beyond $z\approx1.5$, where we find $M_\mathrm{eff}$ to be on the order of a few $\times 10^{13} M_\odot$. 

\begin{figure}
\includegraphics[width=3.25in]{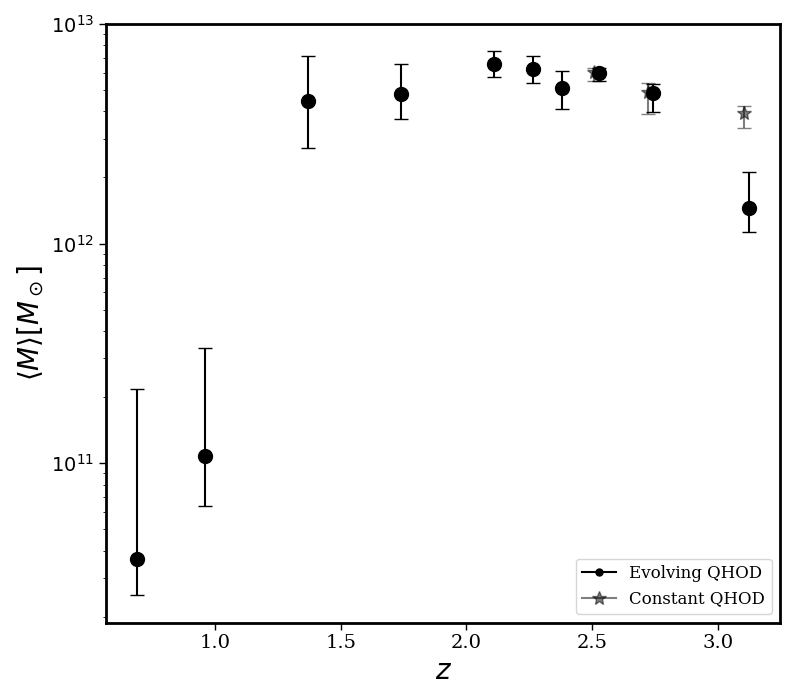}
\caption{Average halo mass of DSFGs as a function of redshift as determined from Equation~(\ref{Mmean}). Error bars are $16^{th}$ and $84^{th}$ percentiles. Black points are the results from our fiducial model in which the quasar HOD parameters evolve at $z>2.5$ and gray stars are the results of keeping the quasar HOD parameters constant across all redshifts. \label{Mmeanz}}
\end{figure}

\begin{figure}
\includegraphics[width=3.25in]{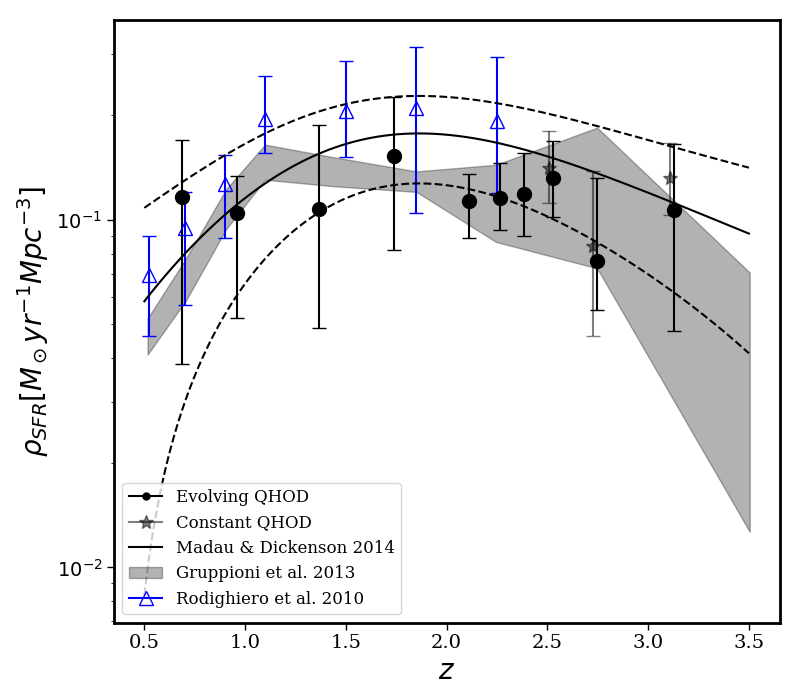}
\caption{Cosmic star formation rate density as a function of redshift as computed from the results of our halo model. Error bars are $16^{th}$ and $84^{th}$ percentiles. Black points are the results from our fiducial model in which the quasar HOD parameters evolve at $z>2.5$ and gray stars are the results of keeping the quasar HOD parameters constant across all redshifts. Also shown are the derived cosmic star formation rate densities from \citet{rodi10}, \citet{grup13}, and \citet{mada14} with which the derived values agree within $\sim 1 \sigma$ uncertainty with the exception of the second and last redshift bins. The dashed lines indicate the 1$\sigma$ boundaries of the Gaussian prior. \label{rhoSFR}}
\end{figure}

The value of $M_\mathrm{eff}$ in our lowest redshift bin can be compared to the results of \citet{serr14} who find for DSFGs at $z\sim0.5$ that $\log(M_\mathrm{eff})$ = 12.84 $\pm$ 0.15.  
This is greater than our value of $\log(M_\mathrm{eff})$ = 10.7$^{+1.0}_{-0.2}$ by two orders of magnitude, though with our large uncertainty on the upper limit it is just over 2$\sigma$ away from our value.
In their analysis, they use luminous red galaxies as their tracer population and also find a lower dust temperature $T_d=26$~K for the DSFG SEDs. 
If we constrain our $T_d$ to 26~K in our $z=0.5-0.81$ bin we find a corresponding decrease in $L_0$ to maintain the same total model amplitude as the fiducial model, the same best-fitting values for $M_\mathrm{eff}$ as our fiducial parameters, and the same $\chi^2_\mathrm{red}$ statistic.
\citet{vier13} and \citet{plan14} implement the same halo model for auto- and cross-correlations of the \textit{Herschel}-SPIRE maps and \textit{Planck} maps and find the peak halo mass integrating the model over $z=0-4$ and $z=0-6$ to be $\log(M_\mathrm{peak}/M_\odot) = 12.1 \pm 0.5$, and $\log(M_\mathrm{peak}/M_\odot) = 12.6 \pm 0.1$, respectively.
Both of these values are within $\sim$2$\sigma$ of our $z=0.5-0.81$ and $z=0.81-1.15$ samples. 
A potential source of the discrepancy between our lowest redshift $M_\mathrm{eff}$ values and those found by \citet{serr14}, \citet{vier13}, and \citet{plan14} lies in the quasar HOD parameters. 
Our two lowest redshift bin $M_\mathrm{eff}$ values can be increased by $\sim$1.5 dex by increasing the minimum mass to host a satellite quasar by factors of $\sim$2.5 and $\sim$1.8, respectively, which effectively removes satellite quasars from all but the most massive dark matter haloes at these redshifts. 

At $z>1.5$ we see an increase in the most efficient halo mass such that its value is constrained to be above $10^{13} M_\odot$ out to $z\sim2.9$.
Our derived most efficient halo masses in the redshfit bins spanning $z=1.56-2.9$ reach values that are at  least 0.5 dex higher than the previous results obtained by integrating over all redshifts. 
Two obvious differences between our method and others' are our division into smaller redshift bins allowing for a unique redshift evolution in the model parameters, and our constraint on the overall normalization parameter $L_0$. 
In \citet{vier13} and \citet{serr14} the normalization is restricted by the intensity of the CIB in each respective frequency band as previously derived (e.g., \citealt{laga00}). 
In our model, we use the integrated infrared luminosity density via the cosmic star formation rate density from \citet{mada14} to constrain all three of our HOD parameters.
Furthermore, we find that our values of efficient halo masses agree with model case 4 in \citet{shan12} in which they allow their normalization parameter to vary.
The normalization parameter alone is not physically meaningful, and we find our halo mass results are robust against changes in $L_0$. 

As a point for further comparison to other studies and a means of better understanding the population of sources that make up the CIB, we use the results of our model to calculate the luminosity-weighted mean halo mass in each redshift bin. 
In general, the mean halo mass can be calculated by:
\begin{equation}
\langle M \rangle = \frac{\int \frac{dn}{dM}M \langle N \rangle dM}{\int \frac{dn}{dM} \langle N \rangle dM}
\end{equation}
For our parameterization of the number of central and satellite galaxies weighted by the luminosity density, this becomes:
\begin{equation}
\langle M \rangle = \frac{\int \left[ N_{cen} \Sigma(M) + \int \frac{dn}{dm} \Sigma(m) dm \right] \frac{dn}{dM}dM}{\int \left[\frac{N_{cen} \Sigma(M)}{M}+ \int \frac{dn}{dm}\frac{\Sigma(m)}{m} dm\right] \frac{dn}{dM} dM}
\label{Mmean}
\end{equation}
The mean halo masses of star-forming galaxies producing correlated signal around quasars are plotted in Figure~\ref{Mmeanz}. 
The shape of the redshift evolution of $\langle M \rangle$ is similar to that of the most effective halo mass.  
We find that the DSFGs at $z\ga1.5$ occupy haloes in the range of masses $\sim10^{12}M_\odot - 10^{13} M_\odot$. 
The values of $\langle M \rangle$ are consistent with expected halo masses of star forming and submillimeter galaxies (\citealt{hick12}; \citealt{chen16}; \citealt{mani18}).
Our finding is consistent with that of \citet{coch17} for star forming galaxies identified in HiZELS sample, however our results hint at a redshift evolution that is not seen in their sample between $z=0.8-2.2$.  

We further investigate the average physical properties of DSFGs clustered around quasars by computing the cosmic star formation rate density in each of our redshift bins and comparing with other measurements. 
Using the $L-M$ relation given by our best-fitting parameters, we compute the emissivity density (Eq.~\ref{jnu}), convert this to flux density, and use the derived SED to compute an estimate of the average bolometric infrared luminosity $L_{IR}$ of DSFGs clustered around quasars. We compute $L_{IR}$ per comoving volume within each redshift bin and integrate over redshift to obtain the far infrared luminosity density of DSFGs within each redshift bin.
We then use the scaling relation between $L_{IR}$ and star formation rate given by \citet{kenn98} to compute the star formation rate density $\rho_{SFR}(z)$.
Figure~\ref{rhoSFR} displays these results with comparisons to other observationally-derived star formation rate densities.
We use the curve from \citet{mada14}, plotted as the solid black line, as a Gaussian prior on our HOD parameters with a standard deviation of 0.05 $M_\odot$/yr.
These one sigma boundaries are plotted as dashed lines.
For comparison, we plot the findings of \citet{rodi10} and \citet{grup13} resulting from studies of infrared luminosity functions from \textit{Spitzer} and \textit{Herschel} data, respectively.. 
The star formation rate densities implied by the best-fitting halo model parameters of our evolving HOD model agree with the average star formation rate density of the others within our propagated 1 $\sigma$ uncertainties at all redshifts.

The decrease in the normalization of our fiducial halo model $L_0$ at $z\ga2.9$ and the 5$\sigma$ decrease in $M_\mathrm{eff}$ at $z=2.9-3.5$ provides reason to believe that the assumed HOD parameters of high-z quasars do not adequately capture the clustering signal.
When altering the high redshift quasar model parameters to remain constant with the low redshift values, we find $L_0$ has a more smooth evolution that continues increasing with redshift, and the $\rho_{SFR}$ values remain in agreement with expectations from other results. 
Fixing the quasar HOD parameters to the low-z values causes the evolution of the both the one halo and two halo terms in the highest redshift bin $z\sim3.1$ to decrease with increasing $M_\mathrm{eff}$. 
In order to maintain the correct ratio of one halo to two halo amplitude, it requires a larger $M_\mathrm{eff}$ as compared to the fiducial model. 
This is compensated by an increase in $L_0$ and a decrease in $T_d$, but the overall emissivity is lower. 
Our results indicate a need for a better understanding of the high redshift ($z\geq2.9$) quasar halo occupation distribution.
A recent study of high-redshift ($2.9<z<5.1$) quasar clustering by \citet{timl17} find the quasar bias to be 7.32 $\pm$ 0.12, consistent with the value used in our fiducial model, but they do not implement a full halo occupation model. 
Instead they convert the bias to a characteristic mass scale for high redshift quasars and find $M_\mathrm{halo}\sim6\times10^{12}M_\odot$, consistent with our $\langle M \rangle$ at $z\sim3.1$. 

We expect the average star formation rate density of the sources we probe clustered around quasars to be in agreement with the average $\rho_{SFR}$ of the universe as long as our model appropriately considers the clustering of dark matter haloes and how those haloes are populated by quasars and dusty star-forming galaxies.
Despite our many assumptions and the simplicity of our model, we find $\sim 1 \sigma$ consistency between the derived cosmic star formation rate density of the DSFGs probed by our analysis and the cosmic star formation rate density found in other studies.
While our study probes highly clustered dark matter haloes, the favorable comparison between our derived $\rho_{SFR}$ and other results suggests that our analysis probes representative star-forming haloes.
This is a non-trivial result. 
For example, if star-formation properties of dark matter haloes of $10^{12}-10^{13} M_\odot$ were strongly affected by their Mpc-scale environment, then extrapolating from the haloes used in this work to $\rho_{SFR}$ would not have produced a favorable comparison. 

Using the $L-M$ relation, we also calculate the contribution to the emissivity density from satellite DSFGs. 
This model, also used in \citet{shan12} and \citet{vier13}, is unique in that it allows for more power in the one halo term without increasing the number of satellite galaxies, but instead allowing for brighter satellite galaxies. 
Using our best-fitting parameters in each redshift bin we calculate that $\sim2-21\%$ of the emissivity density is from satellite DSFGs. 
This fraction is consistent with the $(14\pm8)\%$ satellite fraction derived from clustering measurements of resolved SPIRE sources with flux limits greater than 30 mJy \citep{coor10}.

\subsection{Downsizing of the halo masses hosting dusty star-forming galaxies}

We find evidence of the peak halo mass in the $L-M$ relation $M_\mathrm{eff}$ increasing with redshift up to $z=2.9$ in a manner that is consistent with cosmic downsizing (\citealt{cowi96}; \citealt{font09}); the finding that stellar growth or star formation undergoes a shift toward lower characteristic mass scales as a function of cosmic time.
More specifically, we consider the definition in which the most intense star formation occurs in increasingly massive dark matter haloes as a function of increasing redshift.
This is evident in Figure~\ref{Meffz} for our redshift bins between $z=0.5$ and $z=2.9$ where the most effective halo mass at hosting the most IR-luminous galaxies increases from $M_\mathrm{eff}\sim10^{10.7} M_\odot$ to $M_\mathrm{eff}\sim10^{13.9} M_\odot$.
This trend in halo mass was found by \citet{conr09} in two different models where stellar growth is dominated by either star formation or merger activity. 
The presence of this trend in two extreme modes of star formation, which must both contribute to stellar growth, suggests that such downsizing is a general characteristic of the cosmic evolution of stellar mass growth.
In our paper, we find observational evidence for this trend.

This trend in infrared sources is supported by clustering studies of sources in the \textit{Herschel} PACS Evolutionary Probe (PEP) survey. 
\citet{magl13} investigate the large scale (two-halo) clustering of highly star forming ($\ga 100 M_\odot yr^{-1}$) infrared-selected galaxies at $z\la1$ and derive a minimum halo mass $M_\mathrm{\min}\sim10^{11}M_\odot$.
They compare these results to the clustering of \textit{Herschel}-PEP sources at $z\sim2$ \citep{magl11}, which are much more strongly clustered and indicate a minimum halo mass $M_\mathrm{min}\sim10^{13.9}M_\odot$.

When examining the mean halo mass of the population of DSFGs, we find that the evolution with redshift increases over the first few bins, then flattens to a few $\times 10^{12} M_\odot$ around $z\sim1.5$. 
These results are consistent with \citet{behr13} who find that at $z\sim2$ the most actively star forming galaxies have halo masses near $10^{13} M_\odot$, but conclude based on instantaneous baryon conversion efficiency (average star formation rate divided by the baryon accretion rate) that haloes of mass $\sim 10^{12} M_\odot$ are the most efficient at converting gas into stars at all epochs.
The average halo mass result is consistent with a second definition of comic downsizing: more massive haloes host galaxies that formed their stars earlier. 
This is known as "archaeological downsizing" as it is evidenced in the star formation histories of $z=0$ elliptical galaxies \citep{thom05}. 
A flat or increasing $\langle M \rangle$ dependence on redshift is consistent with this type of downsizing because haloes of mass $10^{12} M_\odot$ at $z\sim2$ will grow to be more massive in the present day universe than those of equal mass at $z\sim1$.

Our findings are also consistent with \citet{ishi16} who  perform HOD modelling of the star-forming $gzK$ galaxy sample at $z\sim2$. 
They construct an auto-correlation function and model the HOD using the number of sources. 
Our choice of minimum halo mass for the DSFGs is in good agreement with their best-fitting minimum halo mass to host a satellite DSFG. They determine mean halo masses from their analysis with values ranging from $\sim 5\times10^{12} - 6\times10^{13} M_{\odot}$. 
They discover that the luminosity dependence of the mass parameters is the same at $z\sim2$ as at $z\sim0$ but that the high redshift halo masses are larger over all magnitude ranges. 
They thus conclude that galaxies formed in more massive haloes at $z\sim2$ than they do at $z\sim0$, consistent with our results.  

The halo masses and trends with redshift are similar to those found for other sub-mm galaxies. \citet{wilk17} perform a clustering analysis on 850 $\mu$m-selected sub-millimeter galaxies in the UKIRT Infrared Deep Sky Survey (UKIDSS), UltraDeep
Survey (UDS) field. 
They find that these galaxies occupy average halo masses $> 10^{13} M_\odot$ at $z>2.5$, while their samples between $z=2.0-2.5$ and $z=1.0-2.0$ occupy haloes of a few $\times 10^{12} M_\odot$ and a few $\times 10^{11} M_\odot$, respectively. 
Their study provides strong evidence of downsizing of the dark matter haloes hosting sub-millimeter galaxies, and our results are in agreement with theirs.

Further supporting evidence of our result of downsizing in \textit{Herschel}-detected sources can be found in \citet{skli17}. 
They investigate the star formation histories of \textit{Herschel} sources over the redshift range $1.2\leq z \leq4$.
They find that for main sequence galaxies at a given stellar mass, the higher redshift galaxies both have higher star formation rates and their star formation histories indicate faster rising star formation rates than galaxies at the same stellar mass at lower redshift. 
The general picture is that the more massive galaxies in their sample build up their mass earlier and faster than lower mass DSFGs.
Though this is a different definition of downsizing as it pertains to stellar mass, it is still consistent with our picture of how DSFGs evolve with redshift. 

\section{Summary and conclusions}
\label{discuss}

In this paper, we have presented a stacking analysis of \textit{Herschel}-SPIRE far-infrared data on the locations of 11,235 quasars from the SDSS in 10 redshift bins spanning $z=0.5-3.5$. 
We detect Herschel counterparts to quasars and their host galaxies in the stacks, and in addition there is a detection of extended emission on scales of a few Mpc  associated with DSFGs clustered around the optically-selected quasars. 
The detection of large scale correlated signal is significant (above what is expected from stacking on random positions), and there is an increase in extended surface brightness between $z\sim1.5$ and $z\sim2.5$. 
This extended emission allows us to recover the angular cross-correlation function for sources that make up the CIB clustered around quasars as a function of redshift owing to the known redshift distribution of the spectroscopic quasar catalog. 

We start with a model in which we attempt to explain the peak infrared surface brightness in our dataset at $z\sim 2$ as due to the peak in star formation rate density and find that the scenario is likely more complicated. 
Due to cosmological surface brightness dimming, such model would result in the peak signal at $z\sim1$, regardless of choice of spectral energy distribution.
We implement a physically-motivated halo occupation distribution model for recovering information about the clustered sources. 
The HOD model is used to determine the mass of the most effectively star-forming haloes, as well as  to probe the luminosity-halo mass relationship for these objects \citep{shan12}.
In the HOD model, we incorporate the luminosity-weighted number density of DSFGs as a function of halo mass, so that the model is dependent on the IR luminosity, or star formation rate, of the occupying galaxies as a function of halo mass and redshift. 
We hypothesize that the reason for our observed peak in extended surface brightness between $z\sim 1.5$ and $z\sim2.5$ is associated with the redshift evolution of the dark matter halo clustering of infrared background sources around quasars and/or the downsizing of these star-forming galaxies.

A key advantage of our dataset and analysis is our ability to directly probe the evolution of the DSFG HOD parameters over a broad and finely-sampled redshift range. 
Unlike the other clustering studies we have mentioned, we are not restricted to integrating our model over all relevant epochs of CIB emission. 
Furthermore, the derivation of our halo masses hosting DSFGs as a function of redshift has the unique property of being associated with the same types of objects (quasars) in each of the redshift bins.
This allows us to incorporate all the known information about the dark matter haloes and clustering of the tracer population (quasars), allowing us to focus more extensively on the population which is more difficult to probe (DSFGs). 
We are also not limited by small sample sizes, as the stacking enables a probe of the average properties of the sources contributing to the CIB.
In this way our results are a novel and consistent probe of the physical properties of DSFGs as a function of redshift.

We investigate two possible scenarios that drive the observed redshift evolution of the detected correlated emission as it is inconsistent with DSFGs clustering around quasars in the same way at all redshifts. 
(1) If high redshift quasars occupy more massive dark matter haloes than the low redshift ones, they may have many more galaxies clustered around them producing excess signal at high redshift. 
(2) The number of objects clustered around quasars is approximately equal at all redshifts, but the massive high redshift objects are more conducive for star formation than their low redshift counterparts. 
The evolution of the dark matter haloes hosting quasars is still a matter of open debate (\citealt{rich12}; \citealt{shen13}; \citealt{wang15}), and the confusion limit of current far infrared surveys is too high for accurate number counts of clustered DSFGs.
Nonetheless, in both scenarios, using the quasars as a tracer of DSFG clustering enables us to investigate the dark matter halo masses hosting some of the most actively star forming galaxies in the universe at all epochs, and our findings are consistent with cosmic downsizing providing evidence for hypothesis (1).
Furthermore, the evolution of the amplitude of the $L-M$ relation $L_0$ with redshift indicates an increase in the average IR luminosity of dark matter haloes hosting DSFGs, consistent with hypothesis (2). 
It is likely that the answer to how DSFGs and quasars cluster is a combination of both these hypotheses and further studies of quasar and DSFG HODs can help quantify our understanding of star formation and stellar growth of galaxies and the dark matter haloes in which they inhabit.

Our study provides a unique probe of DSFGs at observed wavelengths of 250, 350, and 500 $\micron$ across a wide range in redshift $z=0.5-3.5$. We find evidence for the downsizing in mass of the most effective dark matter haloes at hosting these objects with a decrease in mass from $10^{13.9} M_\odot$ at $z\sim2.9$ to $10^{10.7} M_\odot$ at $z\sim0.5$ (Figure~\ref{Meffz}). 
Simultaneously, this result provides an additional observational constraint on the characteristic mean mass $\langle M \rangle \approx 10^{12} M_\odot$ of dark matter haloes in which star formation occurs (Figure~\ref{Mmeanz}). 
Figures \ref{Tdz} and \ref{rhoSFR} show that, within broadly drawn priors, our data constrain the temperature and total star formation rate evolution of DSFGs. 
Furthermore, the average star formation rates of DSFGs clustered in and around massive quasar-hosting dark matter haloes are consistent with the cosmic star formation rate density.
Thus clustering analyses such as ours can provide a unique view of the cosmic evolution of high-redshift galaxies.
Our results are robust to variations in the otherwise fixed parameters in the $L-M$ relation describing the redshift evolution and the width of halo mass distribution. 
They are somewhat more sensitive to the quasar HOD in a manner that is worth exploring more as a means for an alternative explanation for the redshift evolution of our cross-correlation results, but this is beyond the scope of this work.
Nonetheless, we have shown (Section~\ref{quasartests}) that the general downsizing behavior observed in this work is robust against modifications to our clustering model.

\section*{Acknowledgements}
Herschel is an ESA space observatory with science instruments provided by European-led Principal Investigator consortia and with important participation from NASA. 
This research has made use of data from HerMES project (http://hermes.sussex.ac.uk/). 
HerMES is a Herschel Key Programme utilising Guaranteed Time from the SPIRE instrument team, ESAC scientists and a mission scientist.
The HerMES data was accessed through the Herschel Database in Marseille (HeDaM - http://hedam.lam.fr) operated by CeSAM and hosted by the Laboratoire d'Astrophysique de Marseille.
Part of this research project was conducted using computational resources at the Maryland Advanced Research Computing Center (MARCC).
NLZ acknowledges support by the JHU Catalyst Award and by the Deborah Lunder and Alan Ezekowitz Founders' Circle Membership at the Institute for Advanced Study.
DC acknowledges the financial assistance of the South African SKA Project (SKA SA) towards this research (www.ska.ac.za).
KRH thanks Graeme Addison for productive conversations and valuable feedback, and for providing the linear matter power spectrum data generate using Code for Anisotropies in the Microwave Background (CAMB).
KRH also thanks Megan Gralla for useful feedback on the analysis.

\bibliographystyle{mnras}
\bibliography{cib_kh}

\appendix

\section{Fiducial Model fits to profiles for all redshifts}
\label{appendixA}

This section contains the profiles with best-fitting models and marginalized parameter constraints for all redshift bins using the fiducial model with an evolving quasar HOD.

\begin{figure*}
\centering
  \includegraphics[width=5in]{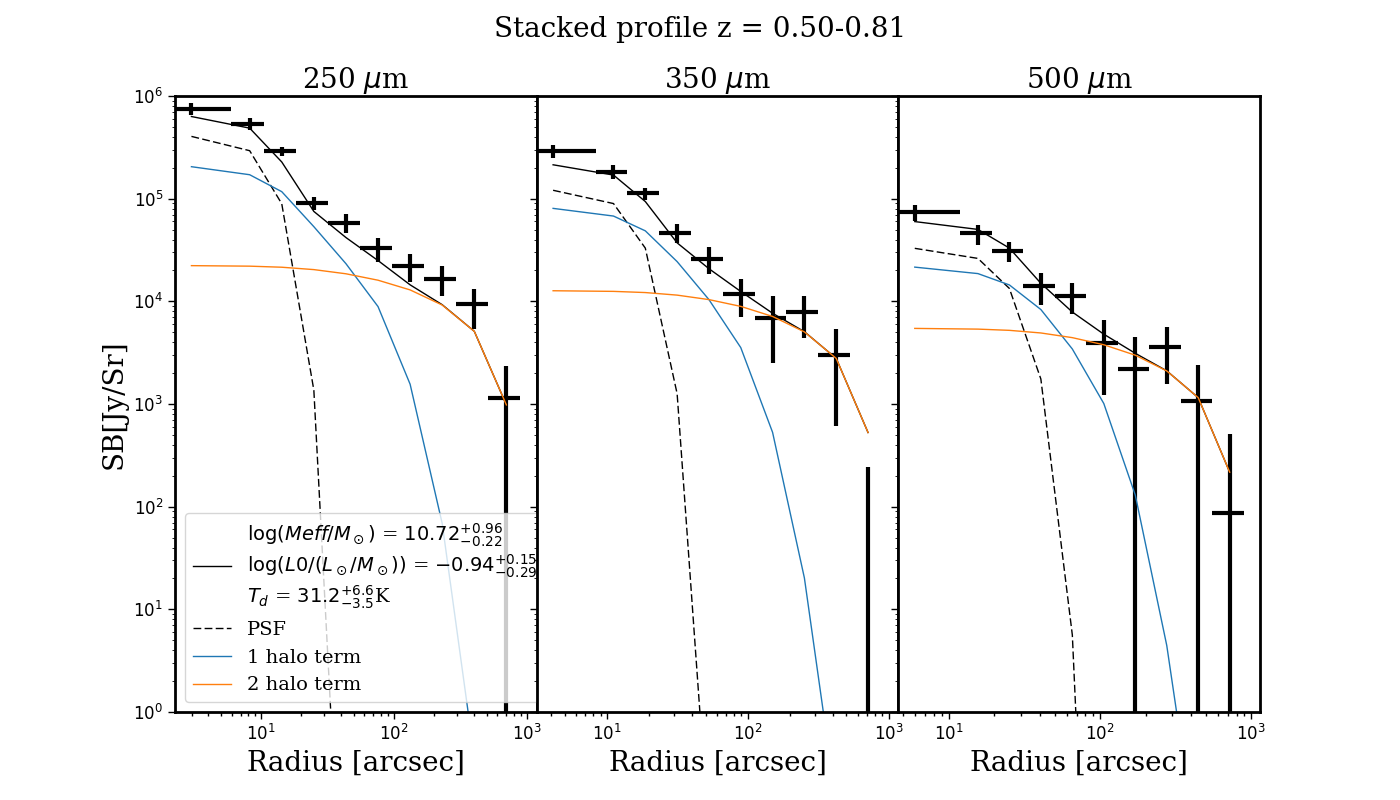}
  \includegraphics[width=5in]{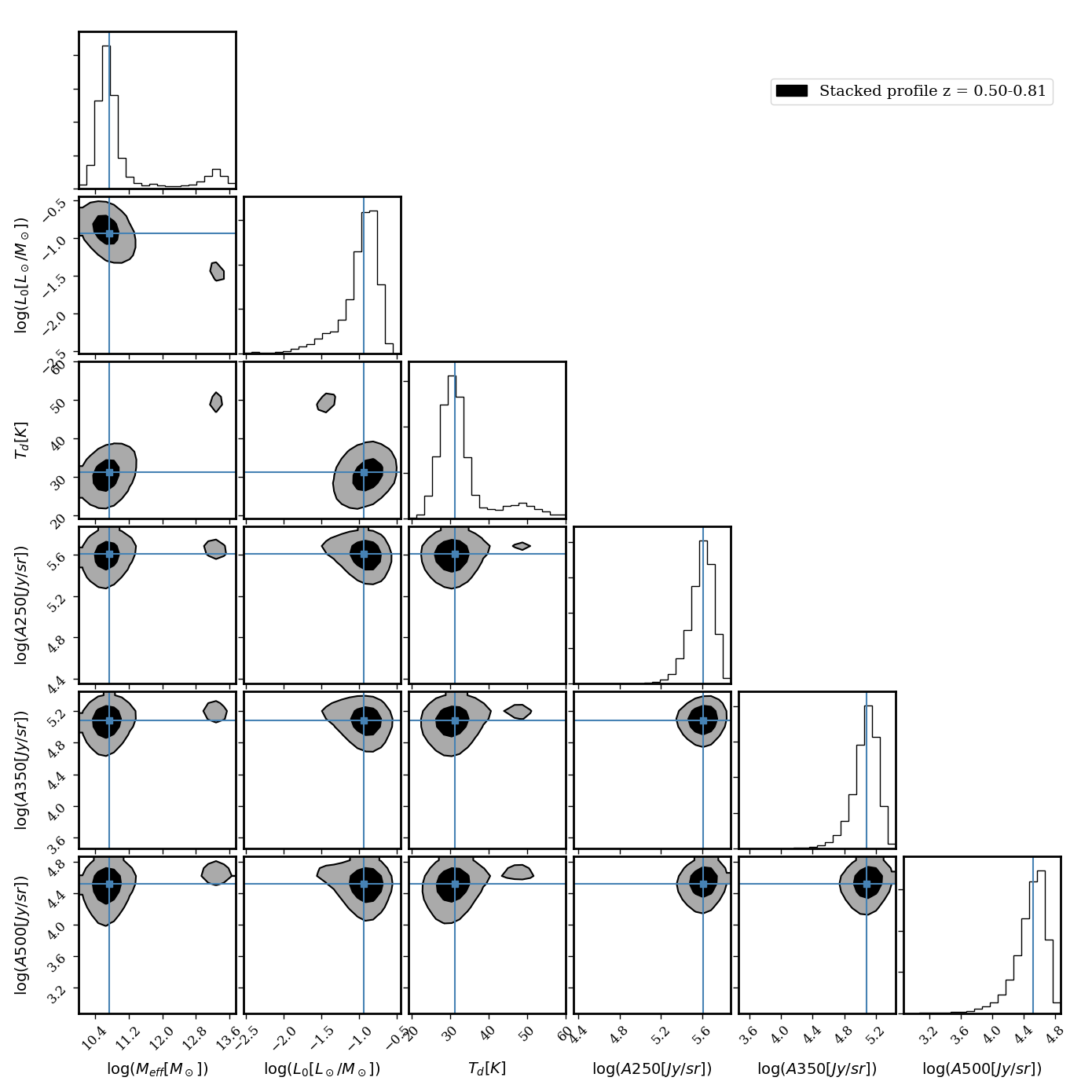}
\caption{Top: Angular cross correlation function and complete model, including the PSF (dashed line), one-halo term (small scales), and two-halo term (large scales), using marginalized best-fitting parameter results for quasars and DSFGs in the redshift range $z=0.5-0.81$. Bottom: Contours of the 68$^{th}$ and 95$^{th}$  percentiles of the parameters in our fiducial HOD model plus the three amplitudes fitting the quasar emission at each wavelength for the same range in redshift. The histograms show the marginalized posterior distributions, and the lines mark the median values. }
\label{profz0}
\end{figure*}

\begin{figure*}
\centering
  \includegraphics[width=5in]{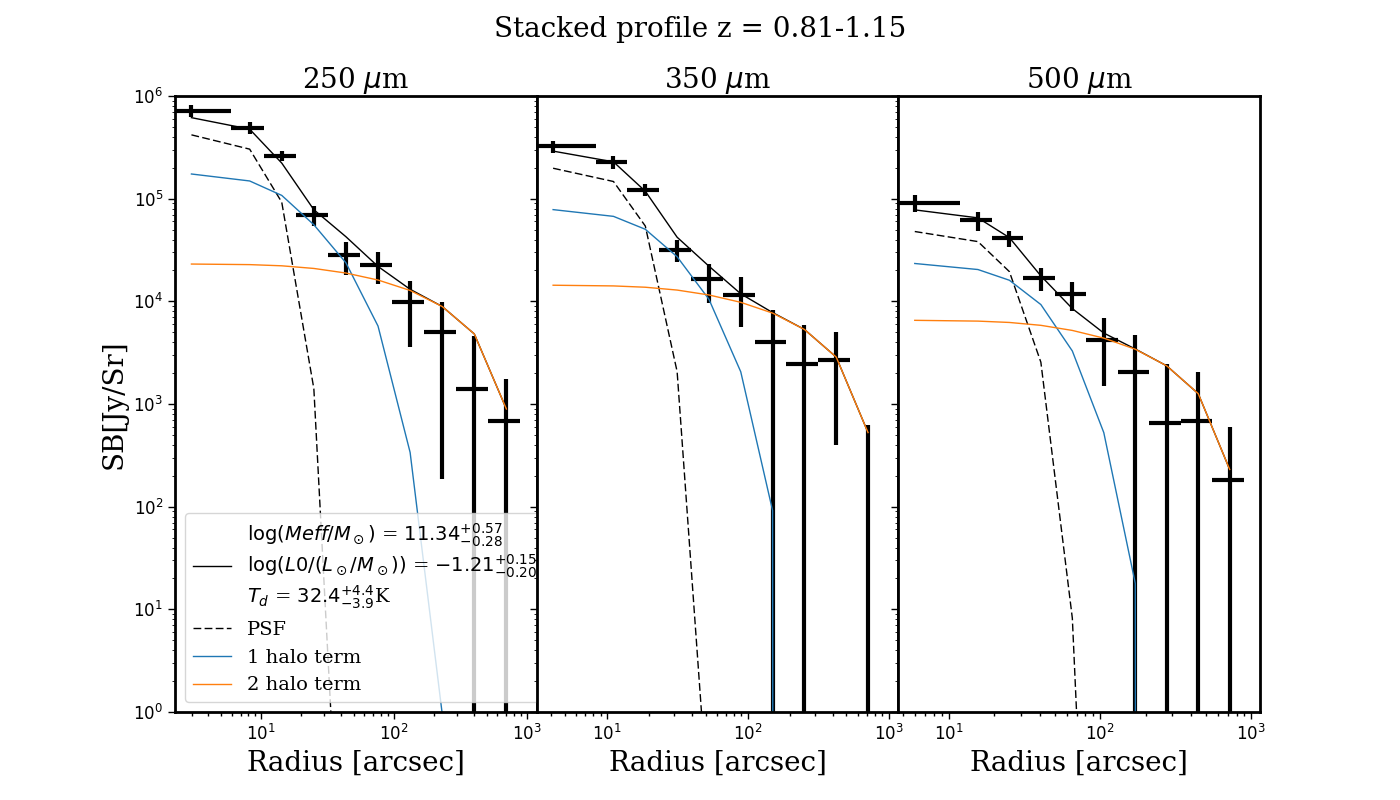}
  \includegraphics[width=5in]{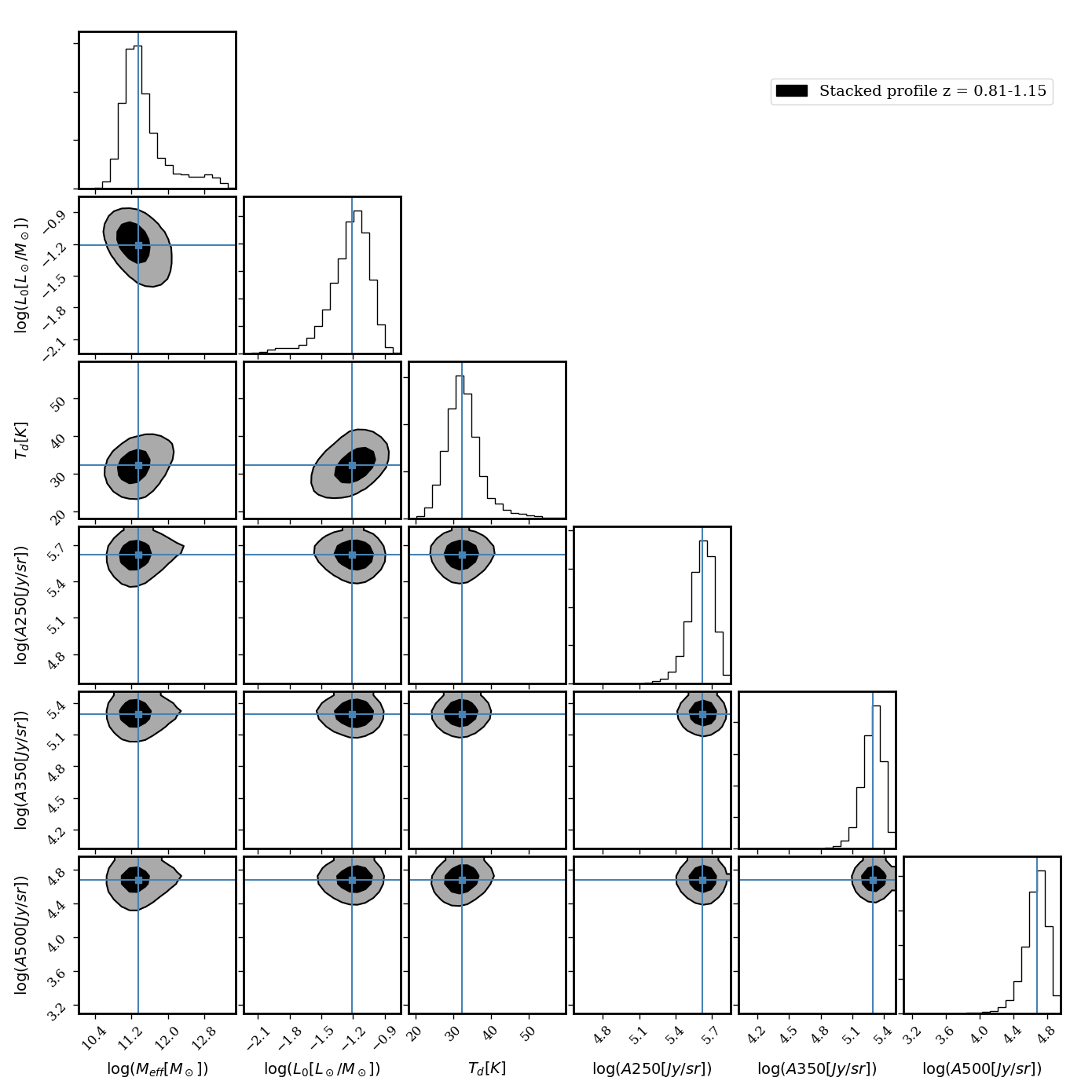}
\caption{Top: Angular cross correlation function and complete model, including the PSF (dashed line), one-halo term (small scales), and two-halo term (large scales), using marginalized best-fitting parameter results for quasars and DSFGs in the redshift range $z=0.81-1.15$. Bottom: Contours of the 68$^{th}$ and 95$^{th}$  percentiles of the parameters in our fiducial HOD model plus the three amplitudes fitting the quasar emission at each wavelength for the same range in redshift. The histograms show the marginalized posterior distributions, and the lines mark the median values.}
\label{profz1}
\end{figure*}

\begin{figure*}
\centering
  \includegraphics[width=5in]{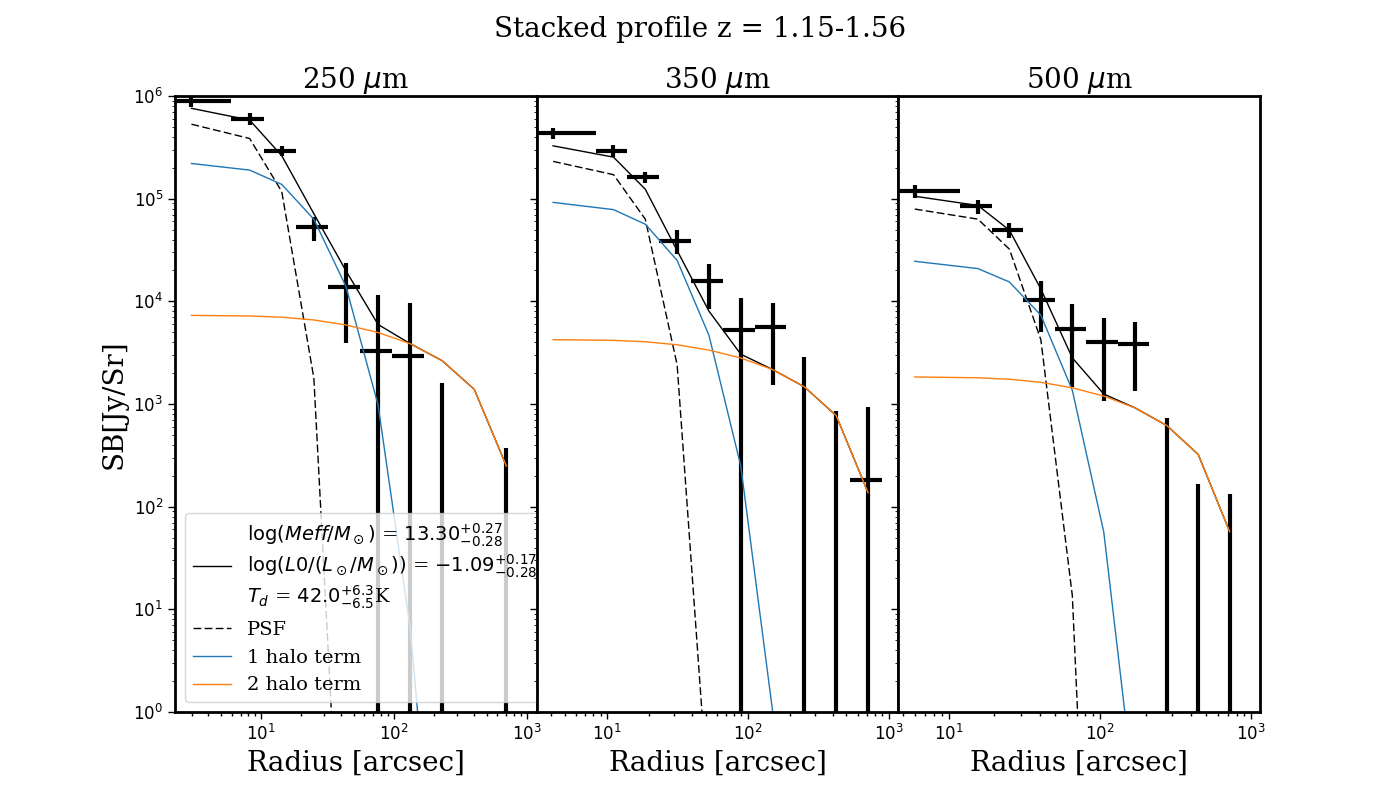}
  \includegraphics[width=5in]{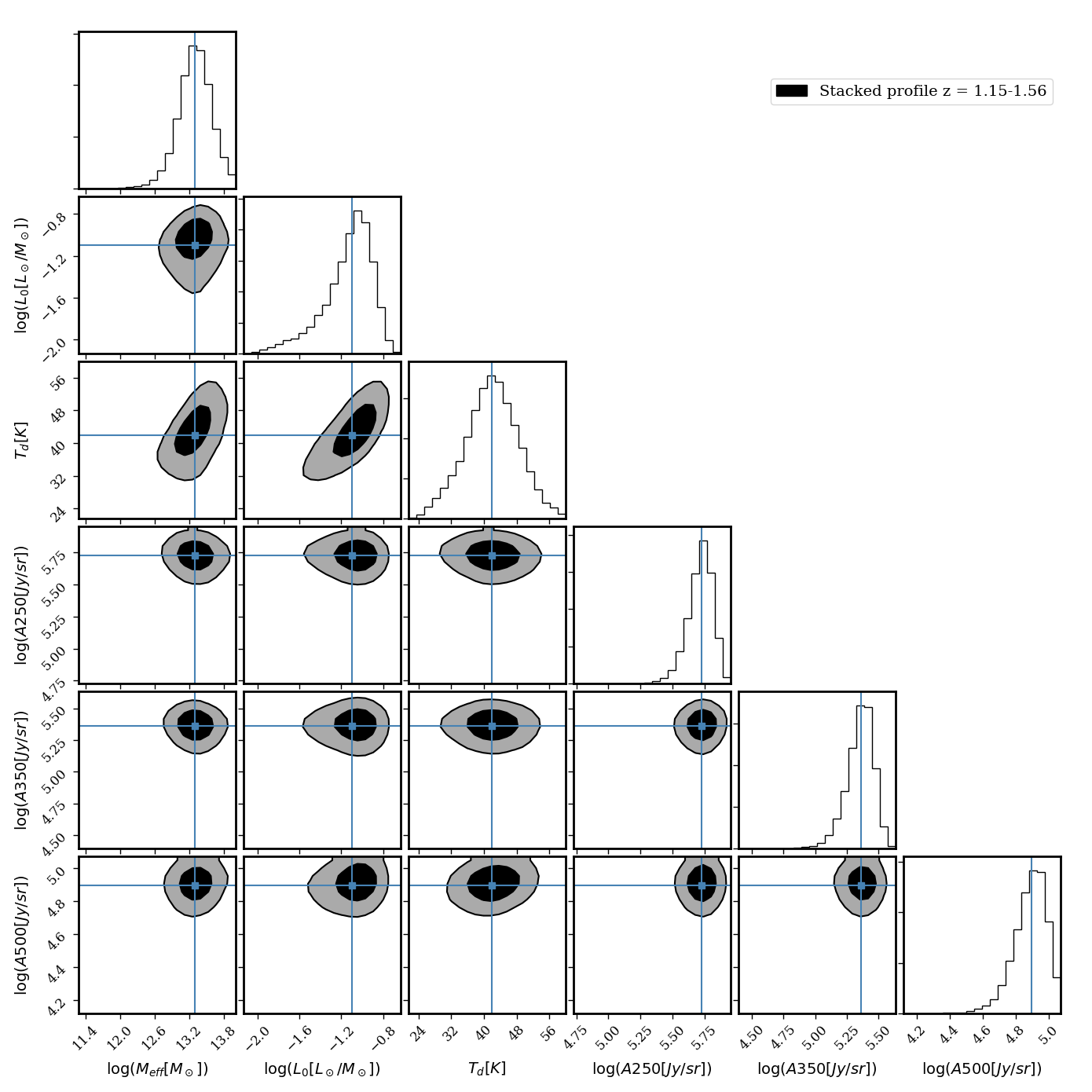}
\caption{Top: Angular cross correlation function and complete model, including the PSF (dashed line), one-halo term (small scales), and two-halo term (large scales), using marginalized best-fitting parameter results for quasars and DSFGs in the redshift range $z=1.15-1.56$. Bottom: Contours of the 68$^{th}$ and 95$^{th}$  percentiles of the parameters in our fiducial HOD model plus the three amplitudes fitting the quasar emission at each wavelength for the same range in redshift. The histograms show the marginalized posterior distributions, and the lines mark the median values.}
\label{profz2}
\end{figure*}

\begin{figure*}
\centering
  \includegraphics[width=5in]{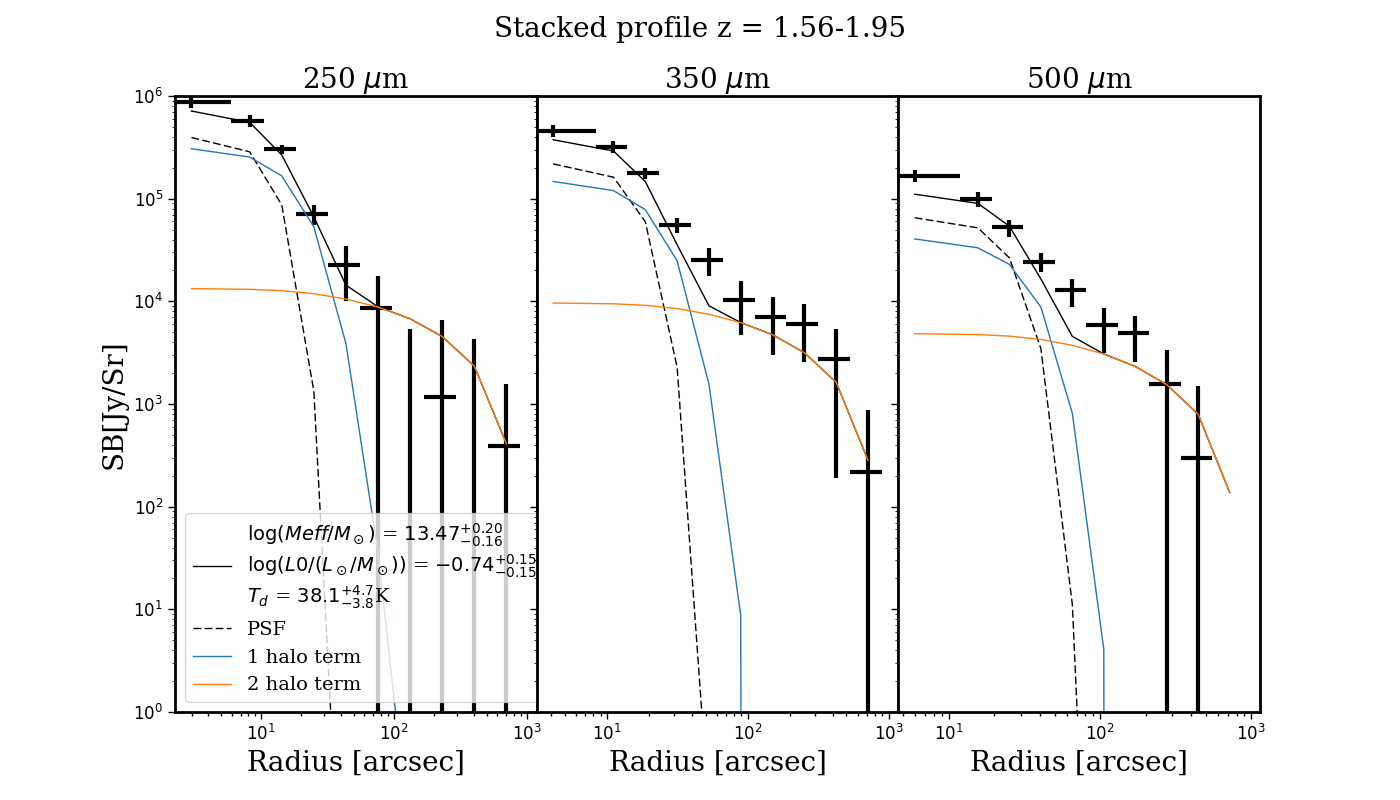}
  \includegraphics[width=5in]{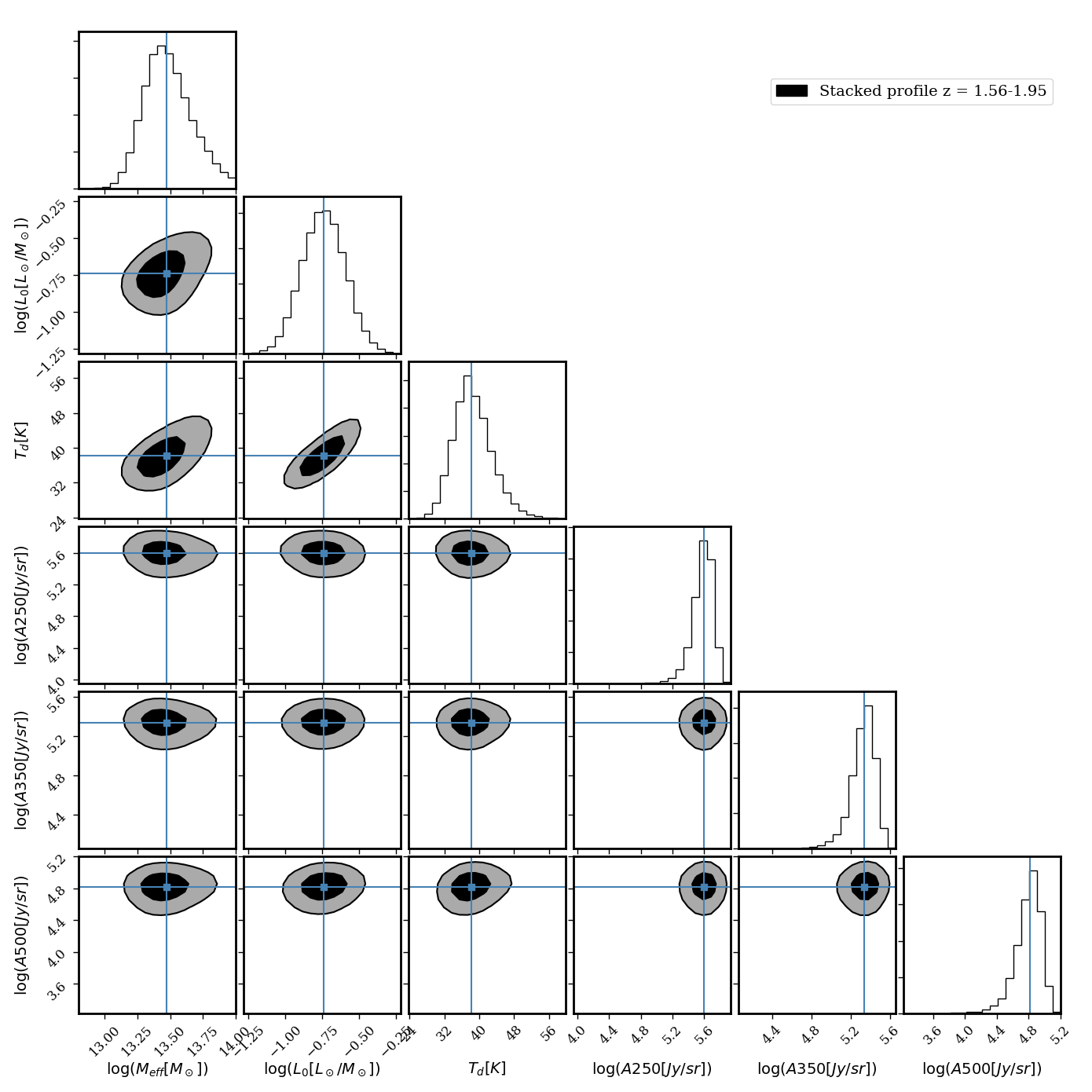}
\caption{Top: Angular cross correlation function and complete model, including the PSF (dashed line), one-halo term (small scales), and two-halo term (large scales), using marginalized best-fitting parameter results for quasars and DSFGs in the redshift range $z=1.56-1.95$. Bottom: Contours of the 68$^{th}$ and 95$^{th}$  percentiles of the parameters in our fiducial HOD model plus the three amplitudes fitting the quasar emission at each wavelength for the same range in redshift. The histograms show the marginalized posterior distributions, and the lines mark the median values.}
\label{profz3}
\end{figure*}

\begin{figure*}
\centering
  \includegraphics[width=5in]{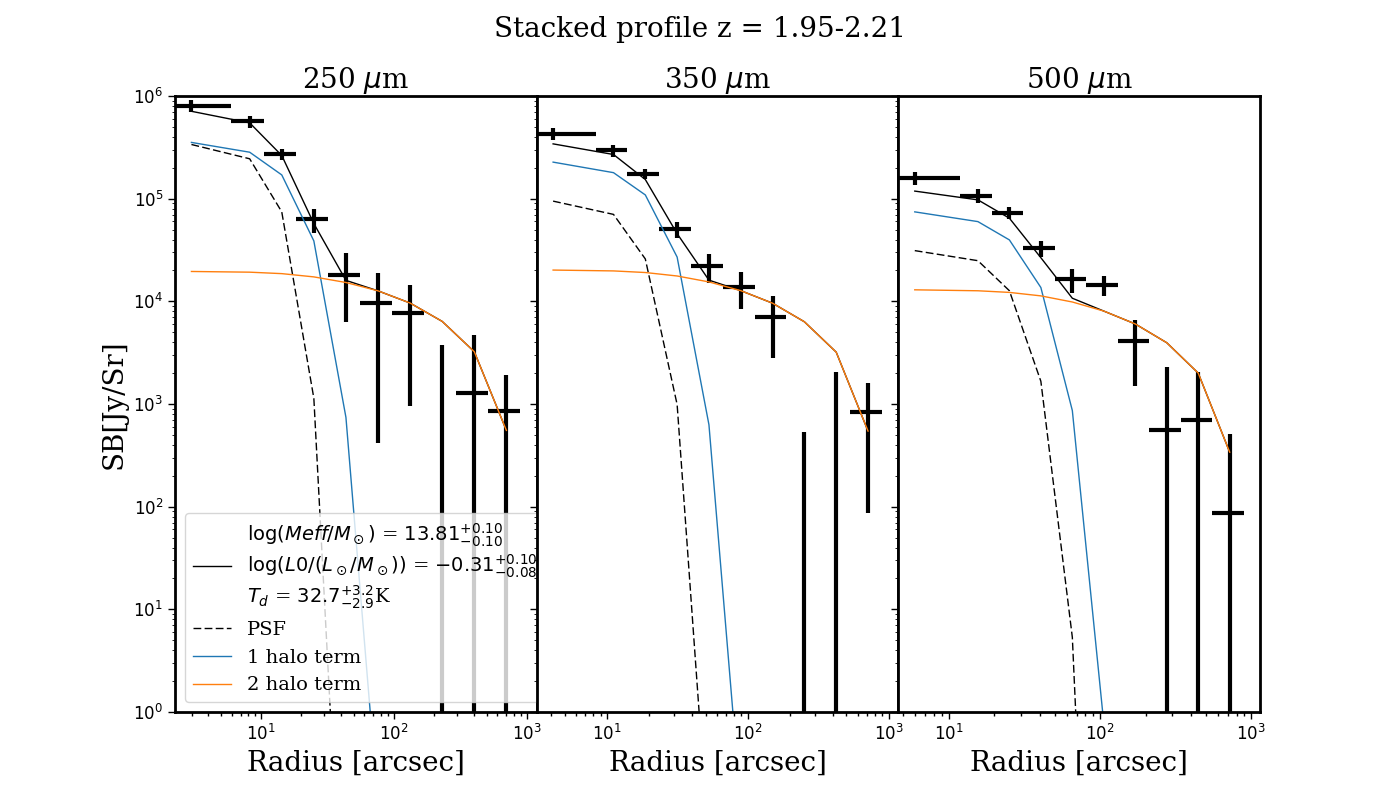}
  \includegraphics[width=5in]{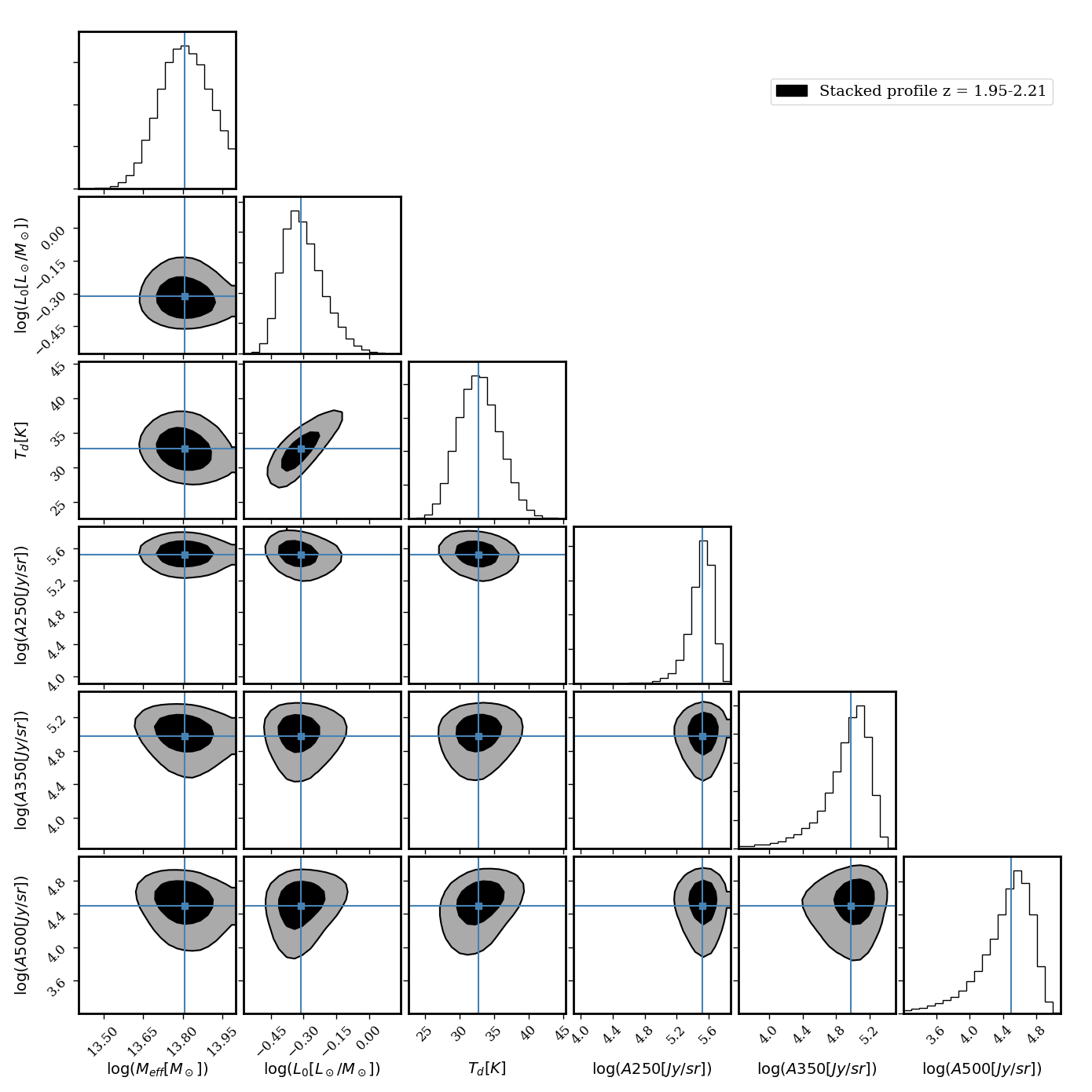}
\caption{Top: Angular cross correlation function and complete model, including the PSF (dashed line), one-halo term (small scales), and two-halo term (large scales), using marginalized best-fitting parameter results for quasars and DSFGs in the redshift range $z=1.95-2.21$. Bottom: Contours of the 68$^{th}$ and 95$^{th}$  percentiles of the parameters in our fiducial HOD model plus the three amplitudes fitting the quasar emission at each wavelength for the same range in redshift. The histograms show the marginalized posterior distributions, and the lines mark the median values.}
\label{profz4}
\end{figure*}

\begin{figure*}
\centering
  \includegraphics[width=5in]{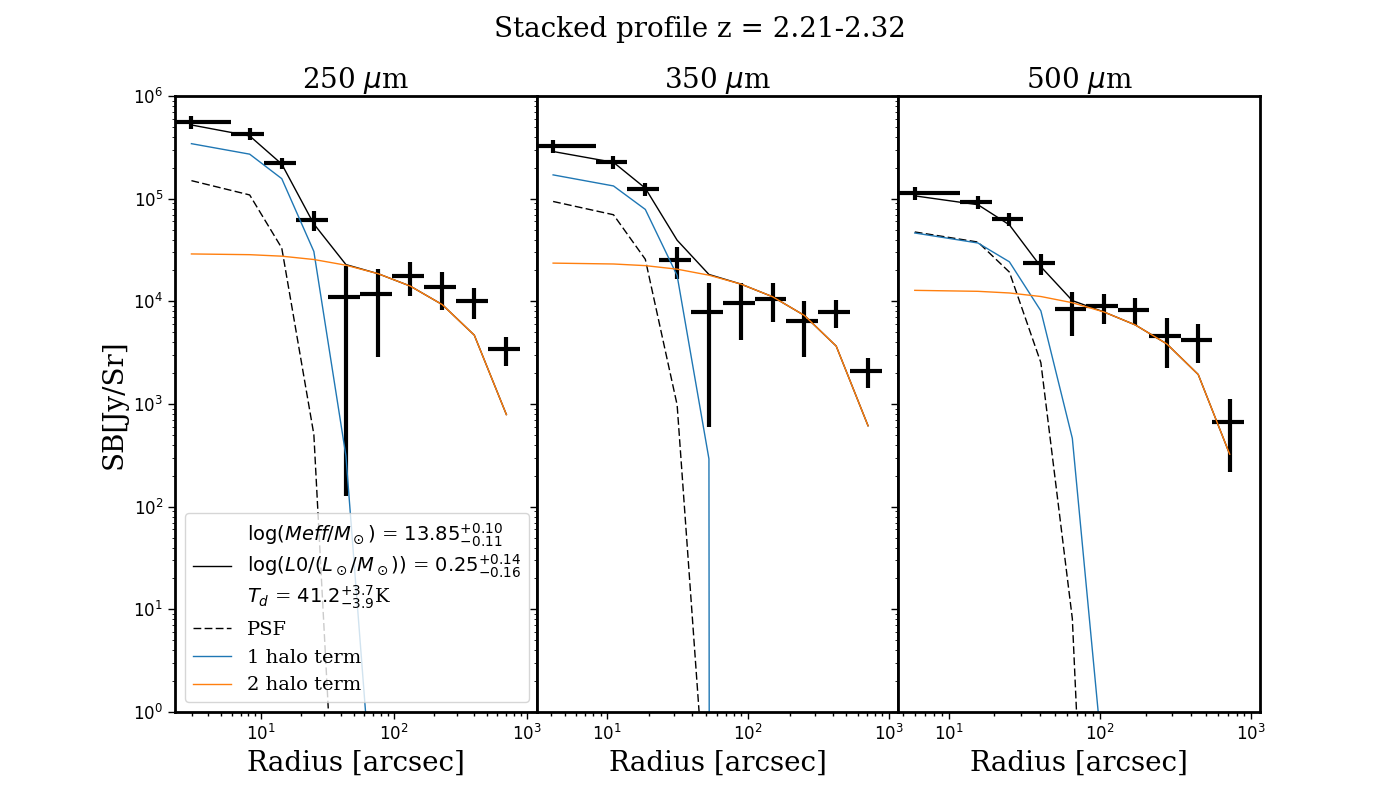}
  \includegraphics[width=5in]{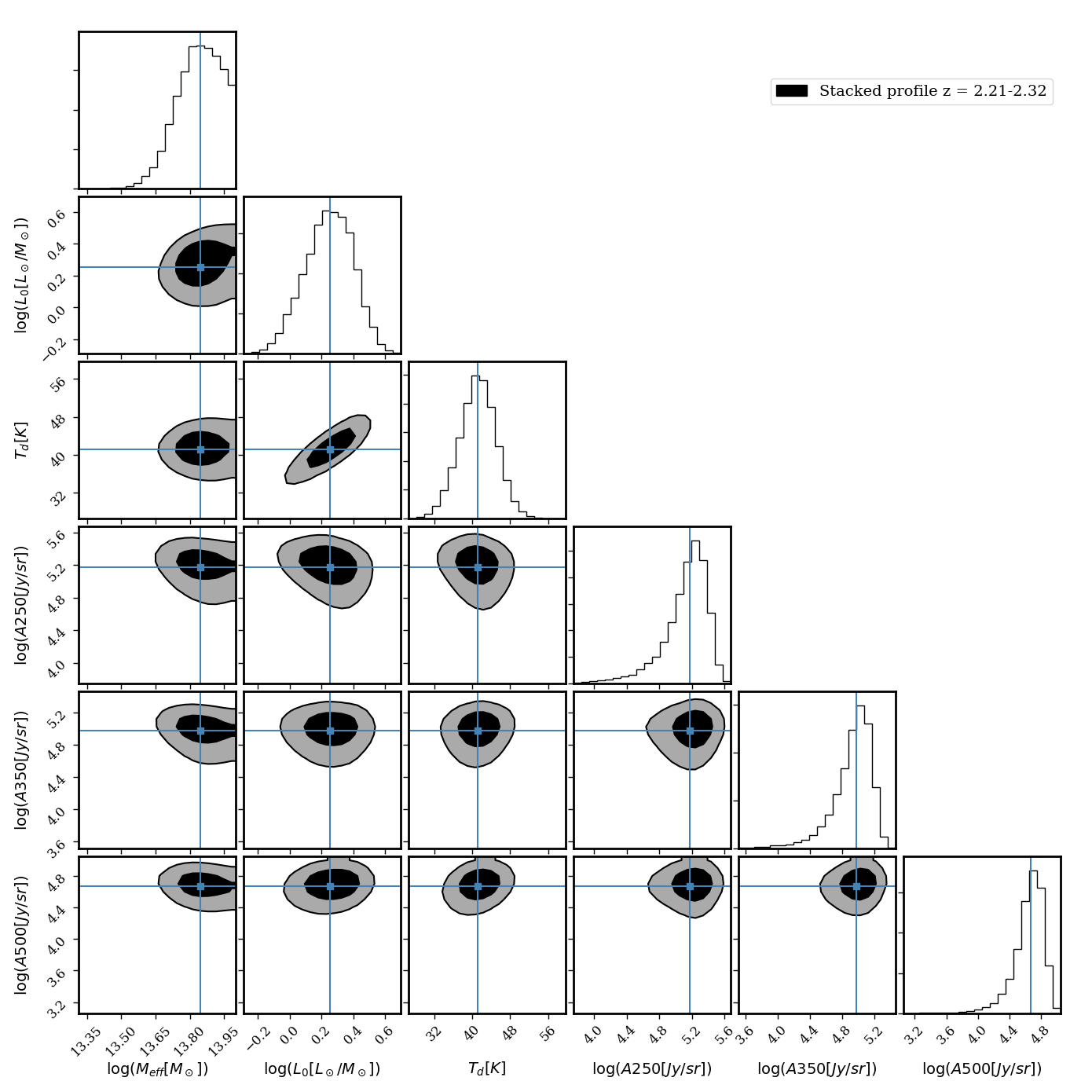}
\caption{Top: Angular cross correlation function and complete model, including the PSF (dashed line), one-halo term (small scales), and two-halo term (large scales), using marginalized best-fitting parameter results for quasars and DSFGs in the redshift range $z=2.21-2.32$. Bottom: Contours of the 68$^{th}$ and 95$^{th}$  percentiles of the parameters in our fiducial HOD model plus the three amplitudes fitting the quasar emission at each wavelength for the same range in redshift. The histograms show the marginalized posterior distributions, and the lines mark the median values. The posterior distribution does not constrain the upper limit of the effective mass $M_\mathrm{eff}$ in this bin.}
\label{profz5}
\end{figure*}

\begin{figure*}
\centering
  \includegraphics[width=5in]{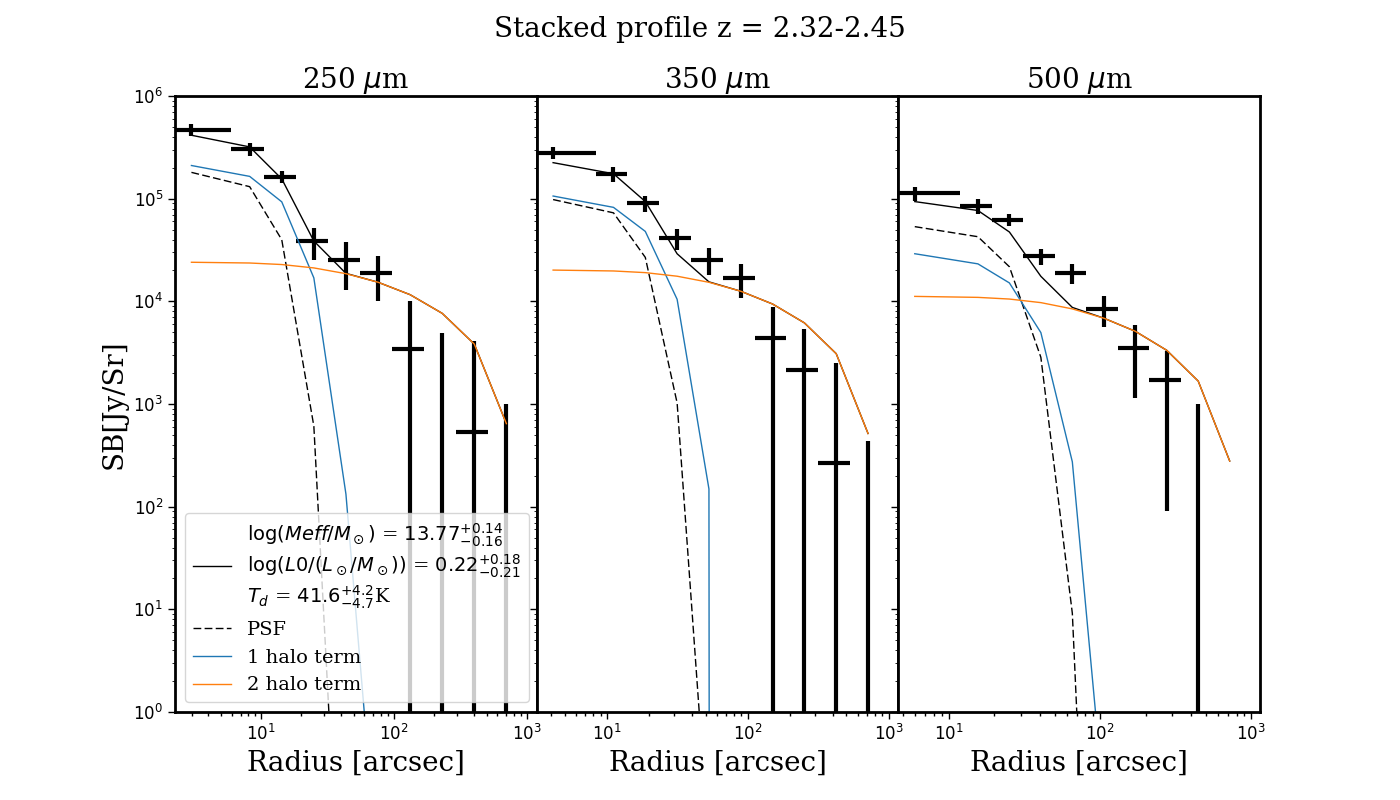}
  \includegraphics[width=5in]{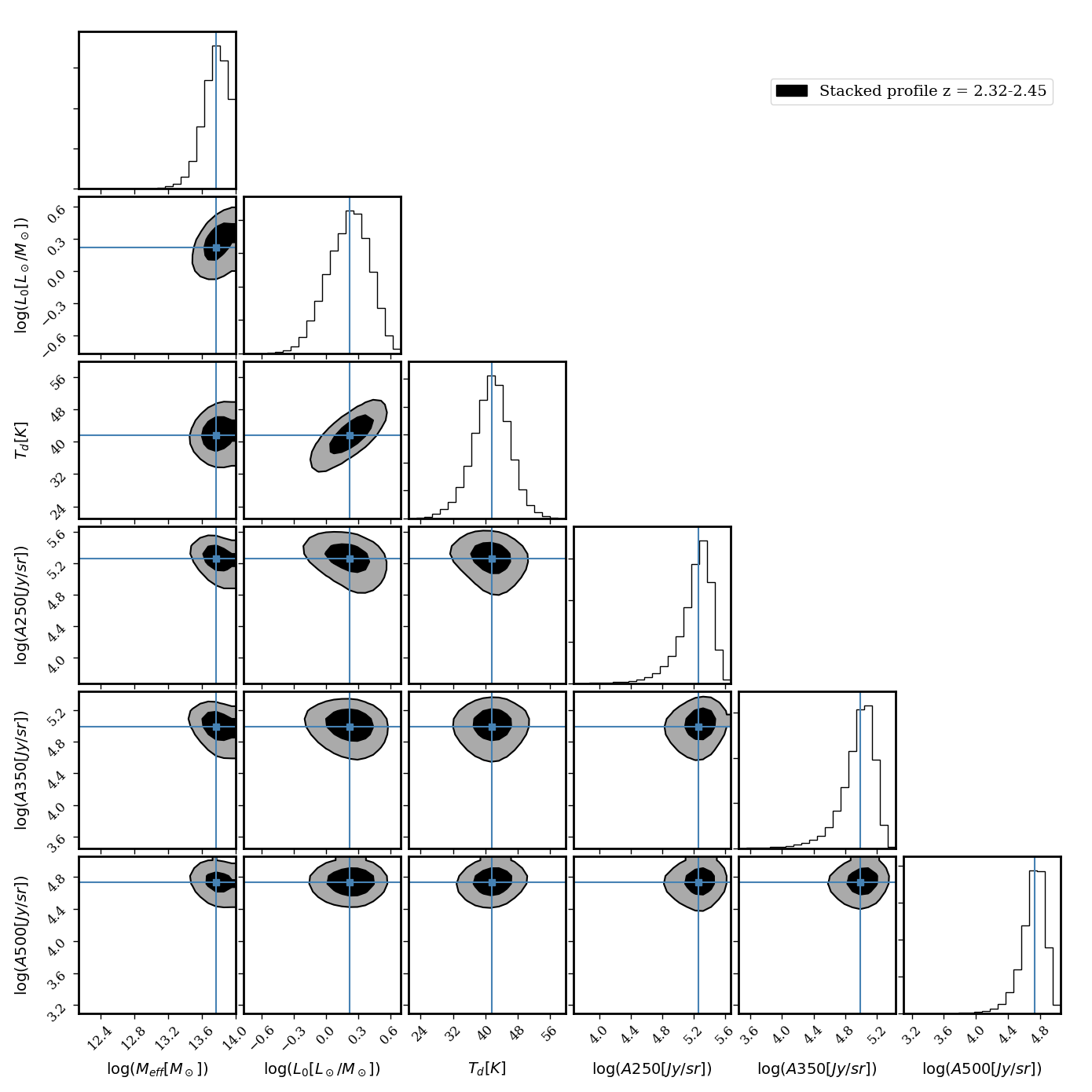}
\caption{Top: Angular cross correlation function and complete model, including the PSF (dashed line), one-halo term (small scales), and two-halo term (large scales), using marginalized best-fitting parameter results for quasars and DSFGs in the redshift range $z=2.32-2.45$. Bottom: Contours of the 68$^{th}$ and 95$^{th}$  percentiles of the parameters in our fiducial HOD model plus the three amplitudes fitting the quasar emission at each wavelength for the same range in redshift. The histograms show the marginalized posterior distributions, and the lines mark the median values. The posterior distribution does not constrain the upper limit of the effective mass $M_\mathrm{eff}$ in this bin.}
\label{profz6}
\end{figure*}

\begin{figure*}
\centering
  \includegraphics[width=5in]{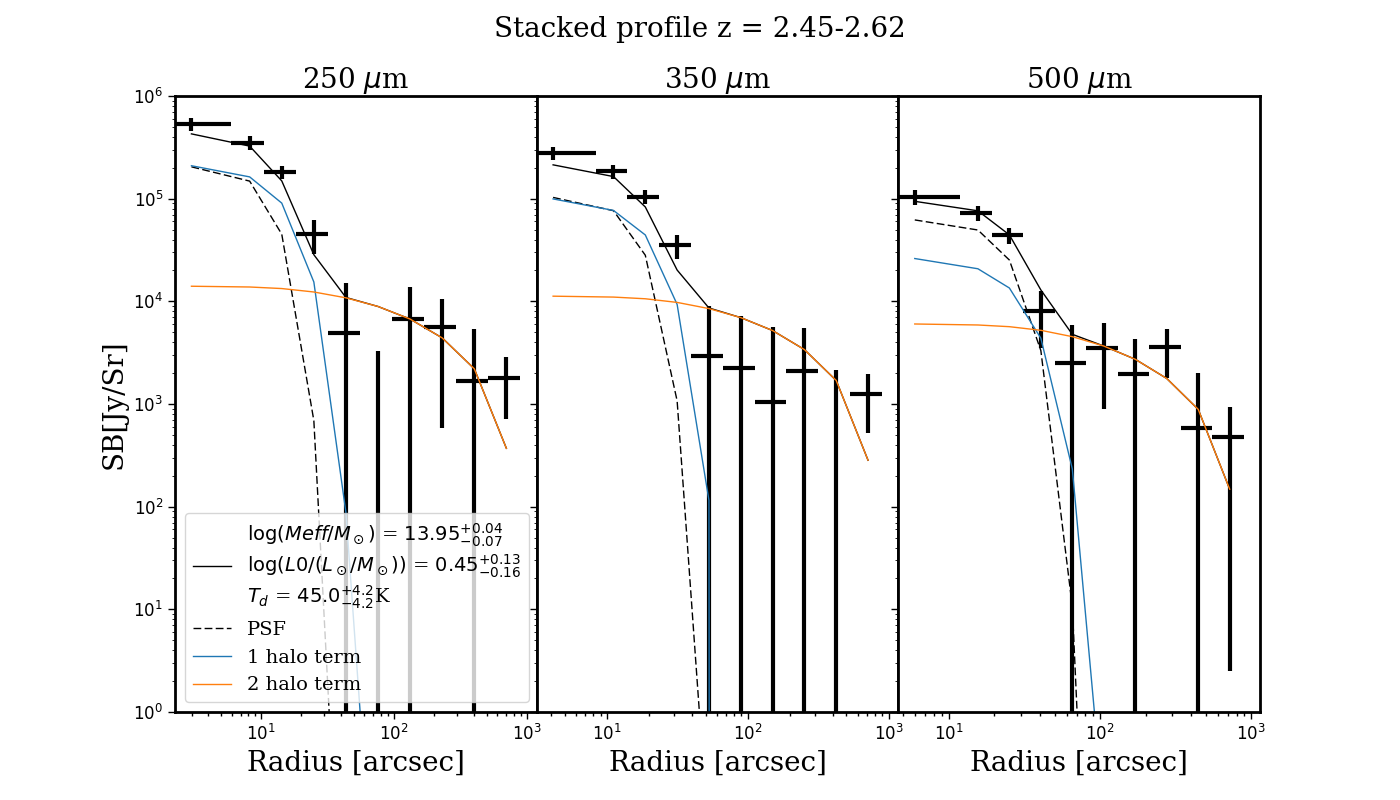}
  \includegraphics[width=5in]{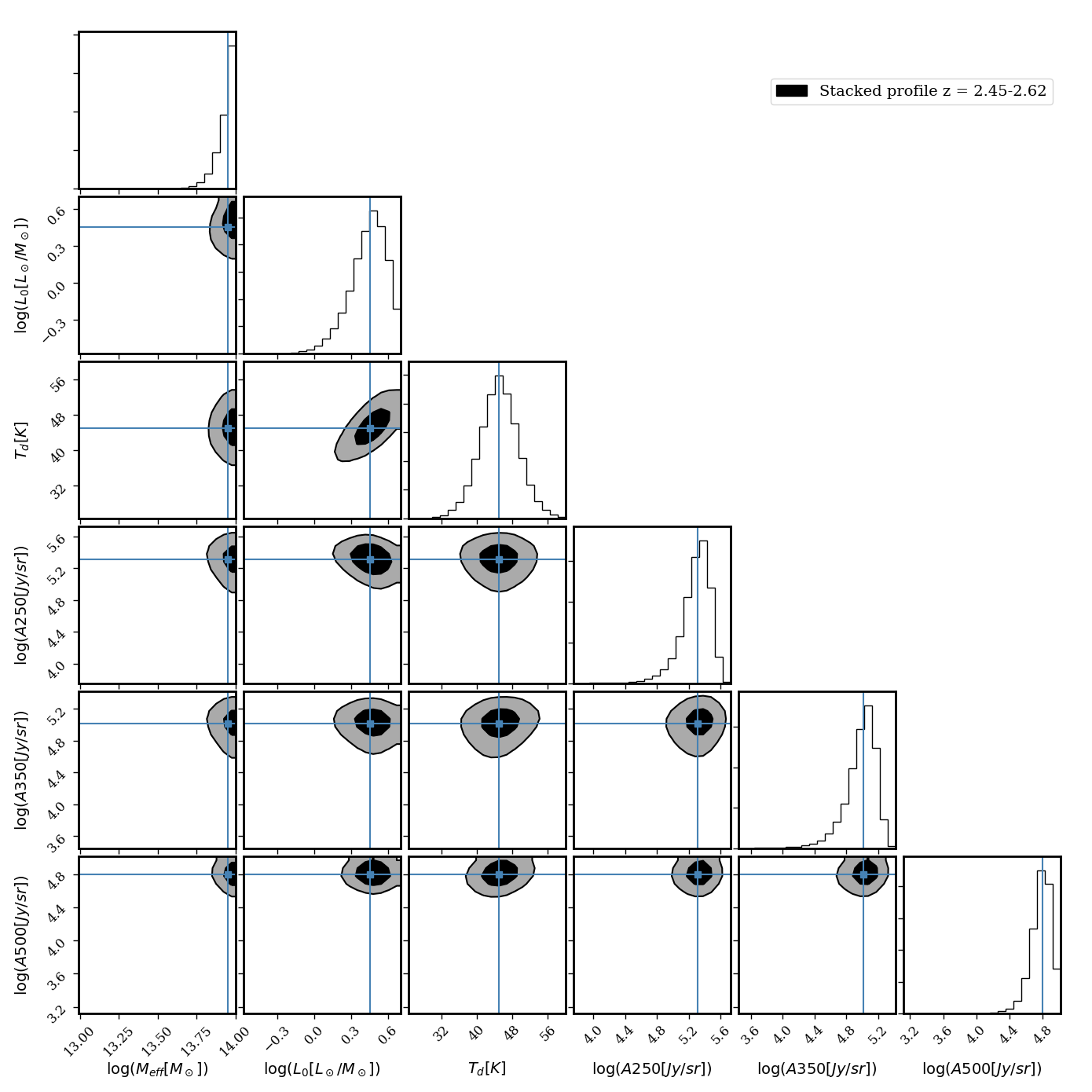}
\caption{Top: Angular cross correlation function and complete model, including the PSF (dashed line), one-halo term (small scales), and two-halo term (large scales), using marginalized best-fitting parameter results for quasars and DSFGs in the redshift range $z=2.45-2.62$. Bottom: Contours of the 68$^{th}$ and 95$^{th}$  percentiles of the parameters in our fiducial HOD model plus the three amplitudes fitting the quasar emission at each wavelength for the same range in redshift. The histograms show the marginalized posterior distributions, and the lines mark the median values. The posterior distribution does not constrain the upper limit of the effective mass $M_\mathrm{eff}$ in this bin.}
\label{profz7}
\end{figure*}

\begin{figure*}
\centering
  \includegraphics[width=5in]{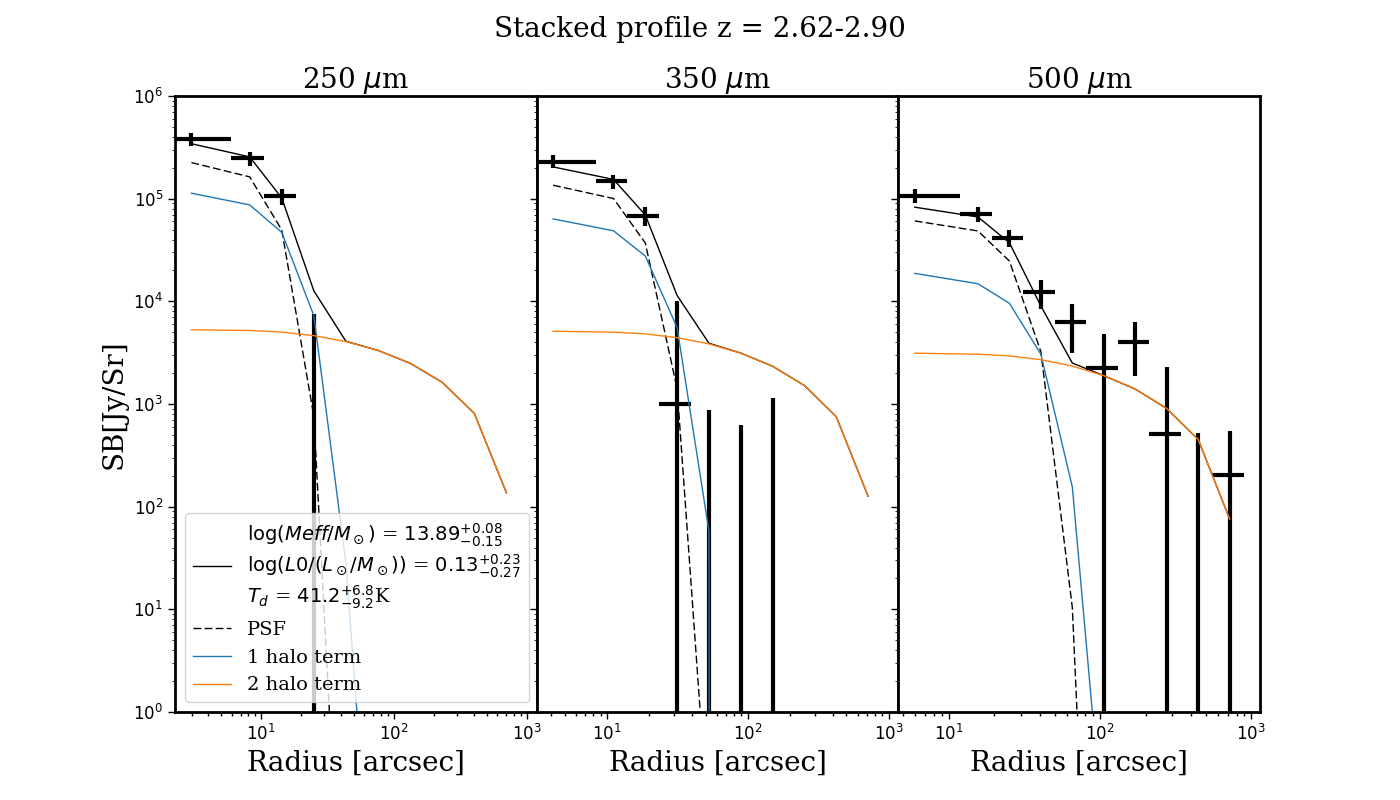}
  \includegraphics[width=5in]{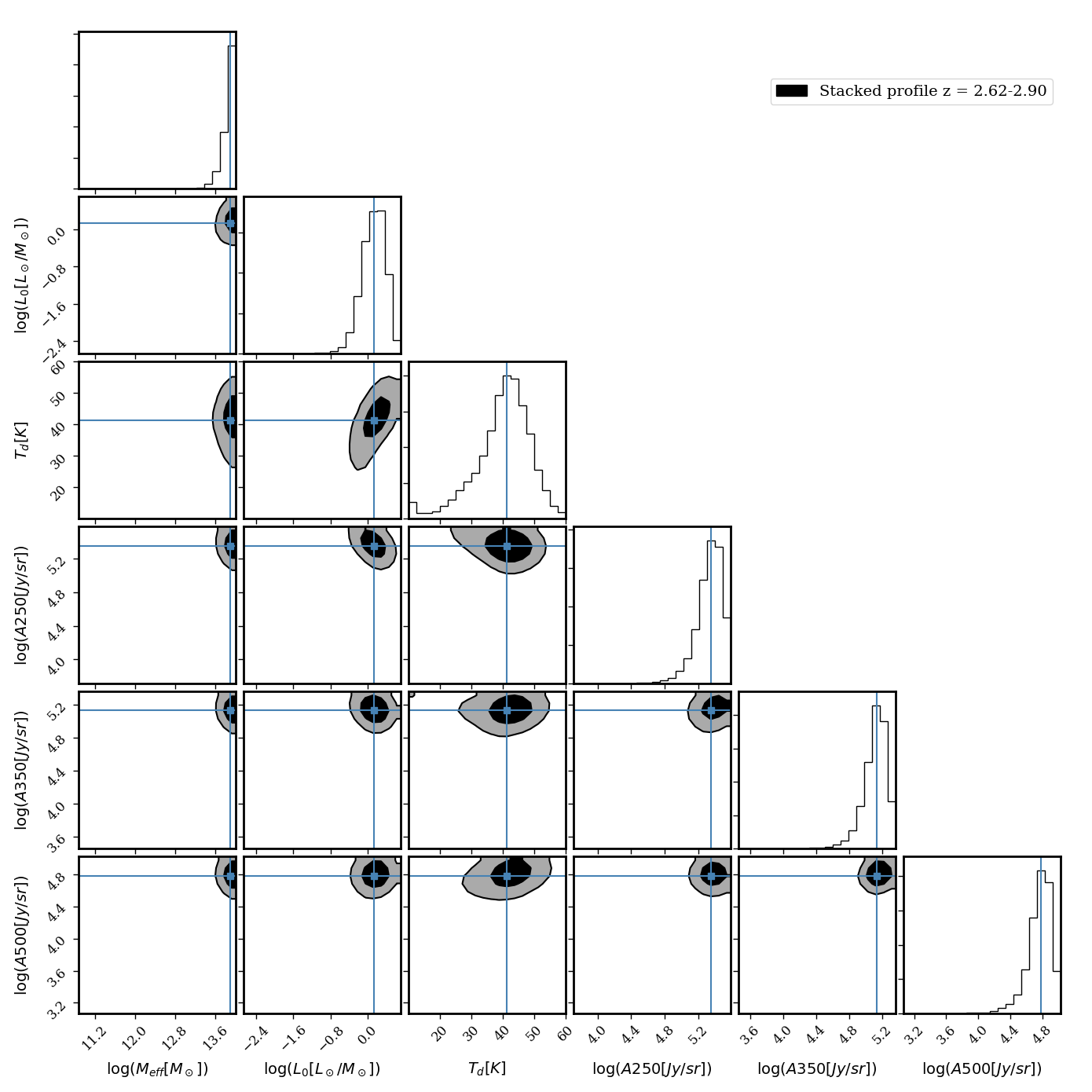}
\caption{Top: Angular cross correlation function and complete model, including the PSF (dashed line), one-halo term (small scales), and two-halo term (large scales), using marginalized best-fitting parameter results for quasars and DSFGs in the redshift range $z=2.62-2.90$. Bottom: Contours of the 68$^{th}$ and 95$^{th}$  percentiles of the parameters in our fiducial HOD model plus the three amplitudes fitting the quasar emission at each wavelength for the same range in redshift. The histograms show the marginalized posterior distributions, and the lines mark the median values. The posterior distribution does not constrain the upper limit of the effective mass $M_\mathrm{eff}$ in this bin.}
\label{profz8}
\end{figure*}

\begin{figure*}
\centering
  \includegraphics[width=5in]{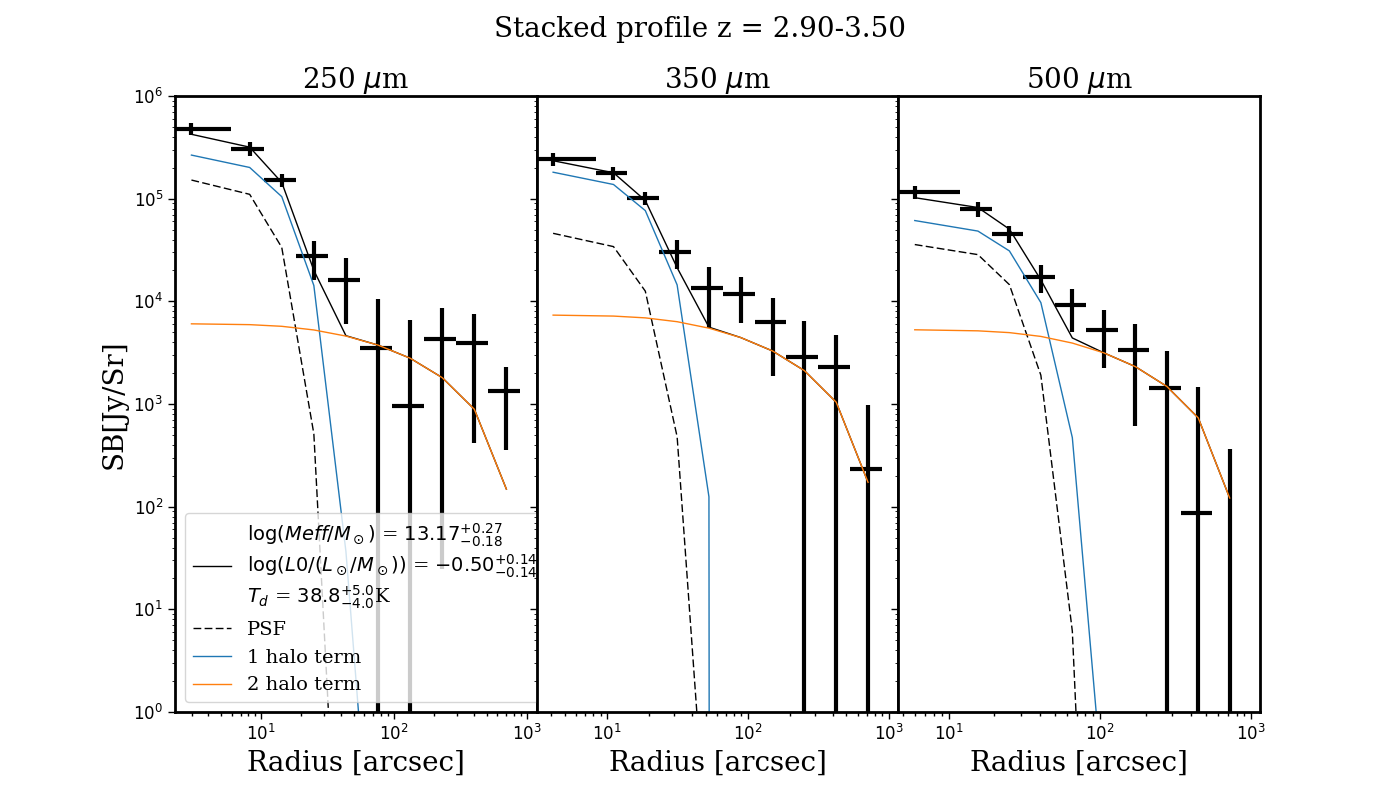}
  \includegraphics[width=5in]{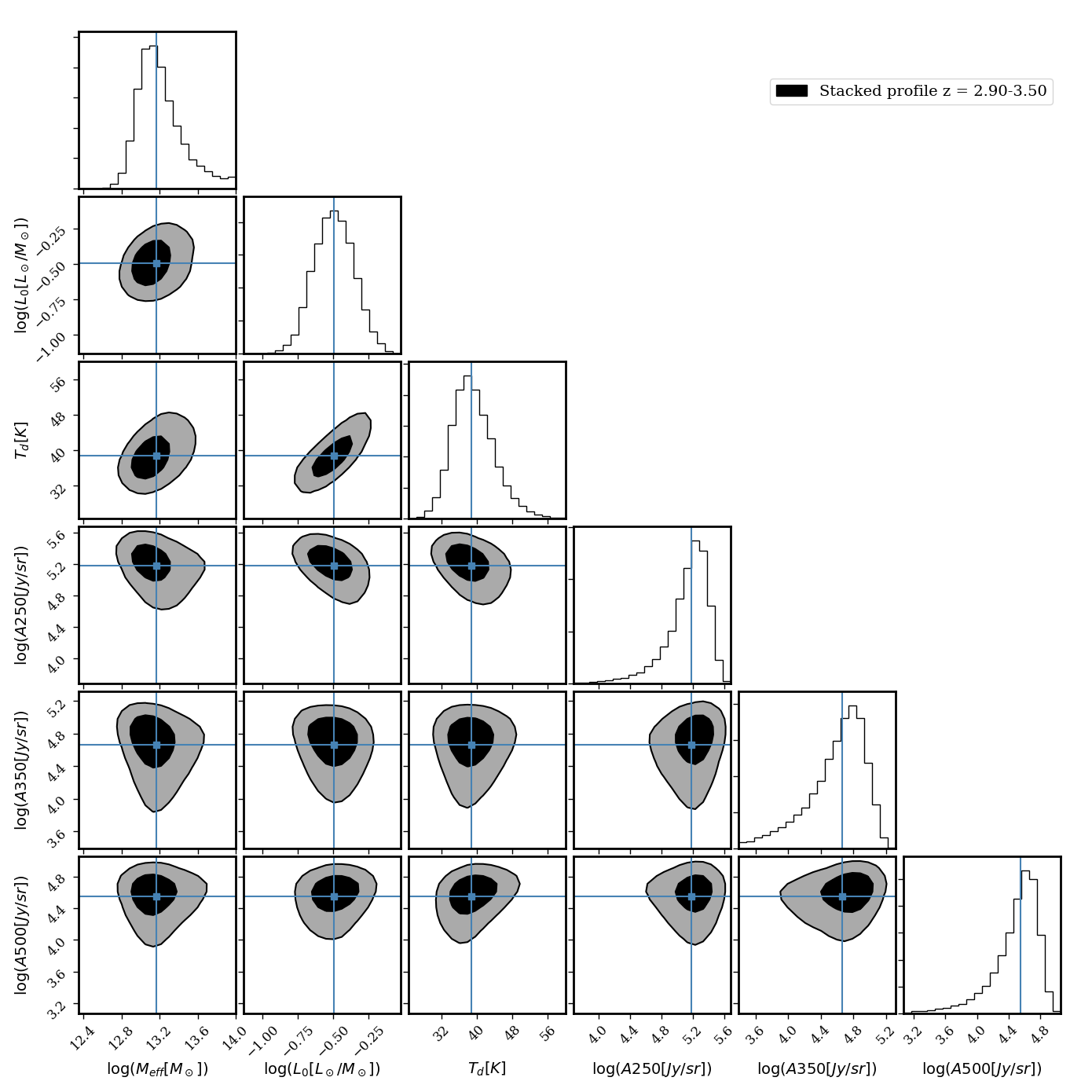}
\caption{Top: Angular cross correlation function and complete model, including the PSF (dashed line), one-halo term (small scales), and two-halo term (large scales), using marginalized best-fitting parameter results for quasars and DSFGs in the redshift range $z=2.90-3.50$. Bottom: Contours of the 68$^{th}$ and 95$^{th}$  percentiles of the parameters in our fiducial HOD model plus the three amplitudes fitting the quasar emission at each wavelength for the same range in redshift. The histograms show the marginalized posterior distributions, and the lines mark the median values. The results of the DSFG HOD parameters are sensitive to the parameters describing the quasar halo occupation function at $z>2.5$.}
\label{profz9}
\end{figure*}

\section{Parameter comparisons between the fiducial model and fits with no star formation rate density constraint}
\label{appendixB}

\begin{figure}
\includegraphics[width=3.25in]{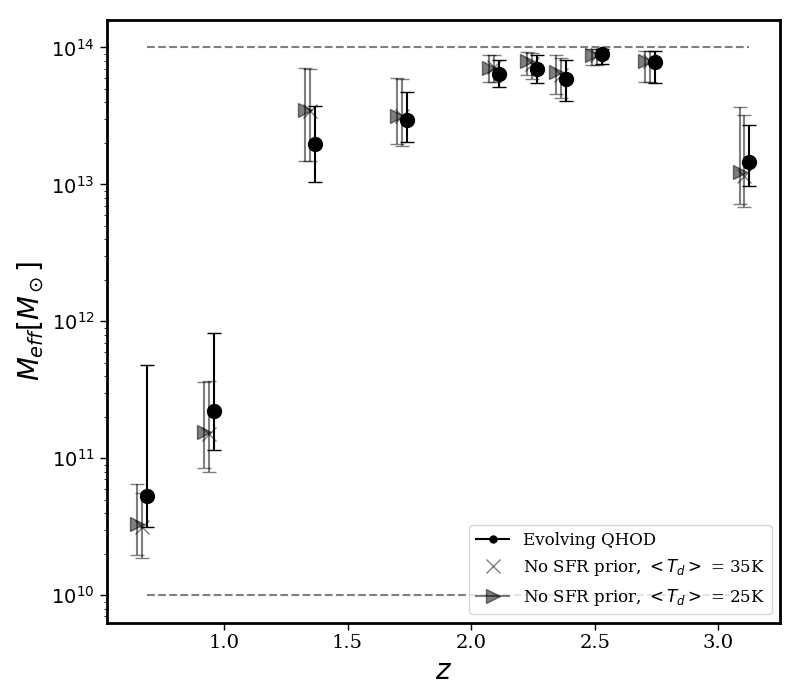}
\caption{Most efficient halo mass at hosting star formation as a function of redshift as determined from the $50^{th}$ percentile value from the posterior distribution of the Markov Chains. Error bars are $16^{th}$ and $84^{th}$ percentiles. Black points are the results from our fiducial model in which the quasar HOD parameters evolve at $z>2.5$. Gray triangles are the results of fitting for the parameters with no prior on star formation and a mean dust temperature prior of 25~K. Gray X's are the results of fitting for the parameters with no prior on star formation and a mean dust temperature prior of 35~K. \label{Meffz2}}
\end{figure}

\begin{figure}
\includegraphics[width=3.25in]{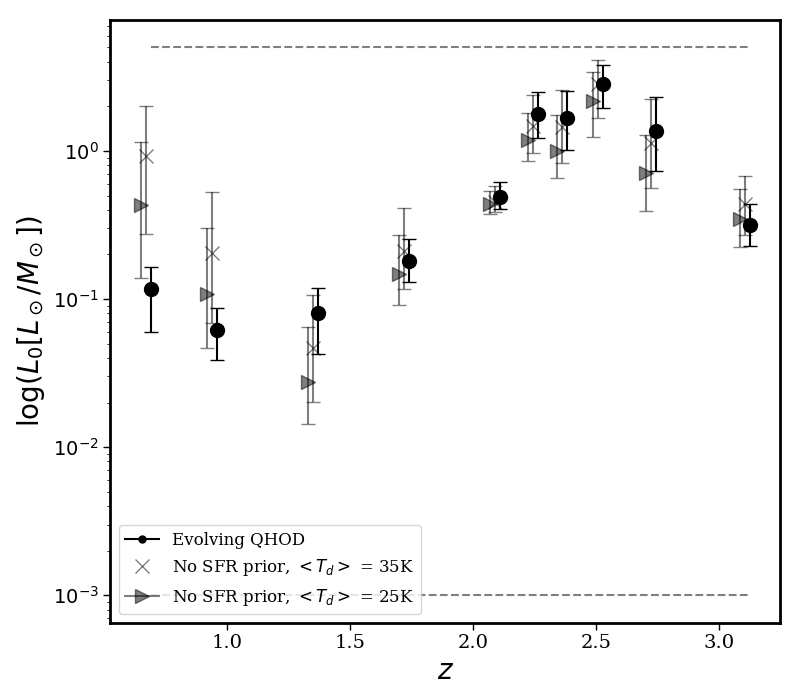}
\caption{Normalization of the $L-M$ relation for DSFGs as a function of redshift as determined from the $50^{th}$ percentile value from the posterior distribution of the Markov chains. Error bars are $16^{th}$ and $84^{th}$ percentiles. Black points are the results from our fiducial model in which the quasar HOD parameters evolve at $z>2.5$. Gray triangles are the results of fitting for the parameters with no prior on star formation and a mean dust temperature prior of 25~K. Gray X's are the results of fitting for the parameters with no prior on star formation and a mean dust temperature prior of 35~K.  \label{L0z2}}
\end{figure}

\begin{figure}
\includegraphics[width=3.25in]{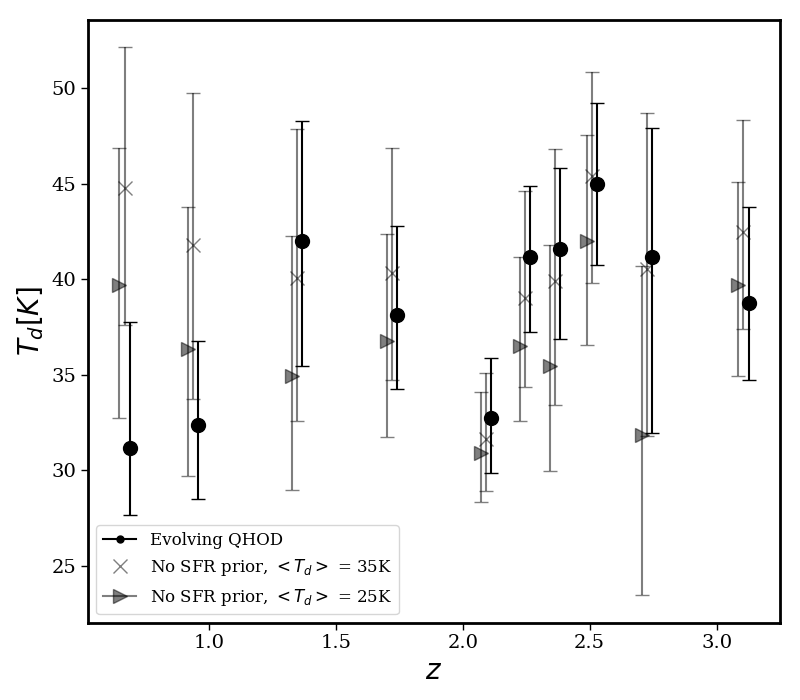}
\caption{Dust temperature of DSFG SEDs as a function of redshift as determined from the $50^{th}$ percentile value from the posterior distribution of the Markov chains. Error bars are $16^{th}$ and $84^{th}$ percentiles. Black points are the results from our fiducial model in which the quasar HOD parameters evolve at $z>2.5$. Gray triangles are the results of fitting for the parameters with no prior on star formation and a mean dust temperature prior of 25~K. Gray X's are the results of fitting for the parameters with no prior on star formation and a mean dust temperature prior of 35~K. \label{Tdz2}}
\end{figure}

\begin{figure}
\includegraphics[width=3.25in]{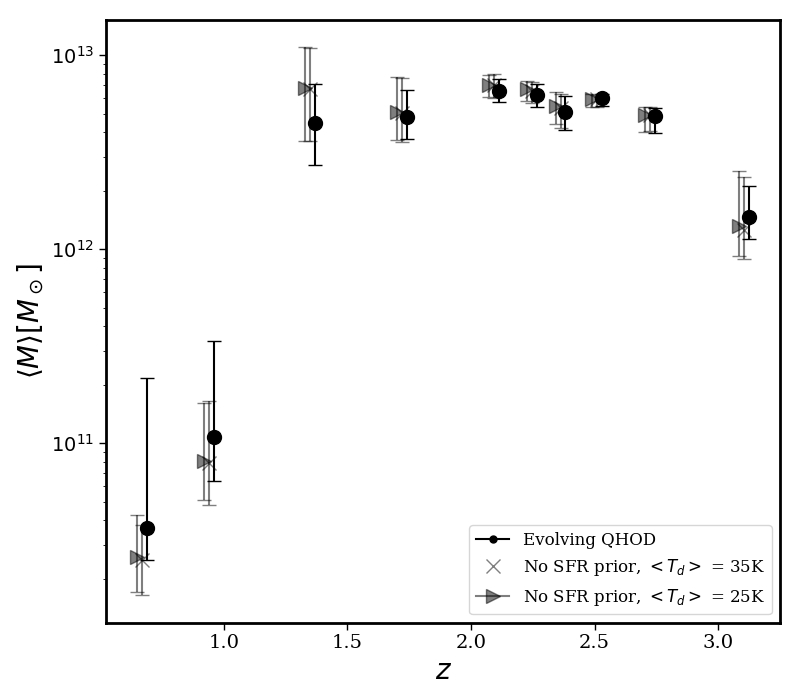}
\caption{Average halo mass of DSFGs as a function of redshift as determined from Equation~(\ref{Mmean}). Error bars are $16^{th}$ and $84^{th}$ percentiles. Black points are the results from our fiducial model in which the quasar HOD parameters evolve at $z>2.5$. Gray triangles are the results of fitting for the parameters with no prior on star formation and a mean dust temperature prior of 25~K. Gray X's are the results of fitting for the parameters with no prior on star formation and a mean dust temperature prior of 35~K. \label{Mmeanz2}}
\end{figure}

\begin{figure}
\includegraphics[width=3.25in]{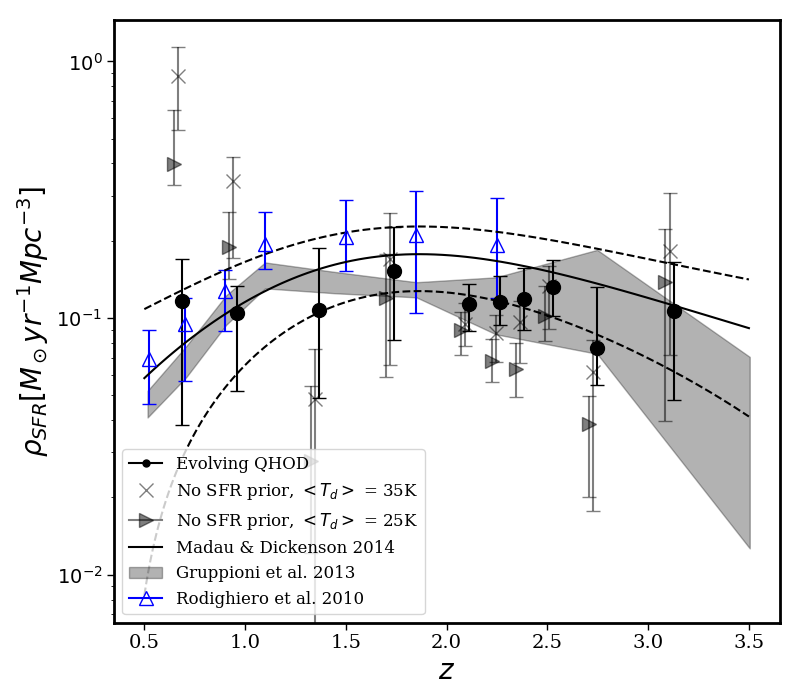}
\caption{Cosmic star formation rate density as a function of redshift as computed from the results of our halo model. Error bars are $16^{th}$ and $84^{th}$ percentiles. Black points are the results from our fiducial model in which the quasar HOD parameters evolve at $z>2.5$. Gray triangles are the results of fitting for the parameters with no prior on star formation and a mean dust temperature prior of 25~K. Gray X's are the results of fitting for the parameters with no prior on star formation and a mean dust temperature prior of 35~K. Also shown are the derived cosmic star formation rate densities from \citet{rodi10}, \citet{grup13}, and \citet{mada14} with which the derived values agree within $\sim 1 \sigma$ uncertainty with the exception of the second and last redshift bins. The dashed lines indicate the 1$\sigma$ boundaries of the Gaussian prior. \label{rhoSFR2}}
\end{figure}

This section contains the comparisons between the results of the fiducial model in which we put a Gaussian prior on $\rho_{SFR}$ when fitting for our model HOD parameters vs. when we put no prior on $\rho_{SFR}$ but alter the mean of the Gaussian prior on the dust temperature. 
The standard deviation of the Gaussian prior on $T_d$ is 10~K as in the fiducial model.
In Figures~\ref{Meffz2}-\ref{rhoSFR2} the black dots indicate the results from the fiducial model with evolving quasar HOD parameters, the gray triangles indicate the results of fitting with a mean dust temperature prior of 25~K, and the gray X's indicate the results of fitting with a mean dust temperature prior of 35~K.
Motivation for these fits is explained in Section~\ref{quasartests}.
The take away from these results is that imposing a Gaussian prior on $\rho_{SFR}$ has no effect on the primary result of the observed downsizing trend in the halo masses hosting DSFGs. 
Rather, the effect of imposing such a prior is to obtain more physical constraints on the dust temperature of the SEDs describing DSFGs.

\bsp	
\label{lastpage}
\end{document}